\documentclass{emulateapj}
\usepackage{amsmath}
\usepackage{changepage}
\usepackage{booktabs}
\usepackage{longtable}
\usepackage{multirow}

\shortauthors{Jun et al.}
\begin{document}
\title{Rest-frame Optical Spectra and Black Hole Masses of $3 < z < 6$ Quasars}
\author{Hyunsung David Jun\altaffilmark{1,2}, Myungshin Im\altaffilmark{2,3,14}, Hyung Mok Lee\altaffilmark{3}, 
Youichi Ohyama\altaffilmark{4}, Jong-Hak Woo\altaffilmark{3}, Xiaohui Fan\altaffilmark{5}, Tomotsugu Goto\altaffilmark{6}, Dohyeong Kim\altaffilmark{2,3}, Ji Hoon Kim\altaffilmark{2,7}, Minjin Kim\altaffilmark{8,9}, Myung Gyoon Lee\altaffilmark{3}, Takao Nakagawa\altaffilmark{10}, Chris Pearson\altaffilmark{11,12,13}, Stephen Serjeant\altaffilmark{11}}

\altaffiltext{1}{Jet Propulsion Laboratory, California Institute of Technology, 4800 Oak Grove Dr., Pasadena, CA 91109, USA; hyunsung.jun@jpl.nasa.gov}
\altaffiltext{2}{Center for the Exploration of the Origin of the Universe (CEOU), Astronomy Program, Department of Physics 
and Astronomy, Seoul National University, Seoul 151-742, Korea}
\altaffiltext{3}{Astronomy Program, Department of Physics and Astronomy, Seoul National University, Seoul 151-742, Korea}
\altaffiltext{4}{Academia Sinica, Institute of Astronomy and Astrophysics, P.O. Box 23-141, Taipei 10617, Taiwan, China}
\altaffiltext{5}{Steward Observatory, The University of Arizona, Tucson, AZ 85721, USA}
\altaffiltext{6}{Institute of Astronomy and Department of Physics, National Tsing Hua University, No. 101, Section 2, Kuang-Fu Road, Hsinchu, 30013, Taiwan, China}
\altaffiltext{7}{Subaru Telescope, National Astronomical Observatory of Japan, 650 North A'ohoku Place, Hilo, HI 96720, USA}
\altaffiltext{8}{Korea Astronomy and Space Science Institute, Daejeon 305-348, Korea}
\altaffiltext{9}{University of Science and Technology, Daejeon 305-350, Korea}
\altaffiltext{10}{Institute of Space and Astronautical Science, Japan Aerospace Exploration Agency, Sagamihara, Kanagawa 252-5210, Japan}
\altaffiltext{11}{Department of Physical Sciences, The Open University, Milton Keynes, MK7 6AA, UK}
\altaffiltext{12}{RAL Space, CCLRC Rutherford Appleton Laboratory, Chilton, Didcot, Oxfordshire OX11 0QX, UK}
\altaffiltext{13}{Oxford Astrophysics, Denys Wilkinson Building, University of Oxford, Keble Rd, Oxford OX1 3RH, UK}
\altaffiltext{14}{Author to whom correspondence should be addressed; mim@astro.snu.ac.kr}

\begin{abstract}
 We present the rest-frame optical spectral properties of 155 luminous quasars at 3.3\,$<$\,$z$\,$<$\,6.4 taken with the $AKARI$ 
space telescope, including the first detection of H$\alpha$ emission line as far out as $z$\,$\sim$\,6. We extend the scaling relation between the rest-frame optical continuum and line luminosity of active galactic nuclei (AGNs) to the high luminosity, high redshift regime that has rarely been probed before. Remarkably, we find that a single log-linear relation can be applied to the 5100\,$\text{\AA}$ and H$\alpha$ AGN luminosities over a wide range of luminosity (10$^{42}$\,$<$\,$L_{5100}$\,$<$\,10$^{47}$\,ergs\,s$^{-1}$) or redshift (0\,$<$\,$z$\,$<$\,6), suggesting that the physical mechanism governing this relation is unchanged from $z$\,=\,0 to 6, over five decades in luminosity. Similar scaling relations are found between the optical and the UV continuum luminosities or line widths. Applying the scaling relations to the H$\beta$ black hole mass ($M_{\rm BH}$) estimator of local AGNs, we derive the $M_{\rm BH}$ estimators based on H$\alpha$, {\ion{Mg}{2}}, and {\ion{C}{4}} lines, finding that the UV-line based masses are overall consistent with the Balmer-line based, but with a large intrinsic scatter of 0.40\,dex for the {\ion{C}{4}} estimates. Our 43 $M_{\rm BH}$ estimates from H$\alpha$ confirm the existence of BHs as massive as $\sim$\,10$^{10}M_{\odot}$ out to $z$\,$\sim$\,5, and provide a secure footing for previous {\ion{Mg}{2}}-line based studies that a rapid $M_{\rm BH}$ growth has occurred in the early universe. 
\end{abstract}

\section{Introduction}
Quasars, galaxies that are in an active phase due to vigorous accretion of matter toward the central supermassive black hole (BH), has been vastly discovered by many surveys (e.g., \citealt{Sch83}; \citealt{Hew95}; \citealt{Boy00}; \citealt{Yor00}; \citealt{Ric02}; \citealt{Im07}; \citealt{Lee08}; \citealt{Wil10b}; \citealt{Wu10}). Through the discovery of quasars at high redshift (\citealt{Fan00}; \citealt{Coo06}; \citealt{Got06}; \citealt{Ste07}; \citealt{Wil07}; \citealt{Mor11}; \citealt{Ven13}; \citealt{Ban14}), we are witnessing the early stages of supermassive BH growth in the distant universe. The number density of optically luminous quasars at high redshift, quickly increases with cosmic time towards its maximum at $z$\,=\,2\,$\sim$\,3 (\citealt{Dun90}; \citealt{War94}; \citealt{Sch95}; \citealt{Ken95}; \citealt{Ric06b}; \citealt{McG13}).
Accompanied by high accretion rates among luminous quasars at $z$\,$>$\,4 (e.g., \citealt{Wil10a}; \citealt{DeR11}; \citealt{DeR14}), this suggests a rapid BH growth of the luminous population of active galactic nuclei (AGNs) in the early universe. Also, an unusual population of AGNs known as dust-poor quasars -- quasars with little infrared emission from hot and warm dust -- is found to be more common at higher redshift (\citealt{Jia10}; \citealt{Jun13}; \citealt{Lei14}). Luminous dust-poor quasars tend to have lower BH masses ($M_{\rm BH}$) or higher Eddington ratios compared to typical luminous quasars (\citealt{Jia10}; \citealt{Jun13}), indicating the build-up of the AGN dusty substructure during its early mass accretion. 

One of the key findings in the study of high redshift quasars is that there exist extremely massive BHs, with the mass reaching $M_{\mathrm{BH}} \sim 10^{10} M_{\odot}$ at $z$\,=\,2--5, and $\sim$\,10$^{9} M_{\odot}$ at $z$\,=6--7 (\citealt{Jia07}; \citealt{Net07}; \citealt{Kur07}; \citealt{Ves08}; \citealt{She08}; \citealt{Mor11}; \citealt{DeR14}). Under the concordance cosmology, the time gap between the reionization epoch of the universe from the recent $Planck$ study of the cosmic microwave background, $z\sim11.5$ \citep{Pla14}, and $z$\,=\,6 is 0.5 Gyr. Considering the case where a Population II stellar seed $M_{\rm BH}$ starts to grow at $z\sim11.5$, the given time is too short for the seed to become an extremely massive BH at $z$\,=\,6.
Under the Eddington-limited accretion where the mass accretes at a maximal rate with the radiative pressure gradient and gravity in balance, we expect a BH to grow as 
\begin{equation}M(t)=M_{0}\,\exp \Big(\frac{1-\epsilon}{\epsilon}\frac{t-t_{0}}{t_{\mathrm{Edd}}}\Big),\end{equation} 
where $M_{0}$ is the mass at $t_{0}$, $\epsilon$ is the radiative efficiency, and $t_{\mathrm{Edd}}$ is the Eddington limited timescale of 0.45 Gyr. Assuming a typical value of $\epsilon$\,=\,0.1, a BH can grow by 2$\times$10${^{4}}$ times over the time span of 0.5 Gyr, without considering feedback mechanisms that could slow down the BH growth. This maximal growth factor is far too small for a stellar mass BH with a typical seed mass of 10\,$M_{\odot}$, to grow into the extremely massive AGNs that have been observed recently. Consequently, BH seeds that may have started accreting prior to the reionization epoch, which are more massive (e.g., \citealt{Bro03}; \citealt{Beg06}; \citealt{Lod06}; \citealt{Bel11}) or go through super-Eddington accretion (e.g., \citealt{Vol05}; \citealt{Wyi12}; \citealt{Mad14}), are suggested to explain the $M_{\rm BH}$ of quasars at high redshift (also see a review on this subject by \citealt{Nat14}).

Obviously, accurate determination of $M_{\rm BH}$ is an important requirement for understanding the BH growth at high redshift. This is especially true for BHs at the most massive end. A large uncertainty in $M_{\rm BH}$ can scatter the abundant lower mass BHs into the high mass end of $M_{\rm BH}$ distribution, while the effect in the opposite direction is much less significant since higher mass BHs are relatively rare. As a result, the number of extremely massive BHs can be easily overestimated. In principle, the correction to this effect is possible, but it requires a good knowledge on the error of $M_{\rm BH}$ measurements which is, however, rather difficult to obtain. This poses a potential challenge to the understanding of the BH growth at high redshift as described below. 

In most of high redshift quasar studies, BH masses are estimated using the rest-frame ultraviolet (UV)-part of spectra which is redshifted into the rest-frame optical (rest-optical). The velocity widths of broad UV lines such as {\ion{C}{4}} and {\ion{Mg}{2}} are used as measures of the gas motion of the broad line region (BLR), and the continuum or the line luminosity at the rest-frame UV (rest-UV) is used as a proxy for the size of BLR ($R_{\mathrm{BLR}}$). One gets $M_{\rm BH}$ by combining the two pieces of information through a virial mass estimator, $M_{\mathrm{BH}} \propto R_{\mathrm{BLR}} \times \text{FWHM}_{\mathrm{BLR}}^{2}$ (e.g., \citealt{Mcl04}; \citealt{Ves04}; \citealt{Bas05}; \citealt{Sul07}; \citealt{She08}; \citealt{Par13}). While UV-based $M_{\rm BH}$ estimators are useful tools to measure the $M_{\rm BH}$ of AGNs, they are secondary estimators that are derived from rest-optical spectral properties such as the line luminosity or width of H$\beta$ and H$\alpha$, and the continuum luminosity at 5100\,$\text{\AA}$ ($L_{5100}$). Assuming the UV luminosity follows the optical broad line region radius--luminosity ($R_{\mathrm{BLR}}$--$L$) relation with a constant factor, and the UV broad line width follows the optical line width as a power law relation, the UV $M_{\rm BH}$ estimators are derived. Consequently, a number of studies have been carried out to justify the use of {\ion{Mg}{2}} or {\ion{C}{4}}-line based $M_{\rm BH}$ estimators comparing the masses from UV estimators to those from the optical. While some studies suggest that UV-line $M_{\rm BH}$ estimators are reasonably accurate, especially for {\ion{Mg}{2}}, other studies point out a large scatter between {\ion{C}{4}}-based measurements versus H$\beta$-based measurements which can make the {\ion{C}{4}}-based $M_{\rm BH}$ values uncertain by a factor of a few (e.g., \citealt{Net07}; \citealt{She12}, hereafter S12). It has been noted that non-virialized motion (e.g., \citealt{Den12}) or extinction (e.g., \citealt{Ass11}), could severely modify the {\ion{C}{4}} line profile such that the $M_{\rm BH}$ cannot be reliably measured from a simple virial equation, although it is controversial on the exact origin of the discrepant UV-based $M_{\rm BH}$ with that of optical. 
  
Furthermore, for the $M_{\rm BH}$ estimates to be valid, one also needs to justify the application of the low redshift $M_{\rm BH}$ estimators to high redshift, luminous quasars. Although one can expect that the $M_{\rm BH}$ estimators should not evolve in time in any significant way based on physical ingredients of AGN models, this has not been tested at the high redshift, high luminosity regime. An ultimate test would be to perform a reverberation mapping study of high luminosity quasars at high redshift, but such a study would take decades to complete, since the variability timescale is long for luminous quasars and the cosmological time dilation makes it even longer. Another way, albeit less direct than the reverberation mapping method, would be to investigate the correlation between the line and continuum luminosities. At low redshift, the H$\alpha$ or H$\beta$ line luminosities are known to tightly correlate with the optical continuum luminosity (e.g., \citealt{Gre05}). As the radiation energy $L$ from an accretion disk increases, the distance to the broad line region increases as $R\propto L^{0.5}$ from a simple photo-ionization argument, or the energy flux of the radiation incident upon BLR clouds would be independent of the luminosity of the central power source, which has been confirmed observationally (\citealt{Kas00}; \citealt{Kas05}; \citealt{Ben13}). A modification in the $R_{\mathrm{BLR}}$--$L$ relation at high redshift or high luminosity for example, would result in an increase or a decrease in the incident energy flux upon the BLR, thus the correlation between the line and continuum luminosities is likely to be modified. Interestingly, several studies suggest that the $R_{\rm BLR}-L$ relation is not valid for luminous AGNs having massive black holes with low spins (\citealt{Lao11}; \citealt{Wan14}), due to a decrease in the ionizing flux $L_{\rm ion}$ caused by a decrease in the radiation temperature of the accretion disk. Since the line luminosity $L_{\rm line}$ is proportional to $L_{\rm ion}$, such theoretical expectations can be tested by examining the $L_{\rm line}$ -- $L_{\rm 5100}$ relation at luminous end.

In order to estimate the $M_{\rm BH}$ using an optical mass estimator and to test the universality of the scaling relations in the rest-optical for quasars at $z$\,$>$\,3.5, spectroscopic observation is necessary at $\lambda$\,$>$\,2.5\,$\mu$m. This however, is a very challenging task from the ground due to high thermal background at $\lambda$\,$>$\,2.5\,$\mu$m and atmospheric absorptions, limiting such efforts to the study of H$\beta$ line at $z$\,$<$\,3.5 (e.g., \citealt{She04}; \citealt{Net07}; \citealt{Ass11}). Recently, $AKARI$ spectroscopic observations have provided a breakthrough for the study of the rest-optical spectra of distant objects, where its unique 2.5--5.0\,$\mu$m coverage enables the redshifted H$\alpha$ line to be probed from $z$\,=\,3 to 6.5. \citet{Oya07} reported the $AKARI$ detection of the redshifted H$\alpha$ line of a quasar RX J1759.4+6638 at $z$\,=\,4.3, and \citet{Sed13} investigated the star formation rate of submm galaxies at $z$\,$>$\,3.5 based on H$\alpha$. 

With one of the $AKARI$ mission programs (guaranteed time) and also through several small open time programs, we performed a spectroscopic study of quasars at high and low redshifts, with the aim to obtain the rest-optical spectra of high redshift quasars or the rest-frame 2.5--5.0\,$\mu$m spectra of low redshift AGNs. We call all these programs QSONG (Quasar Spectroscopic Observations with NIR Grism) after the name of the mission program, and here, we present the rest-optical spectral properties of 155 type-1 quasars from QSONG, along with scaling relations and $M_{\rm BH}$ estimates based on these spectra.   
    
The contents of the paper are constructed as follows. First, we use the spectro-photometric data including $AKARI$ for $z\gtrsim3$ AGNs (section 2), and derive the continuum and line emission properties (section 3) in order to calibrate the H$\alpha$ $M_{\rm BH}$ for the usage in high redshift. We check the validity of the continuum and broad line luminosity relations and line width relations with respect to the local, update the mass equations for H$\alpha$, H$\beta$, {\ion{Mg}{2}}, and {\ion{C}{4}}, and compare the UV--optical $M_{\rm BH}$'s (section 4). Finally, we discuss on the reliability of single epoch $M_{\rm BH}$ estimators at high redshift, and investigate the massive end evolution of $M_{\rm BH}$ in distant AGNs (section 5). Throughout this paper we adopt a flat $\Lambda$CDM cosmology with parameters of $H_{0}=\mathrm{70\,km\,s^{-1}\,Mpc^{-1}}$, $\Omega_{m}=0.3$, and $\Omega_{\Lambda}=0.7$. For the virial factor in the $M_{\rm BH}$ estimator, we adopt $f=5.1\pm1.3$ based on the calibration of the $M_{\rm BH}$--$\sigma$ relation using the combined sample of AGNs and quiescent galaxies by \citet{Woo13}.

\section{Data}
\subsection{Sample}
The majority of the data comes from the $AKARI$ program QSONG, which is a two-year warm (phase--3) mission program consisting of $\sim$\,900 approved pointings, or $\sim$\,150 hours of observations. The program is aimed to obtain the rest-optical spectra of high redshift AGNs \citep{Jun12} or 2.5--5.0\,$\mu$m spectra of low redshift AGNs \citep{Kim15} containing Brackett and 3.3\,$\mu$m Polycyclic aromatic hydrocarbon lines. The sample for this study is limited to the high redshift, occupying 65\% of the entire QSONG data. It is composed of optically luminous and spectroscopically confirmed type-1 quasars at $z$\,$\gtrsim$\,3, mostly out of SDSS (DR5 catalog, \citealt{Sch07}; and additional discoveries from \citealt{Fan00}; \citealt{Fan01}; \citealt{Fan03}; \citealt{Fan04}; \citealt{Fan06}) and APM-UKST (\citealt{Sto96}; \citealt{Sto01}) surveys. Additional targets are from various references (\citealt{Web88}; \citealt{Gre91}; \citealt{Hen94}; \citealt{Gri95}; \citealt{Ken95}; \citealt{Djo98}; \citealt{Ren04}; \citealt{McG06}).  

\begin{figure}
\centering
\includegraphics[scale=.9]{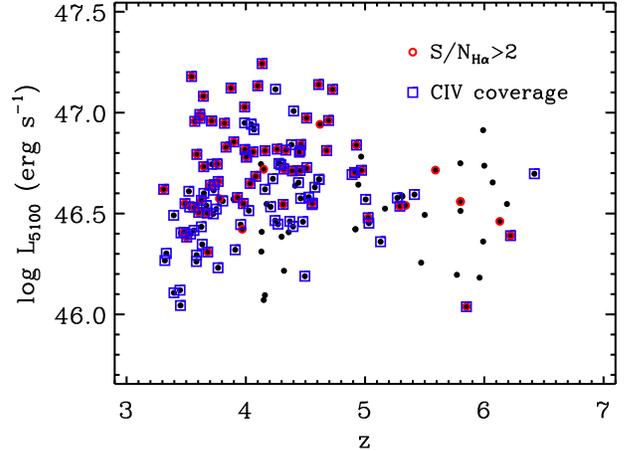}
\caption{5100\,$\text{\AA}$ luminosity--redshift distribution of our $AKARI$ observed quasars (black dots). The subsample with S/N$_{\mathrm{H}\alpha}$\,$>$\,2 are pointed in red, and objects with rest-UV spectral coverage including the {\ion{C}{4}} emission are marked as blue squares.} 
\end{figure}

The targets are type-1 AGNs so that they allow the $M_{\rm BH}$ estimation from the broad line kinematics, through H$\alpha$/H$\beta$ appearing within the $AKARI$ near-IR (NIR) spectral coverage. In order to provide a minimal sensitivity limit to the sample, we first considered the aperture size (68\,cm) of the telescope and the restricted exposure time available at each sky position. After simulating the rest-optical spectra under the expected $AKARI$ signal-to-noise ratio (S/N), the targets were chosen with $z$-band flux limits of $\sim$18.5 and 19 AB magnitudes for bright and faint subsamples, respectively, with longer exposure time assigned to fainter targets for a clear line detection. Moreover, the targets were bounded in 3.3\,$<$\,$z$\,$<$\,6.4 so that the H$\alpha$ emission and the surrounding continuum are placed within the 2.5--5.0\,$\mu$m window of $AKARI$ NIR spectroscopy. Without further constraints the targets were randomly selected in coordinates, redshift, and luminosity. We plot the distribution of $z$--$L_{5100\text{\AA}}$ in Figure 1.

Following the H$\alpha$ observations of distant AGNs from $AKARI$ NIR spectroscopy (\citealt{Oya07}; \citealt{Oya09}), our initial H$\alpha$ survey of 14 quasars at $z$\,$\sim$\,6 under the Helium-cooled (phase--2) open time program HZQSO \citep{Im10}, demonstrated the feasibility of $AKARI$ observations in detecting the redshifted H$\alpha$ emission. QSONG is essentially a phase--3 extension of the survey, for the purpose of vastly expanding the number of targets at the expense of warm phase sensitivity. Thus it probes a lower redshift distribution of quasars than HZQSO, with a peak at $z$\,$\sim$\,4. In addition, two more phase--3 open time programs, HQSO2 and DPQSO, were carried out either to push the redshift limit of QSONG or to detect fainter optical lines (H$\beta$ and [{\ion{O}{3}}]), from deep exposures. Unfortunately, the Helium-dry observations led to significantly higher noise levels than expected, restricting the distinct scientific goals of the phase--3 programs that required better sensitivity. Therefore, we decided to merge all open time programs listed above under the scope of QSONG.

\subsection{Data acquisition}
\begin{figure*}
\begin{center}$
\hspace*{0.15cm}
\begin{array}{ccc}
\vspace*{-0.3cm} \hspace*{-0.6cm}
\includegraphics[scale=.215,trim=291.2 0 0 0,angle=90,clip=true]{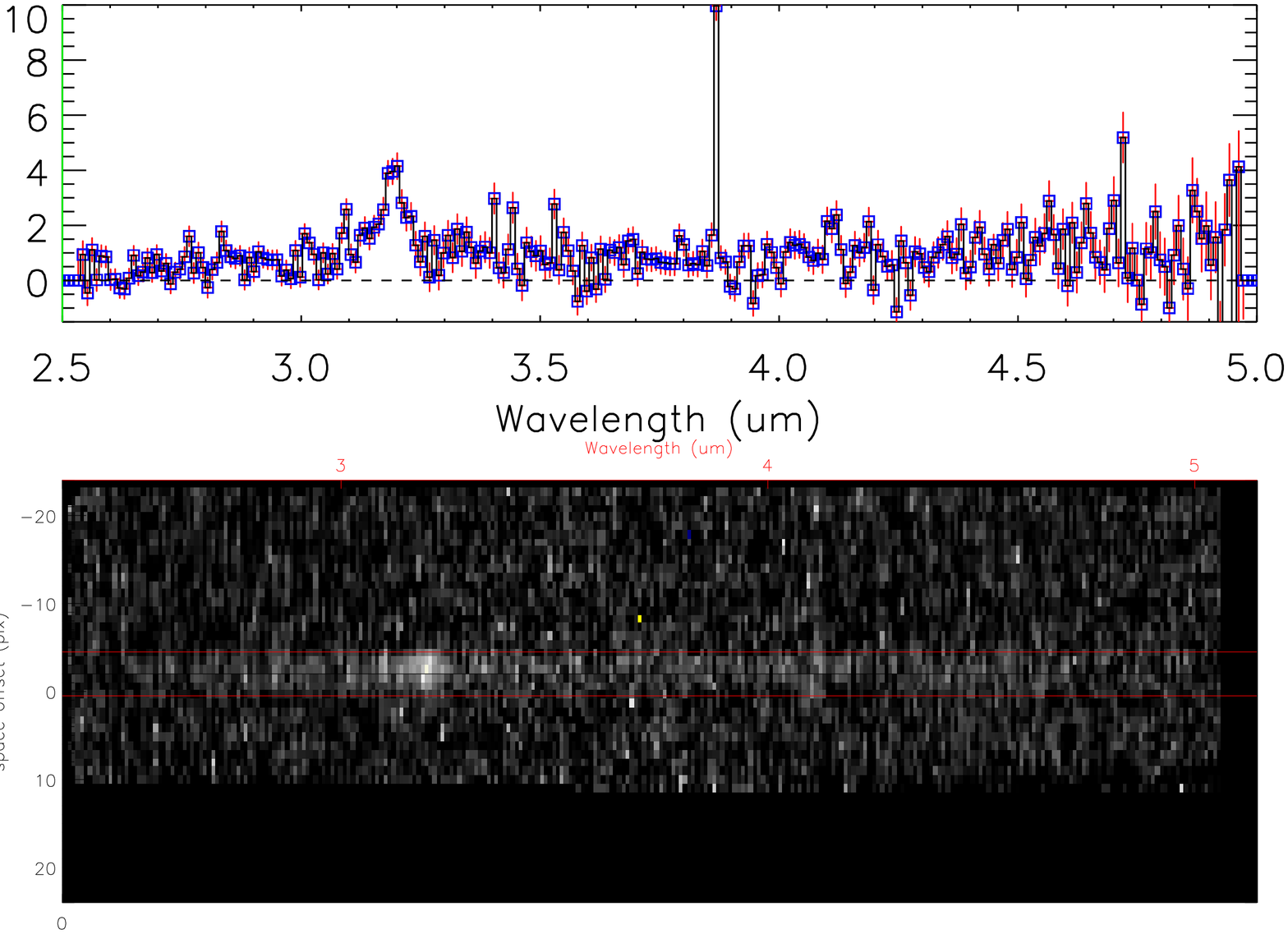}
&\hspace*{-1.75cm}\includegraphics[scale=.215,trim=291.2 0 0 0,angle=90,clip=true]{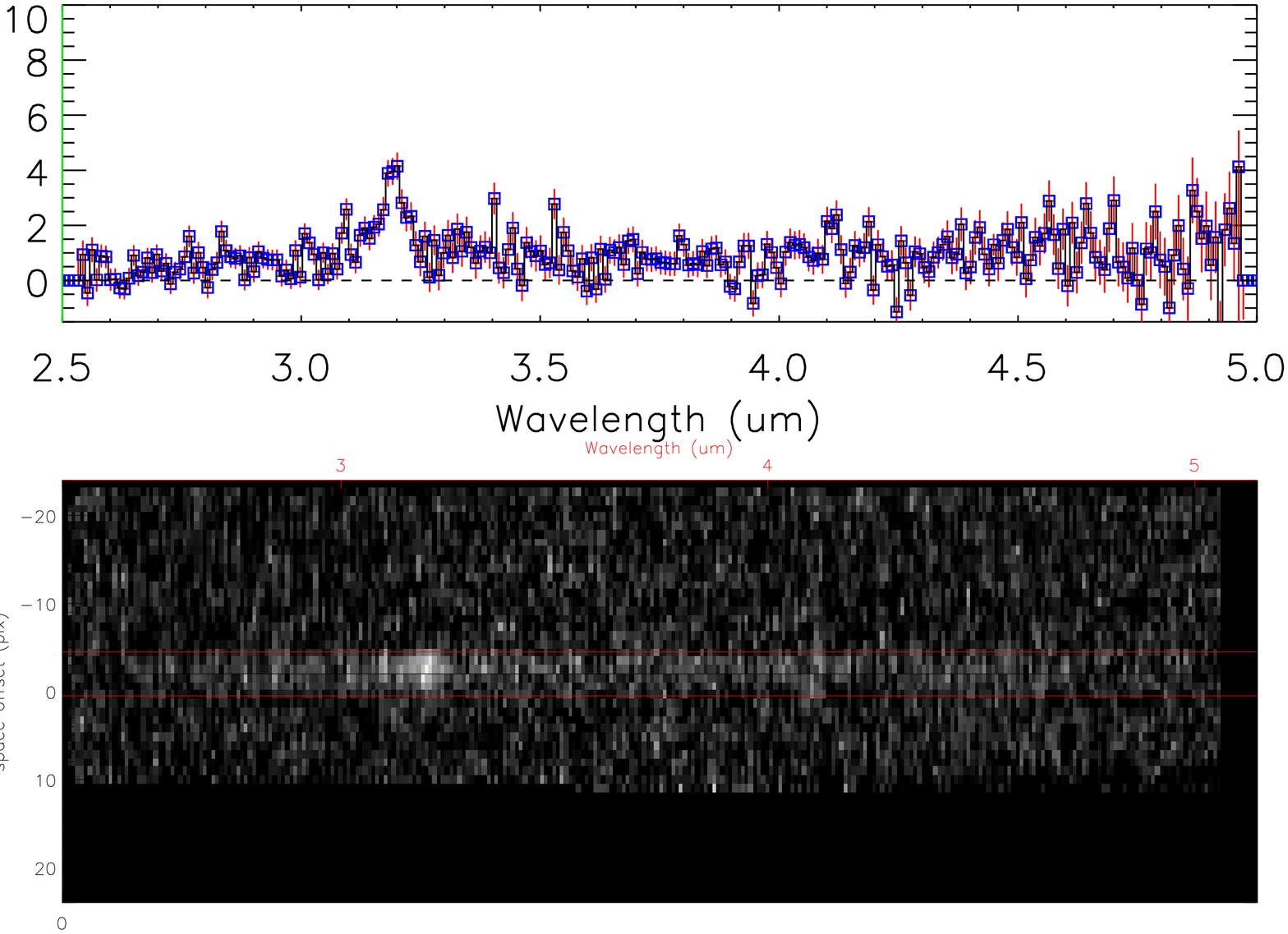}\hspace*{2.6cm}
&\vspace*{-1.75cm}\\\hspace*{.15cm}
\includegraphics[scale=.241,trim=0 0 325 90,angle=90,clip=true]{f2-1.eps}
&\hspace*{-0.6cm}\includegraphics[scale=.241,trim=0 0 325 90,angle=90,clip=true]{f2-2.eps}\hspace*{3cm}
&\vspace*{-0.55cm}\hspace*{-3.75cm}\includegraphics[scale=.635,trim=0 -10 0 0]{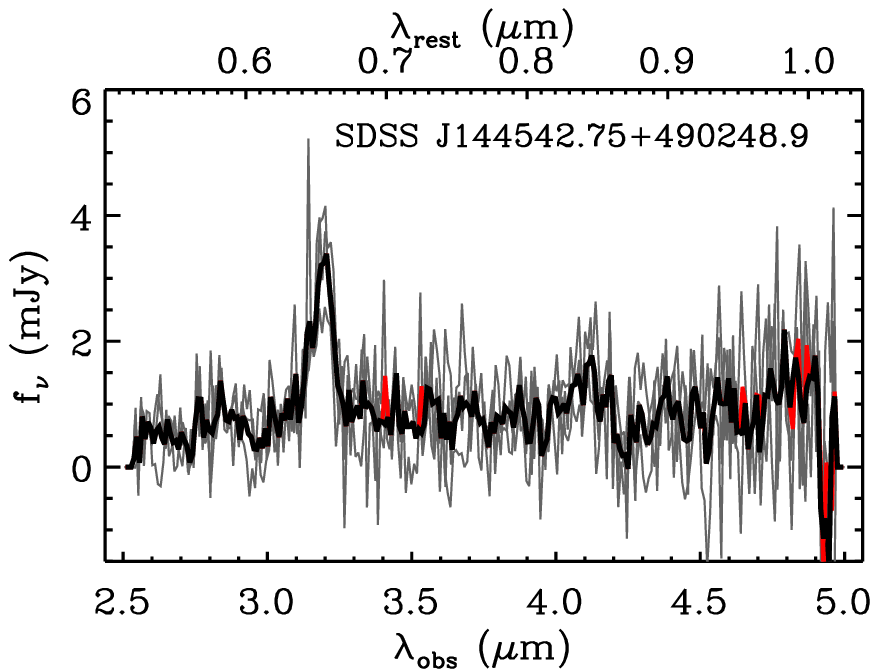}
\end{array}$
\end{center}
\caption{A sequential visualization of the data treatment additional to the pipeline processing. From the pipeline processed spectrum (left, 1D at top, and 2D at bottom), hot pixels were rejected (center) before the spectral extraction within the aperture mask (red lines). The multiple pointings extracted (right, gray lines) were stacked (thick black line) with sigma clipping (red line indicates clipped data).} 
\end{figure*}

\begin{deluxetable*}{ccccccc}
\vspace*{-0.7cm}
\tablecolumns{6}
\tablecaption{Summary of Observations}
\tablewidth{0.95\textwidth}
\tablehead{
\colhead{Program name} & \colhead{Phase} & \colhead{Observed period}  & \colhead{Mode} 
& \colhead{Number of targets} & \colhead{Observed pointings} & \colhead{Rejected pointings}}
\startdata
HZQSO & 2 & Nov 2006--Aug 2007 & NG & 7 & 26 & 2\\
		 &    & 							 & NP & 5 & 16 & 7\\
QSONG, HQSO2, DPQSO & 3 & Jun 2008--Jan 2010 & NG & 147 & 622 & 74\\
		 &   & 							                       & NP & 6 & 11 & 3
\enddata
\tablecomments{NG and NP stand for NIR grism and prism modes, while a pointing is about 10 minutes long. For the QSONG program data, only the high redshift subsample is noted. The total number of targets is 165 excluding 22 rejected sources from confusion, problems in the spectra or data reduction (section 2.3). 10 objects are either duplicated or different in observing mode or program, yielding an effective total of 155 independent objects.} 
\end{deluxetable*} 

We mostly used the NIR grism (NG) mode of the Infrared Camera (IRC, \citealt{Ona07}; \citealt{Ohy07}) onboard the $AKARI$ satellite \citep{Mur07}. It offers a low, wavelength dependent spectral resolution ($R$), where $R$\,=\,120 at 3.6\,$\mu$m. This corresponds to a velocity resolution of 2500\,km\,s$^{-1}$ in full width at half maximum (FWHM), sampled by a pixel scale of 0.0097\,$\mu$m in wavelength. The targets were placed in a 1$\arcmin \times 1\arcmin$ rectangular slit aperture to reduce source confusion. The wavelength dependence of $R$ can be expressed as $R$\,=\,120\,($\lambda/3.6\,\mu \text{m}$), since the dispersion is nearly a constant \citep{Sak12}. Meanwhile, a limited number of NIR prism (NP, $R$\,=\,19 at 3.5\,$\mu$m) observations were performed to better catch the fainter continuum and line luminosities. The angular pixel scale is 1.5$\arcsec$ such that all targets are point-like in our probed redshifts. 

The observations were performed under the Astronomical Observation Template (AOT) mode of AOT04, typical for spectroscopic observations. The number of NG pointings per target was normally 3--5 for QSONG, where one pointing observation corresponds to 355 or 400 sec on-source exposure. The number of pointings were determined based on the $z$-band flux, generally set to be smaller for phase--2 and larger for phase--3 open time programs. The NP observations were performed with usually 1--2 pointings. By the time of termination of the satellite mission, the QSONG program was 85\% complete with 144 high redshift quasars observed. The open time programs were complete before the satellite lifetime, adding another 33 targets. Table 1 summarizes the observations.

To supplement the NIR spectra, we compiled the optical spectra of the $AKARI$ quasars from the Sloan Digital Sky Survey (SDSS) database (DR10 including both the SDSS-I/SDSS-II and the SDSS-III BOSS data, \citealt{Ahn14}), and from observations of APM-UKST quasars (\citealt{Sto96}; \citealt{Per01}) and Q0000--26 \citep{Sch89}, in order to estimate the {\ion{C}{4}} line based $M_{\rm BH}$. Also, we collected broad-band photometric data from optical to mid-infrared (MIR) imaging, for the calculation of the rest-frame UV--optical continuum luminosity of AGNs through SED fitting (section 3.2). The data 
includes SDSS DR9, 2MASS PSC, UKIDSS DR10, and $WISE$ AllWISE releases (\citealt{Ahn12}; \citealt{Skr06}; \citealt{Law07}; \citealt{Wri10}), and existing Pan-STARRS, $Spitzer$, and $AKARI$ imaging (\citealt{Hin06}; \citealt{Jia06}; \citealt{Oya09}; \citealt{Jia10}, \citealt{Lei14}). The Galactic extinction is corrected for these spectro-photometric data, assuming the total-to-selective extinction ratio of R$_{V}$=3.1 and using the corrected form \citep{Bon00} to the extinction map values of \citet{Sch98}. The photometric measurement schemes are different in each survey data, such that host galaxy contamination may not be well subtracted. We keep the diverse magnitude type and imaging resolution of each survey data however, as the high luminosity AGNs yield a compatible set of magnitudes dominated by the central AGN contribution \citep{Jun13}. The optical spectra and the multi-wavelength imaging data are outlined in Table 2.

\subsection{Data reduction}
The data were reduced using the automated IDL pipeline package IRC\_\,SPECRED (versions 20110114, 20111121 for phase--2 and 3, \citealt{Ohy07}), where pre-processing (dark, linearity, flat corrections), image registration and coaddition, flux and wavelength calibration, astrometry, spectral extraction, and aperture correction were the main tasks involved. The standard pipeline configuration was adopted, except for the usage of short exposure when the image taken for registering subframes of spectral data were contaminated by saturated stars. This procedure considerably improved the registration of 2-D spectra in both spatial and wavelength directions. In addition, astrometry of the reference image was upgraded using the 2MASS point source catalog to better extract the faint NIR spectra. Indeed, the zeroth order positions of the spectra and the 2MASS coordinates were visually well aligned for spectral extraction.

Due to the increased number of hot pixels and the background level in the phase--3 data, a non-negligible number of bad pixels remained in the reduced spectral data even after applying the IRC\_\,SPECRED pipeline. To remove the remaining hot pixels, a further data reduction step was taken to obtain a cleaner set of spectra, as depicted in Figure 2. After running the spectroscopic pipeline we subtracted the remaining hot pixels using L.A.Cosmic \citep{van01} with a threshold of 2.5\,$\sigma$, and combined the 1-D spectra with a 2.5-sigma clipping. This threshold was chosen by visually inspecting the reduced spectra so that the chosen threshold removes the spiky hot pixels efficiently without affecting emission lines. Because the typical FWHM of the broad lines in our AGNs is broader than the spectral resolution of NG (2500\,km\,s$^{-1}$) but not that of NP (15,800\,km\,s$^{-1}$ at 3.5\,$\mu$m), the hot pixel rejection and the sigma clipping were applied on the NG data only, while the undersampled NP emission spectra were kept unchanged. We clipped 2.7\% and 2.9\% of the NG spectral data through the hot pixel reduction procedure and the combining process respectively. 
 
We extracted the 1-D spectra of 3 pixel width for the phase--2 data to maximize S/N, but the width was widened to 5 pixels for the phase--3 data since the noisier spectra made it difficult to determine the center of the object spectrum. Host galaxy contamination is negligible at the bright luminosities of the sample quasars (S11), enabling flexible extraction widths. Aperture corrections were automatically carried out from the pipeline to derive the total flux, for given respective extraction widths. The pipeline did a fair job of placing the extraction aperture on the right location, but visual inspection showed it necessary to make a $-$1 pixel shift in spatial direction for 90\% of the sample. For 10\% of the cases, a different shift of $-$2 to 1 pixels was necessary. The wavelength zeropoints were determined from the pipeline taking into account the satellite attitude drift and sub-pixel coordinate rounding effects, and we did not apply any further correction as the zeropoints were confined within a 0.5 pixel scatter.

\begin{deluxetable}{ccccc}
\tablecolumns{6}
\tablecaption{Supplementary Data}
\tablewidth{0.45\textwidth}
\tablehead{
\colhead{Name} & \colhead{Wavelength} & \colhead{N} & \colhead{Exposure}  & \colhead{Reference}}
\startdata
\sidehead{Spectra}
SDSS            & 3800--9200\,\AA   & 111 & $\geq$\,45m & 1\\
BOSS            & 3650--10400\,\AA & 98 & $\geq$\,45m & 1\\
APM--UKST & $\sim$3500--9000\,\AA   & 16 & 15--60m & 2,3\\
Hale            & 4500--9000\,\AA    &  1 & 30m & 4\\

\sidehead{Photometry}
SDSS     & $ugriz$ & 134 & 54s & 5\\
Pan-STARRS & $zy$ & 14 & 6m & 6\\
2MASS  & $JHK$     & 76 & 8s & 7\\ 
UKIDSS & $YJHK$    & 34 & 40s & 8\\
$WISE$     & $W1$--$W4$ & 161 & 200s & 9\\
$Spitzer$ & IRAC, MIPS 24$\mu$m & 27 & 17--23m & 6,10,11,12\\
$AKARI$  & IRC 2--11$\mu$m & 1 & -- & \,13
\enddata
\tablecomments{N is the number of matches to the 165 $AKARI$ objects in Table 1. Exposure times are typical values. The references numbered are 1. \citealt{Ahn14}; 2. \citealt{Sto96}; 3. \citealt{Per01}; 4. \citealt{Sch89}; 5. \citealt{Ahn12}; 6. \citealt{Lei14}; 7. \citealt{Skr06}; 8. \citealt{Law07}; 9. \citealt{Wri10}; 10. \citealt{Hin06}; 11. \citealt{Jia06}; 12. \citealt{Jia10}; 13. \citealt{Oya07}.} 
\end{deluxetable} 
\begin{figure}
\centering
\hspace*{-0.07cm}\includegraphics[scale=.8]{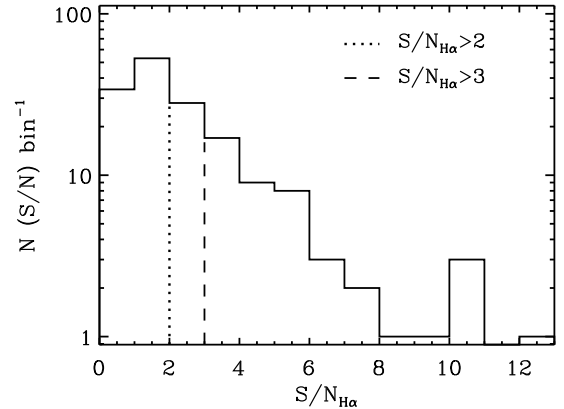}
\caption{The S/N$_{\mathrm{H}\alpha}$ distribution of the 160 $AKARI$ observed quasars with available S/N (section 3.1). The S/N$_{\mathrm{H}\alpha}$\,$>$\,2 and $>$\,3 cuts are marked in dotted and dashed lines, which are given to limit the measurement of $L_{\mathrm{H}\alpha}$ and FWHM$_{\mathrm{H}\alpha}$ respectively (Figure 5).}
\end{figure}
\begin{figure}
\centering
\includegraphics[scale=.485]{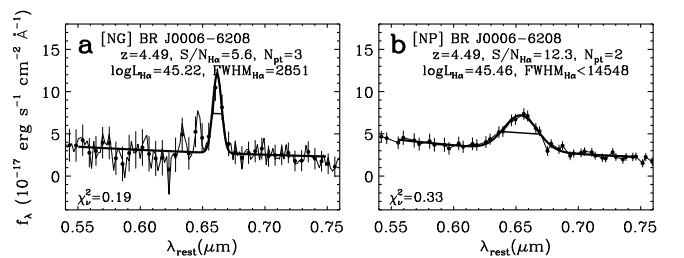}
\caption{Examples of the rest-frame H$\alpha$ emission fitting. On top of the spectra (thin line), the resolution (3 pixel) matched data (dots) and errors, the best-fit to the continuum and H$\alpha$ line emission (thick line), and FWHM are indicated. On the figures, the observation mode, the name of object, the redshift and S/N of the H$\alpha$ emission, the number of $AKARI$ pointings, N$_{\mathrm{pt}}$, H$\alpha$ luminosity (ergs\,s$^{-1}$) and FWHM (km\,s$^{-1}$) are printed. The (a) NG and (b) NP observations performed for BR J0006--6208 gives an idea of the enhanced resolution and sensitivity of each spectroscopic mode.} 
\end{figure}
\begin{figure*}
\centering
\hspace*{-0.02cm}\includegraphics[scale=.675]{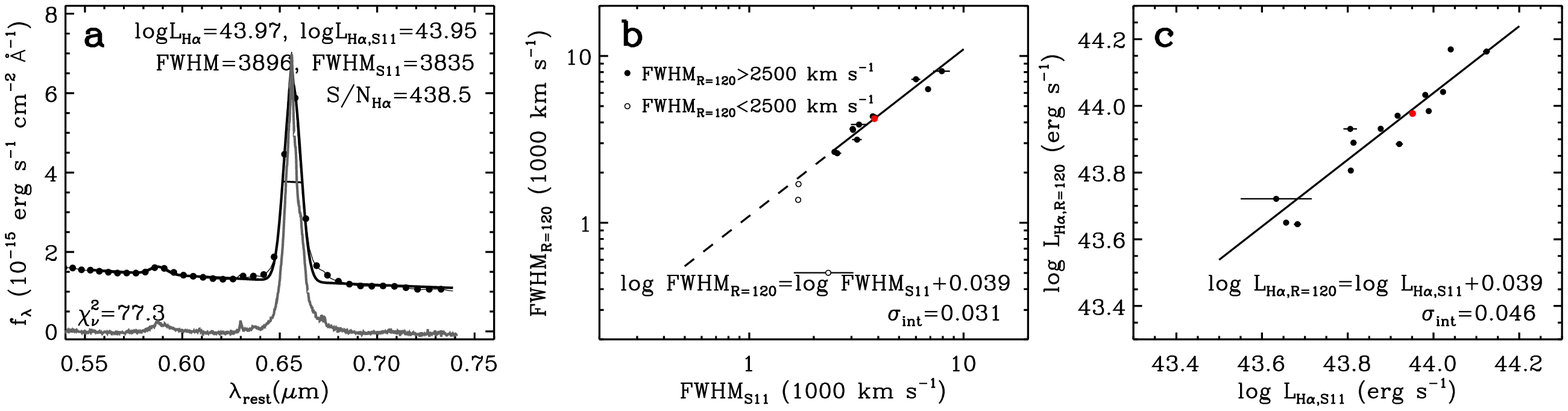}\vspace*{0.2cm}
\hspace*{-0.02cm}\includegraphics[scale=.675]{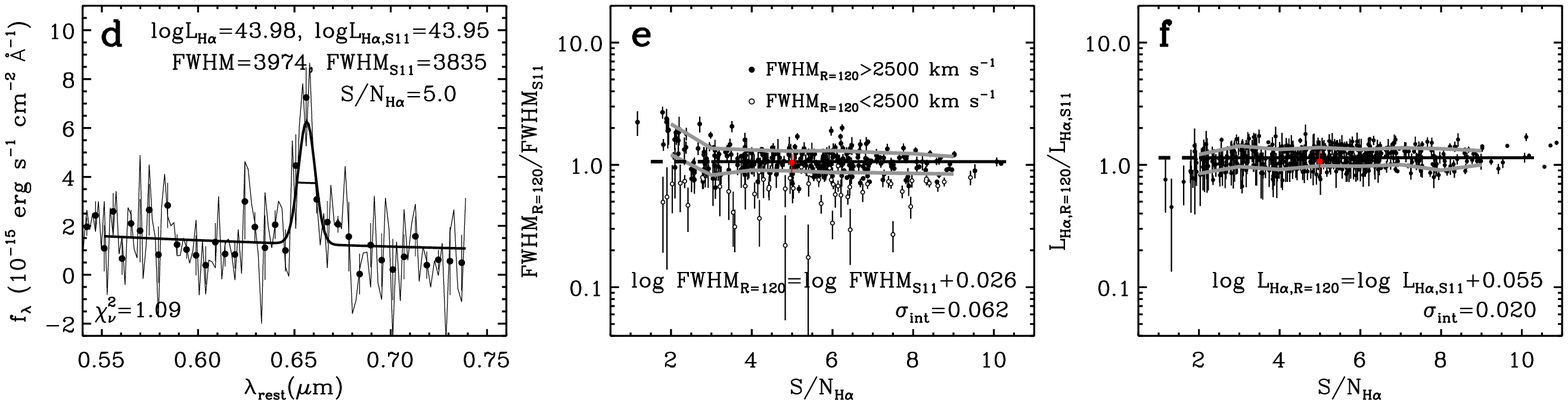}
\caption{Top rows: An example fit to the resolution degraded SDSS spectrum to test the reliability of $M_{\rm BH}$ from $AKARI$ observations (left), where the continuum subtracted original SDSS spectrum (gray), and the $AKARI$ resolution matched spectrum (black dots) with fit to the simulated data (thick line) are plotted. The broad FWHM$_{\mathrm{H}\alpha}$ values of $z$\,$<$\,0.3, $L_{5100}$\,$>$\,10$^{45}$\,ergs s$^{-1}$ SDSS quasars from S11 are compared to our fitting result of the simulated spectrum (center). Offsets from a 1--1 relation and intrinsic scatter are displayed. For the comparison of FWHMs (center), we divide the sample into FWHM$_{\mathrm{H}\alpha}$ larger or smaller than 2500 km\,s$^{-1}$ (filled and open dots) and only use the FWHM$_{\mathrm{H}\alpha}$\,$>$\,2500 km\,s$^{-1}$ data. The red highlighted symbol represents the example on the leftmost panel. Likewise, the H$\alpha$ luminosity from S11 and our single Gaussian fit to the simulated spectra, are compared (right). Bottom rows: we also test the effect of low S/N to the fitting by adding a set of Gaussian random noise to the resolution degraded spectrum (left). The ratio of our FWHM$_{\mathrm{H}\alpha}$, $L_{\mathrm{H}\alpha}$ measurements to that from S11 are plotted along the S/N$_{\mathrm{H}\alpha}$ (center and right). For the comparison of FWHMs (center), we remove 49 data points from the plot with $\Delta$FWHM$_{\mathrm{H}\alpha}$\,$=$\,0 usually at FWHM$_{\mathrm{H}\alpha}$\,$<$\,2500 km\,s$^{-1}$. The mean and 1\,$\sigma$ offsets are shown in black and gray lines. When calculating the mean and intrinsic scatter of the quantities, only the S/N$_{\mathrm{H}\alpha}$\,$>$\,3, FWHM$_{\mathrm{H}\alpha}$\,$>$\,2500 km\,s$^{-1}$ data are used for the comparison of FWHM$_{\mathrm{H}\alpha}$, and the S/N$_{\mathrm{H}\alpha}$\,$>$\,2 and any FWHM$_{\mathrm{H}\alpha}$ data for the $L_{\mathrm{H}\alpha}$.} 
\end{figure*}

Multiple pointings of the 1-D spectra were stacked for each object, where the NG spectra were interpolated to a fixed wavelength grid, flux averaged, and error rescaled assuming Poisson error statistics. Multiple pointings of NP spectra were stacked without modifying the individual spectrum, due to their poor resolution. We provided secondary flux calibration to the stacked spectra by integrating the $AKARI$ fluxes over the $WISE$ filter response curves, to match the $W1$ and $W2$ fluxes together by a constant additive correction. The average and rms scatter of the corrections are $-0.01\pm0.19$\,mJy. Out of 675 pointing observations, we used 589 pointings from 165 objects, since some of the pointing observations were not usable due to contamination of the object spectrum by adjacent sources. We excluded frames from the analysis when there was a source that is brighter than the target and its distance from the target is less than the FWHM in spatial direction. In rare occasions, the reduction pipeline did not run properly and such data were not used. In Figure 3, we plot the histogram of S/N$_{\mathrm{H}\alpha}$, the S/N of the H$\alpha$ emission line measured within $\pm$\,FWHM$_{\mathrm{H}\alpha}$ from the line center.

\section{Analysis}
\subsection{Spectral fitting}
\begin{figure*}
\centering
\includegraphics[scale=.5,clip]{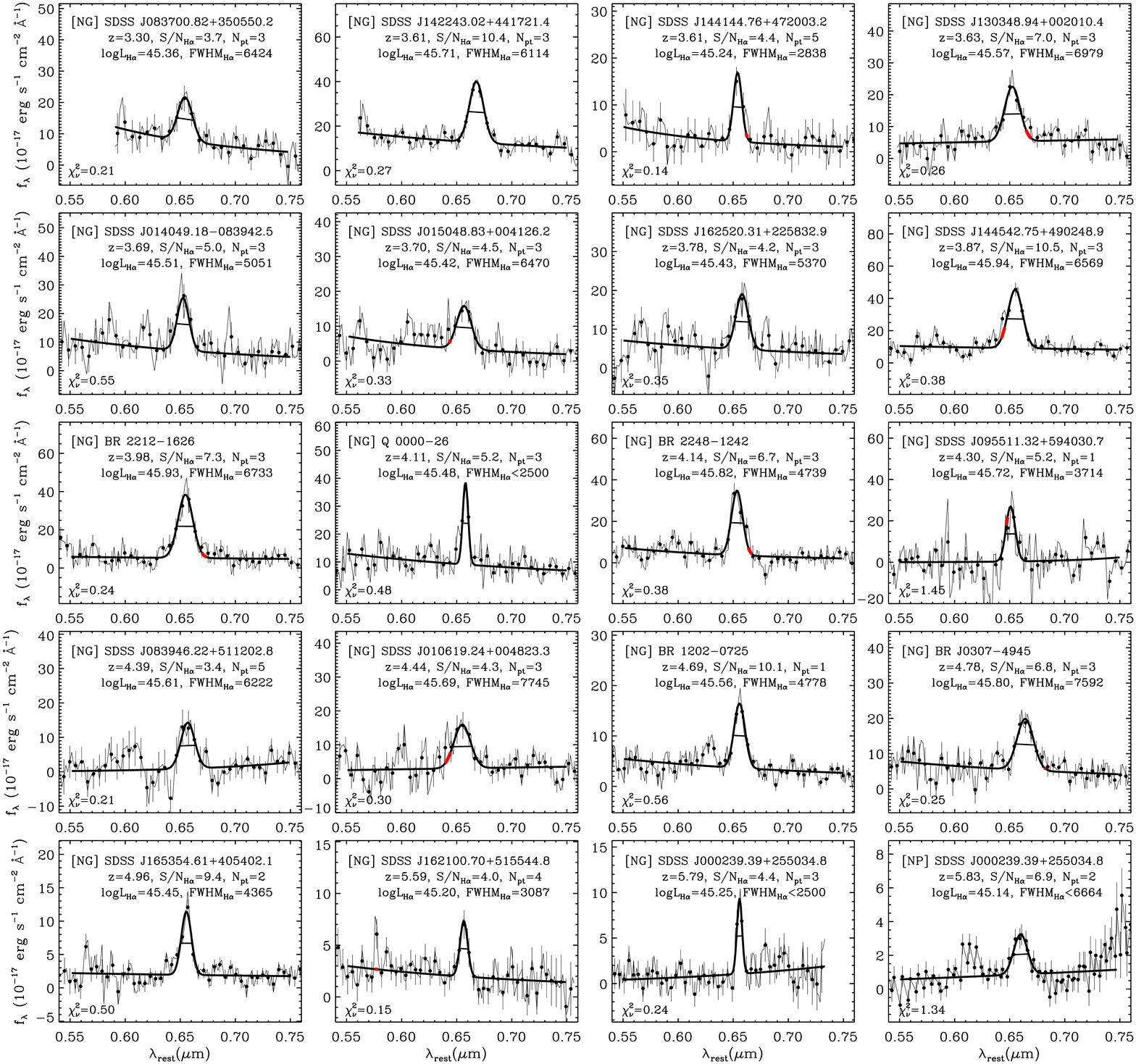}
\caption{Rest-frame H$\alpha$ spectral fitting of selected objects with S/N$_{\mathrm{H}\alpha}$\,$>$\,3, sorted by the H$\alpha$ redshift. The data point symbols and colors follow the meaning of Figure 4. When there were contaminations to the H$\alpha$ emission, we masked out the region (red).}
\end{figure*}
\begin{figure*}
\centering
\includegraphics[scale=.5]{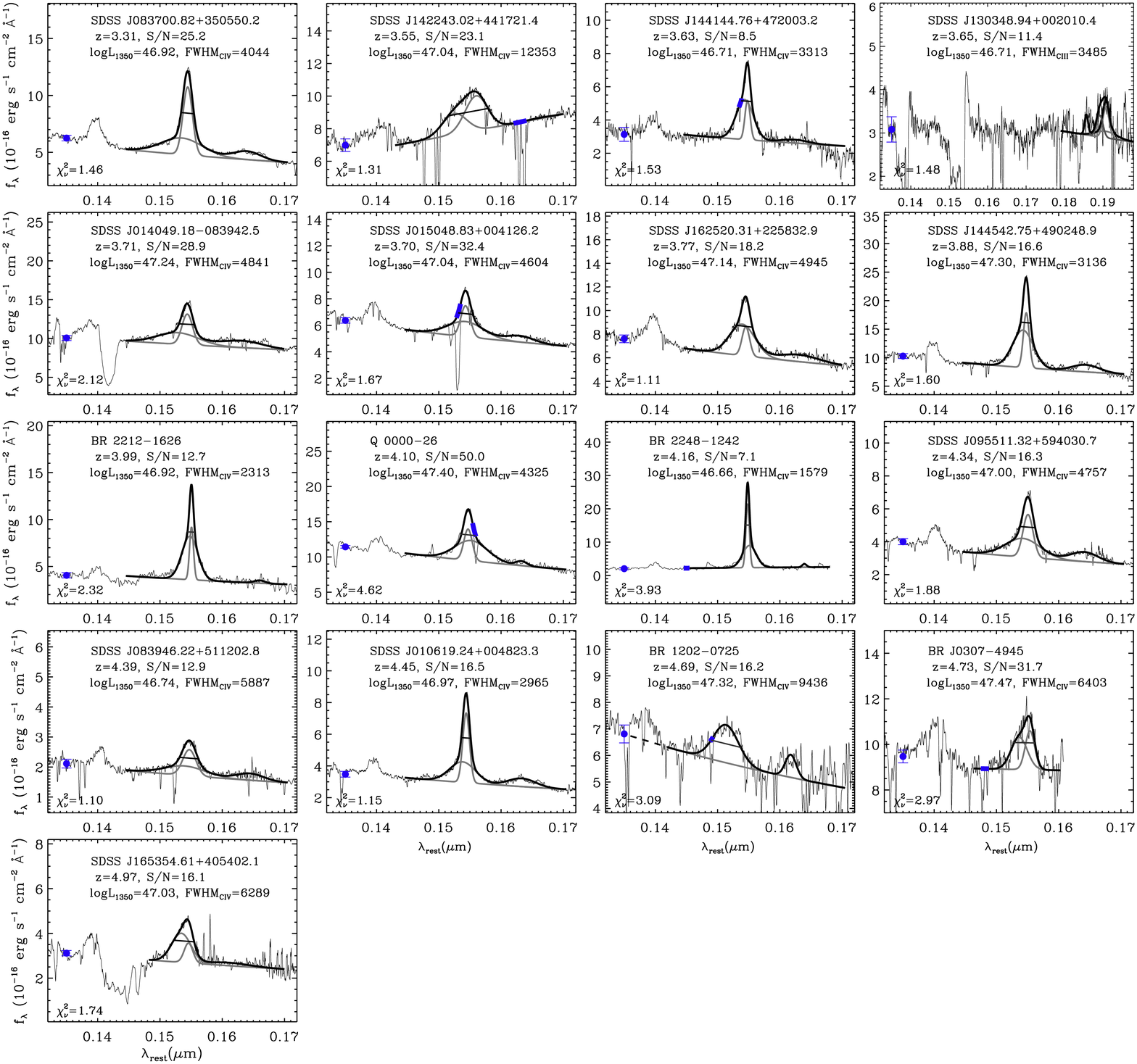}
\caption{Rest-frame {\ion{C}{4}} spectral fitting of the identical objects in Figure 6 plotted at the same relative location, following the format of the plotted data and printed numbers. Additionally shown are the double Gaussian fit (gray) to the {\ion{C}{4}} above the continuum, and the 1350\,$\text{\AA}$ monochromatic flux marked on its wavelength (blue dots), with extrapolated values indicated (dashed lines). The spectra are shown when they are available from the listed references in Table 2. When there were contaminations around the {\ion{C}{4}} emission, we masked out the region (blue). The {\ion{C}{4}} spectra displayed are smoothed down to $R$\,=\,500, to highlight the spectral features better.} 
\end{figure*}

We modeled the rest-frame 5500--7500\,$\text{\AA}$ spectra of our AGNs, as a sum of the power-law continuum $f_{\lambda}$\,=\,$c\,\lambda^{-(2+\alpha)}$ (where $f_{\nu} \propto \nu^{\alpha}$) and the Gaussian H$\alpha$ emission components. We did not attempt to fit the relatively weak emission features far ({\ion{He}{1}}, [{\ion{O}{1}}]), or near the H$\alpha$ ([{\ion{N}{2}}] and [{\ion{S}{2}}] doublets, {\ion{Fe}{2}} complex), as they were not detectable under the $AKARI$ spectroscopic resolution and sensitivity . Likewise, the H$\beta$ and [{\ion{O}{3}}] lines were too weak to be detected in most cases and were not fitted accordingly. Examples of the fitting are given in Figure 4, with fitted parameters shown on each panel. We found that 23\% of the sample show spiky emission/absorption features around the H$\alpha$ line from low S/N spectra, which were manually masked out. The H$\alpha$ line was modeled as a single broad Gaussian with observed FWHMs of 2500--10,000\,km\,s$^{-1}$, for the inability to clearly resolve the narrow or multiple broad components at $R$\,$\sim$\,120. The H$\alpha$ line center was set free within  $\pm$\,2500\,km\,s$^{-1}$ to the UV line-based redshift from references in Table 2. However, we found four (SDSS J143835.95+431459.2, SDSS J142243.02+441721.4, BR J0307--4945, and SDSS J150654.54+522004.6) exceptions whose H$\alpha$ line centers were significantly redshifted from the UV-line based redshifts. In these cases, the velocity range was loosened to $\pm$\,10,000\,km\,s$^{-1}$, where the H$\alpha$ showed velocity shifts of 3600--5600\,km\,s$^{-1}$. Next, the measured broad line width FWHM$_{\mathrm{obs}}$, was subtracted by the instrumental resolution $\mathrm{FWHM}_{\mathrm{ins}}$ (section 2.2) in quadrature, to obtain the intrinsic line width $\mathrm{FWHM}$ = $\sqrt{(\mathrm{FWHM}_{\mathrm{obs}})^{2}-(\mathrm{FWHM}_{\mathrm{ins}})^{2}}$. 

Extracting the broad emission line luminosity and width is important in accurately estimating the $M_{\rm BH}$ of AGNs. The limited $AKARI$ sensitivity (Figure 3) and spectral resolution could produce systematic bias in the measurement of line parameters such as the line luminosity and width (e.g., \citealt{Den09}). Therefore, we investigated how the low resolution, low S/N spectra systematically affect the results of the spectral fitting. This was done by running Monte Carlo simulations on a set of luminous SDSS quasar spectra (DR7, \citealt{Sch10}) of $R$\,$\sim$\,2000 to mimic the quality of $AKARI$ spectra. Given that the host galaxy contamination to the quasar spectrum is negligible at $L_{5100}$\,$>$\,10$^{45}$ ergs\,s$^{-1}$ (Shen et al. 2011, hereafter S11), we collected 15, $L_{5100}$\,$>$\,10$^{45}$\,ergs\,s$^{-1}$, type-1 quasar spectra with continuum sensitivity of S/N\,$>$\,20, where the H$\alpha$ emission region is present ($z$\,$<$\,0.3) and well fit ($\chi^{2}_{\nu}$\,$<$\,2) from S11.

We smoothed the SDSS spectra with a Gaussian function to match the $AKARI$ NG resolution, $R$\,=120, and rebinned the data to match the 3 pixel per resolution sampling of the $AKARI$ spectra. First, we looked into the question of how the results get affected with a single Gaussian fit to the emission line in the low resolution of the AKARI data. For this, without adding extra noise, we followed the same method to measure the H$\alpha$ line FWHM and luminosity as for the $AKARI$ spectra (e.g., Figure 5a). In Figures 5b and 5c we compare the fitted parameters FWHM$_{\mathrm{H}\alpha}$ and $L_{\mathrm{H}\alpha}$ from the smoothed and binned spectra, to that of the measurement from S11. We find that the FWHM at $>$\,2500\,km\,s$^{-1}$ and the line luminosity from the degraded resolution spectra, are remarkably consistent with S11 within $\sim$\,0.04\,dex offset and intrinsic scatter $\sigma_{\text{int}}$, where $\sigma_{\text{int}}^{2}$=$\Sigma_{i=1}^{N}[\{y_{i}-f(x_{i})\}^{2}-\Delta y_{i}^{2}-\beta^{2}\Delta x_{i}^{2}]/(N-1)$ for $f(x)=\alpha+\beta x$ and a set of N data points ($x_{i},y_{i}$) with measurement errors ($\Delta x_{i},\Delta y_{i}$). This could bias the $M_{\rm BH}$ measurements up to $\sim$\,0.1\,dex in offset and $\sigma_{\text{int}}$ when following the $M_{\rm BH}$\,$\sim$\,$L^{0.5}$\,$\times$\,$\text{FWHM}^{2}$ behavior, but this is smaller than the typical $M_{\rm BH}$ measurement uncertainty (e.g., Figure 14c). Exceptions to the consistency between the simulated and observed parameters are the line width measurements at FWHM$_{\mathrm{H}\alpha}$\,$<$\,2500 km\,s$^{-1}$, where the simulated FWHM values fall below the extrapolated linear relation. Therefore, we give a FWHM$_{\mathrm{H}\alpha}$\,$>$\,2500\,km\,s$^{-1}$ limit to our $AKARI$ data to restrict the sample with less biased line width measurements. Meanwhile, the narrow H$\alpha$ and [{\ion{N}{2}}] doublet luminosities of the fiducial SDSS AGNs add up to the broad $L_{\mathrm{H}\alpha}$, by less than 0.01\,dex altogether. 
The weakness of the narrow emission lines in luminous AGNs guarantees that the narrow line contaminations to the degraded resolution spectra are negligible.

\begin{figure*}
\centering
\includegraphics[scale=.5]{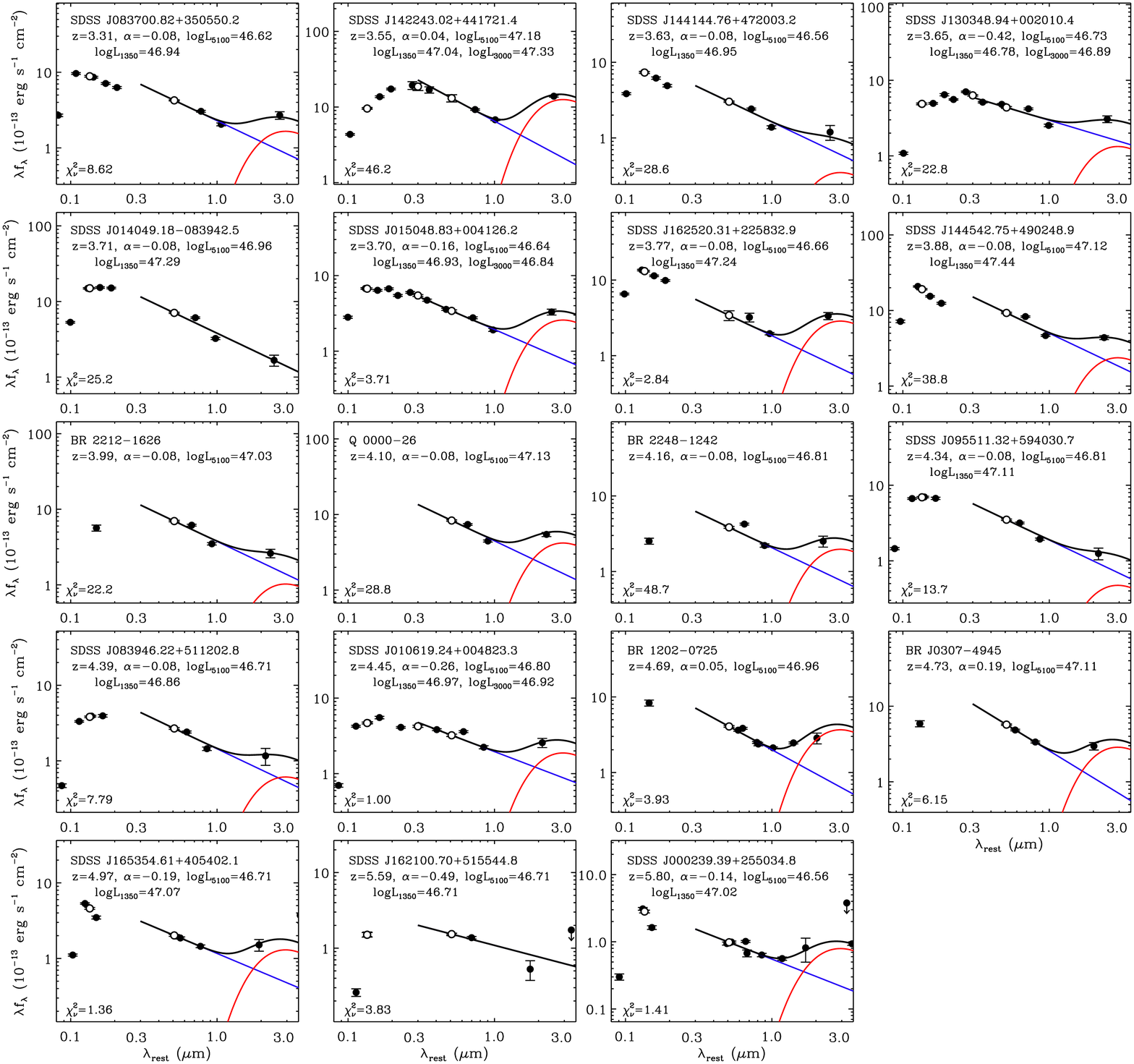}
\caption{Rest-frame UV--NIR broad-band SED of the objects in Figure 6 plotted at the same relative location. The figure shows the observed data points (filled circles) and $WISE$ 2\,$\sigma$ upper limits (arrows), model fits of the accretion disk (blue line) and the T=1250\,K dust components (red curve). Also, the monochromatic 1350, 3000, and 5100\,$\text{\AA}$ fluxes are drawn (open circles) when available.} 
\end{figure*}
\begin{figure*}
\centering
\includegraphics[scale=.825]{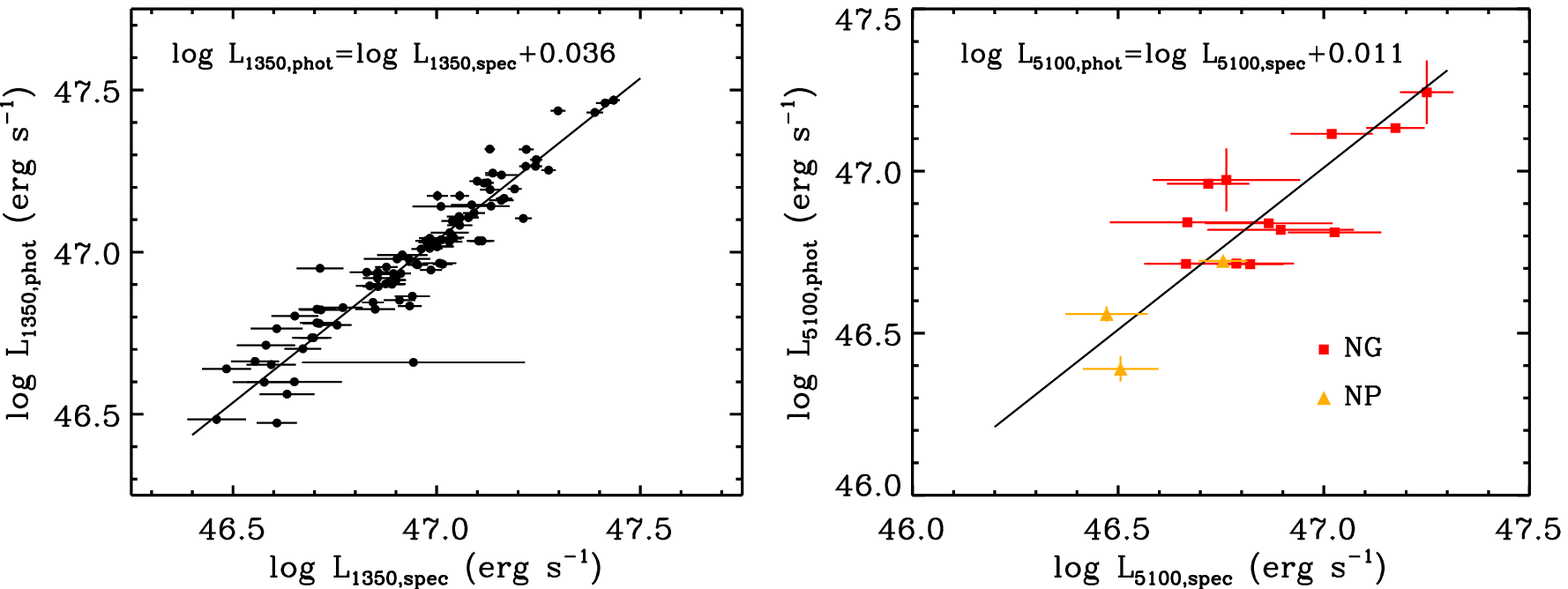}
\caption{Comparison of spectroscopically and photometrically derived monochromatic luminosities, for $L_{1350}$ (left) and $L_{5100}$ (right). The $AKARI$ NG/NP data points for the $L_{5100}$ are highlighted with red squares and yellow triangles, respectively. Offsets to the luminosities from a 1--1 relation is displayed.}
\end{figure*}

Second, to investigate the effect of low S/N to the fitted results, we added a set of random Gaussian noises on top of the degraded resolution spectra. Each SDSS spectrum was repeatedly simulated 30 times adding the random errors, for the S/N$_{\mathrm{H}\alpha}$ to be distributed down to the level of $AKARI$ S/N. Again, we followed the same method to measure the H$\alpha$ line FWHM (only for the FWHM$_{\mathrm{H}\alpha}$\,$>$\,2500\,km\,s$^{-1}$ objects from Figure 5b) and luminosity as for the $AKARI$ spectra (e.g., Figure 5d). In Figures 5e and 5f we plot the ratios of FWHM$_{\mathrm{H}\alpha}$ and $L_{\mathrm{H}\alpha}$ from the degraded resolution and S/N spectra, to that from S11, along S/N$_{\mathrm{H}\alpha}$. Overall, we find both FWHM$_{\mathrm{H}\alpha}$ and $L_{\mathrm{H}\alpha}$ to be within 0.02\,dex in the mean offset to the values before adding the noises (Figure 5b and 5c), at FWHM$_{\mathrm{H}\alpha}$\,$>$\,2500\,km\,s$^{-1}$, S/N$_{\mathrm{H}\alpha}$\,$>$\,3 for the FWHM$_{\mathrm{H}\alpha}$ and S/N$_{\mathrm{H}\alpha}$\,$>$\,2 for the $L_{\mathrm{H}\alpha}$. Also, the $\sigma_{\text{int}}$ in FWHM$_{\mathrm{H}\alpha}$ and $L_{\mathrm{H}\alpha}$ under low S/N are within 0.03\,dex to those of the noise-free, effectively unchanging the bias in the $M_{\rm BH}$ at a similar $\sim$\,0.1\,dex. We note that below the S/N or FWHM limit the fitted quantities systematically diverge from those calculated with the original spectrum. Therefore, we conclude that for a given selection of data neither the poor resolution nor sensitivity biases the fitted results by greater than $\sim$\,0.1\,dex level of systematic offset or scatter in $M_{\rm BH}$, and we give the corresponding FWHM and S/N cuts to the $AKARI$ data.  

Having tested the reliability of fitted quantities under possible systematic biases, we come back to the fitting of the $AKARI$ data and find the fit to converge for 160 out of 165 objects. Five failures show noisy spectra near the H$\alpha$ and were removed. The goodness of the spectral fitting is quantified as the reduced chi-square, and it reaches down to $\chi^{2}_{\nu}$\,$=$\,0.33 in median. The fraction of AGNs passing the reliability limit for the FWHM measurements (S/N$_{\mathrm{H}\alpha}$\,$>$\,3) are 67\,\% (N=8) and 25\,\% (N=37) for the phase--2 and 3 data respectively. Also, we flagged the five NG objects of FWHM$_{\mathrm{int}}$\,$<$\,2500\,km\,s$^{-1}$ with an upper limit of 2500\,km\,s$^{-1}$, and put upper limits on the measured FWHM of the four NP sources (e.g., Figure 4b). In total, there are 43 FWHM$_{\mathrm{H}\alpha}$ measurements including seven out of nine upper limits, where the excluded two upper limits are NP measurements with overlapping coverage in NG. 

Next, we computed the H$\alpha$ line luminosity $L_{\mathrm{H}\alpha}$ and the 5100\,$\text{\AA}$ continuum luminosity $L_{5100}$ from the spectra, converting the measured rest-frame fluxes using the luminosity distance and assuming isotropic radiation. The $L_{\mathrm{H}\alpha}$ was derived from the Gaussian fit to the observed flux. The reliability limit for $L_{\mathrm{H}\alpha}$ (S/N$_{\mathrm{H}\alpha}$\,$>$\,2) is satisfied for 45\,\% (N=72) of the data, while the S/N$_{\mathrm{H}\alpha}$\,$<$\,2 spectra were provided with 2\,$\sigma$ upper limits from their given noise levels collected within $\pm$\,4000\,km\,s$^{-1}$ of the H$\alpha$ line center. Meanwhile, the $L_{5100}$ was calculated from the average of the rest-frame fluxes at 5000--5200\,$\text{\AA}$ to reduce the measurement uncertainty. The $L_{5100}$ measurements were kept only when S/N$_{5100}$\,$>$\,2, and the rest of the data were given with 2\,$\sigma$ upper limits alike $L_{\mathrm{H}\alpha}$. We have less $L_{5100}$ measurements than $L_{\mathrm{H}\alpha}$ where only 25\,\% (N=41) meet S/N$_{5100}$\,$>$\,2, not to mention the limited number of spectra (53\,\%, N=88) covering the rest-frame 5100\,$\text{\AA}$. Thus, we also derived the $L_{5100}$ alternatively by the photometric SED fitting (section 3.2). 

In addition, we fitted the {\ion{C}{4}} region (rest-frame 1445--1705\,$\text{\AA}$) of 121 objects with a SED model containing a power law component and double broad Gaussians to model the {\ion{C}{4}} emission. Also, a single broad Gaussian was used to fit the 1600\,$\text{\AA}$ feature \citep{Lao94}, and the {\ion{He}{2}} and {\ion{O}{3}}] around 1650\,$\text{\AA}$ altogether since these emission are blended but relatively detached from the {\ion{C}{4}}. This component was not regarded as a part of the {\ion{C}{4}}, consistent with the previous studies (e.g,. prescription A of \citealt{Ass11}; S12). We do not subtract the broad {\ion{Fe}{2}} complex around the {\ion{C}{4}} emission, as it does not change the FWHM$_{\mathrm{C_{IV}}}$ meaningfully (S11). The optical spectra were fitted after carefully masking out the absorption features around the {\ion{C}{4}} line for 25\,\% of the spectra. Meanwhile, 16 spectra without error information were fitted assuming the flux error is uniform at all wavelengths, and the rms scatter of the best-fit solution is chosen to be the flux error afterward. Out of five spectra with severe broad absorption line (BAL) features, we fitted the {\ion{C}{3}}]$\lambda$1908 and used its line width as an effective FWHM$_{\mathrm{C_{IV}}}$ surrogate (S12) for two objects, while excluding the remaining three objects from the UV line analysis. In total, we derived FWHM$_{\mathrm{C_{IV}}}$ and $L_{1350}$ for 118 objects. The $L_{1350}$ and its error were calculated from the average of the rest-frame 1350\,$\pm$\,15\,$\text{\AA}$ fluxes to avoid contamination from narrow absorption, while for seven BAL quasars we extrapolated the continuum around the {\ion{C}{4}} emission to 1350\,$\text{\AA}$. When spectra of an object were available from both SDSS-I/SDSS-II and BOSS, we performed the fit to the spectra from each dataset separately, and took the average of the parameter values from the independent fits. We plot examples of the spectral fitting of the H$\alpha$ region in Figure 6, and of the {\ion{C}{4}} in Figure 7.

\subsection{Broad-band SED fitting}
\renewcommand{\tabcolsep}{2pt}
\begin{deluxetable*}{ *{10}{c} }
\tablecolumns{10}
\tabletypesize{\scriptsize}
\tablecaption{Continum and Line Based Properties of the Sample}
\tablehead{
\colhead{Name} & \colhead{$z_{\mathrm{ref}}$} & \colhead{$z_{\mathrm{H}\alpha}$} & \colhead{log $L_{1350}$} & \colhead{log $L_{5100}$} & \colhead{log $L_{\mathrm{H}\alpha}$} &  \colhead{FWHM$_{3,\rm C_{IV}}$} & \colhead{FWHM$_{3,\rm H\alpha}$} & \colhead{log $M_{\mathrm{BH,C_{IV}}}$} & \colhead{log $M_{\mathrm{BH,H\alpha}}$}
\\ (1) & (2) & (3) & (4) & (5) & (6) & (7) & (8) & (9) & (10)}
\startdata
SDSS J000239.39+255034.8 & 5.80 & 5.79 & 99.00$\,\pm\,$99.00 & 46.56$\,\pm\,$0.02 & 45.14$\,\pm\,$0.11 & 99.00$\,\pm\,$99.00 & 2.50$\,\pm\,$-1.00 & 99.00$\,\pm\,$99.00 & 9.25$\,\pm\,$-1.00\\
Q 0000-26 & 4.10 & 4.11 & 47.40$\,\pm\,$0.01 & 47.13$\,\pm\,$0.01 & 45.48$\,\pm\,$0.13 & 4.33$\,\pm\,$0.28 & 2.50$\,\pm\,$-1.00 & 9.88$\,\pm\,$0.22 & 9.56$\,\pm\,$-1.00\\
SDSS J000552.34-000655.8 & 5.85 & 5.85 & 99.00$\,\pm\,$99.00 & 46.04$\,\pm\,$0.08 & 44.75$\,\pm\,$0.32 & 99.00$\,\pm\,$99.00 & 99.00$\,\pm\,$99.00 & 99.00$\,\pm\,$99.00 & 99.00$\,\pm\,$99.00\\
BR J0006-6208 & 4.45 & 4.49 & 46.90$\,\pm\,$0.02 & 46.71$\,\pm\,$0.00 & 45.46$\,\pm\,$0.06 & 11.33$\,\pm\,$1.33 & 2.85$\,\pm\,$0.85 & 10.48$\,\pm\,$0.24 & 9.46$\,\pm\,$0.31\\
SDSS J001115.23+144601.8 & 4.97 & 99.00 & 99.00$\,\pm\,$99.00 & 46.78$\,\pm\,$0.02 & 47.97$\,\pm\,$-1.00 & 99.00$\,\pm\,$99.00 & 99.00$\,\pm\,$99.00 & 99.00$\,\pm\,$99.00 & 99.00$\,\pm\,$99.00
\enddata
\tablecomments{Catalog of the properties derived for the $AKARI$ quasars, sorted by right ascension. 
Column 1: Target name; Column 2: Redshift from references; Column 3: Redshift measured from H$\alpha$; Column 4: 1350\,$\text{\AA}$ luminosity and its uncertainty; Column 5: 5100\,$\text{\AA}$ luminosity and its uncertainty; Column 6: H$\alpha$ luminosity and its uncertainty; Column 7: FWHM of the {\ion{C}{4}} line and its uncertainty; Column 8: FWHM of the H$\alpha$ line and its uncertainty; Column 9: $M_{\rm BH}$ from the {\ion{C}{4}} line and its uncertainty; Column 10: $M_{\rm BH}$ from the H$\alpha$ line and its uncertainty. The units for $L$, FWHM, and $M_{\rm BH}$ are ergs\,s$^{-1}$, 1000\,km\,s$^{-1}$, and $M_{\odot}$. Columns 9 and 10 are from Equations (10) and (7), respectively. Empty parameters are entered as 99 and upper limits are given with errors of -1.\\ (This table is available in its entirety in a machine-readable form in the online journal. A portion is shown here for guidance regarding its form and content.)}
\end{deluxetable*}

The photometry datasets in Table 2 cover a wide wavelength range in broad-band filters from $u$-band through 24\,$\mu$m, thus we fitted the broad-band SEDs to provide further information on the AGN continuum luminosities. Under a photometric sensitivity limit of S/N\,$>$\,5 for the observed optical--NIR and S/N\,$>$\,2 in the MIR, and further rejecting the 2MASS data with a single filter detection, we modeled the SED in the rest-frame 0.3--5\,$\mu$m including 5 data points on average, as a sum of a power law continuum and a black body emission from hot dust of T=1250\,K (e.g., \citealt{Jun13}). For 53\% of the case (N=87) for which there were no rest-frame 0.3--0.6\,$\mu$m data points available, we used the average continuum slope $\alpha$\,=\,-0.08 of luminous SDSS quasars \citep{Jun13}. The uncertainty in fixing the continuum slope was tested from objects that cover the rest-frame 0.3--0.6\,$\mu$m, and comparing the $L_{5100}$ with and without fixing the $\alpha$ value. The test yields the $L_{5100}$ to be offset by $-$0.01\,$\pm$\,0.04\,dex when $\alpha$ is fixed, compared to when $\alpha$ is free. Since the offsets are small, we find our method to fix the $\alpha$ when missing the photometric coverage near the rest-5100\,$\text{\AA}$, to be reliable in tracing the $L_{5100}$. In addition, considering the filter bandwidths and the AGN line equivalent widths from \citet{Van01}, we find the H$\alpha$ to be the only line that meaningfully contributes to the rest-optical photometry over the continuum emission (by $>$\,0.03\,dex). Thus, we removed the data point enclosing the H$\alpha$ emission when the $\chi^{2}_{\nu}$ containing that data point became larger than that without. Through this procedure we obtained 164 photometrically derived $L_{5100}$, while removing one object without any detections in the rest-frame 0.3--5\,$\mu$m under our sensitivity limit. Examples of the broad-band SED fitting are shown in Figure 8. The reduced chi-square values have a median of $\chi^{2}_{\nu}$\,$=$\,3.6, is acceptable given the simplification of the SED model that does not take into account the emission line features, and the general agreement of the fit to the data demonstrated in Figure 8. 

\begin{figure*}
\begin{center}$
\begin{array}{ccc}
\vspace*{-0.4cm} 
\includegraphics[scale=.662]{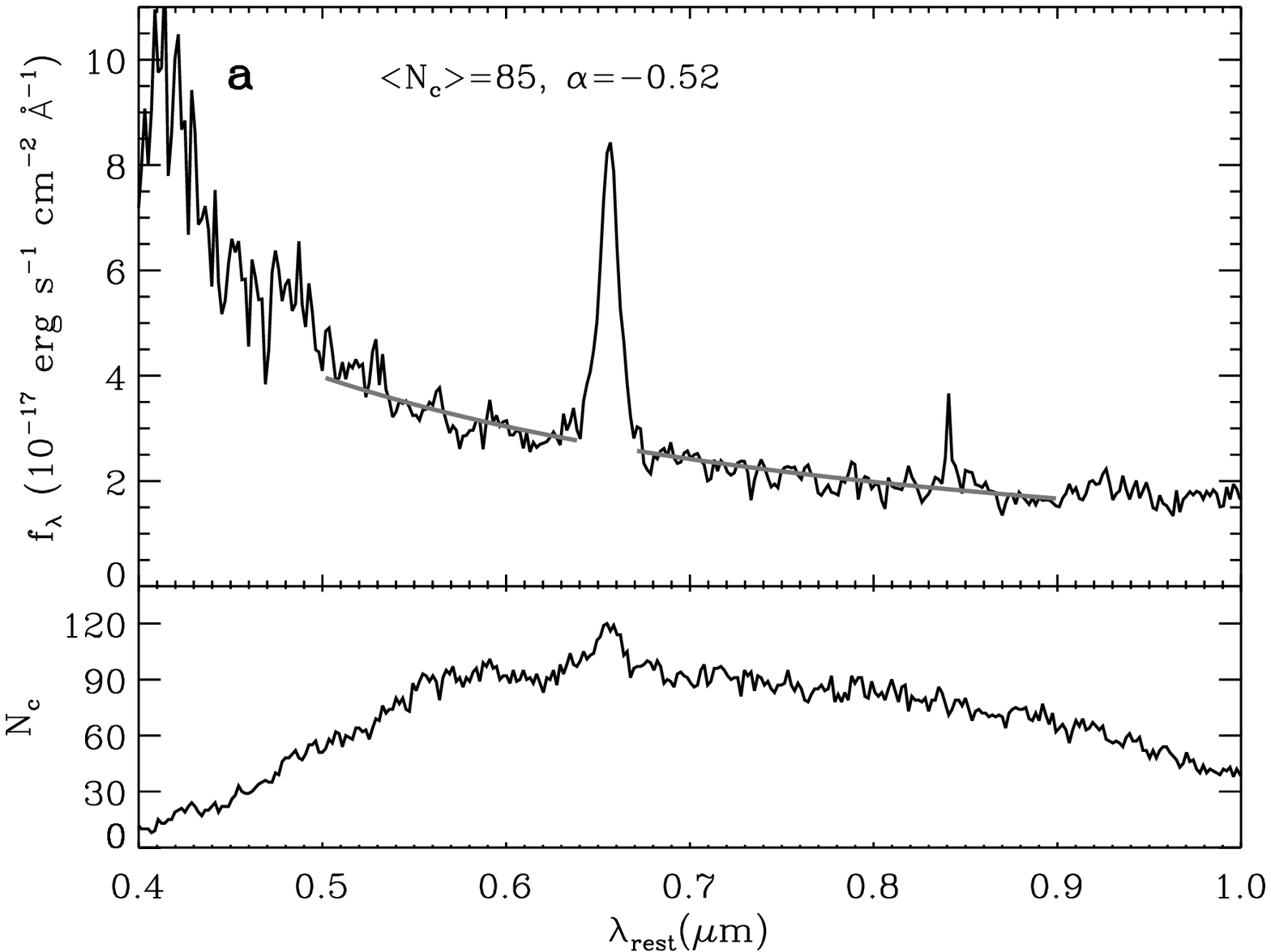} & \hspace*{-0.15cm}
\includegraphics[trim=0 -1.3 0 0, scale=.66]{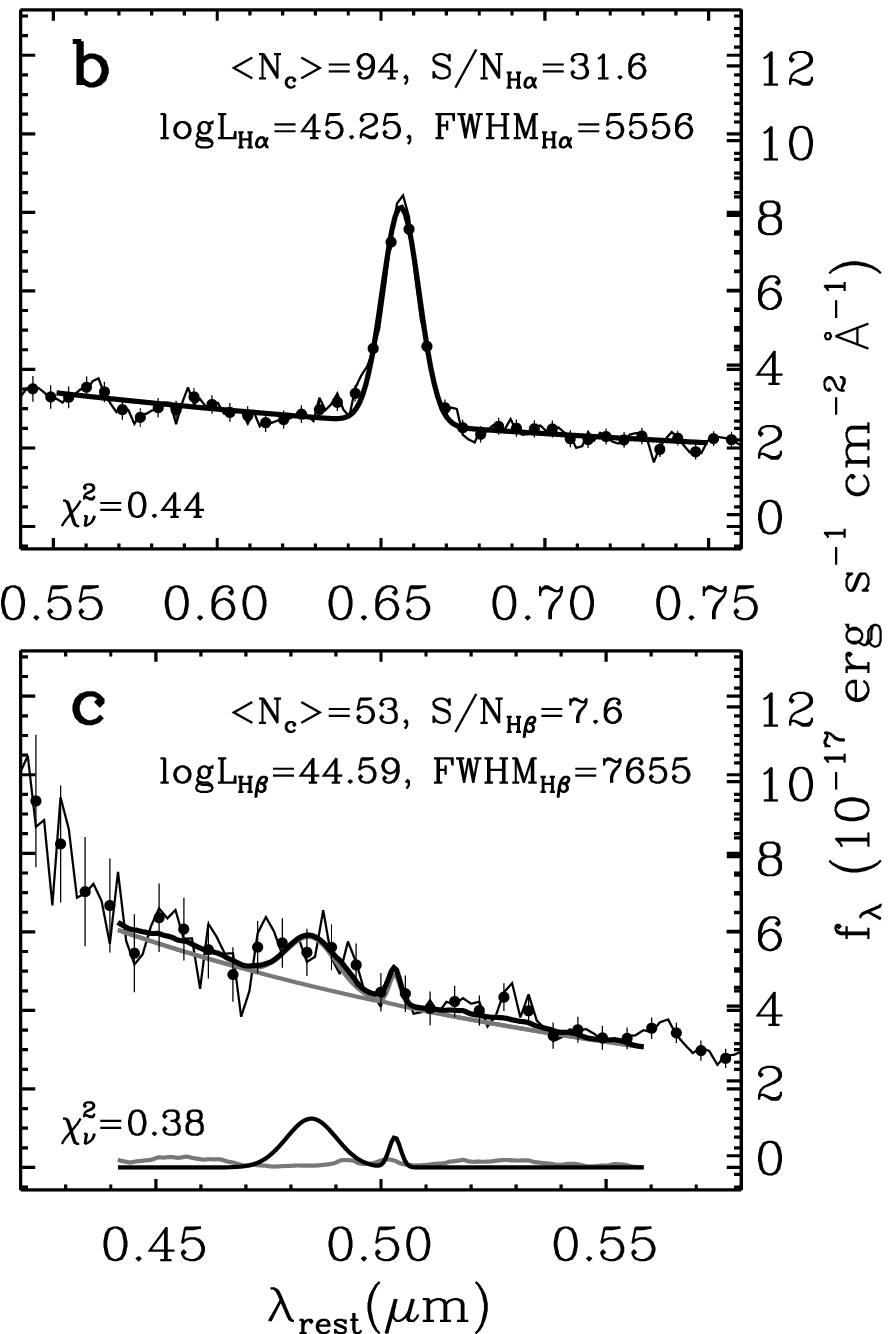}
\end{array}$
\end{center}
\caption{Composite spectra from $AKARI$ observations. (a) The spectra within 0.4--1\,$\mu$m (top), and the number of spectra used to construct the composite, N$_{\mathrm{c}}$, plotted against wavelength (bottom). The mean N$_{\mathrm{c}}$ and the continuum slope from the 0.5--0.9\,$\mu$m region are printed. (b) H$\alpha$ region fit of the composite spectra. (c) H$\beta$ region fit of the composite spectra. The sum of the continuum, H$\beta$ emission, and the {\ion{Fe}{2}} complex (thick black line), and the continuum (gray line) are overplotted on the data, while the H$\beta$/[{\ion{O}{3}}] and the {\ion{Fe}{2}} emission components are separately plotted below the spectrum in black and gray lines.} 
\end{figure*}

Meanwhile, we interpolated the broad-band SED around the rest-frame 1350 and 3000\,$\text{\AA}$ to obtain $L_{1350}$ and $L_{3000}$. For this, we used objects with more than two data points in the rest-frame 500--2500\,$\text{\AA}$ for $L_{1350}$ or 2000--6000\,$\text{\AA}$ for $L_{3000}$. The interpolation is done linearly to the data points, and we obtain 137 $L_{1350}$ and 47 $L_{3000}$ values. The continuum luminosities derived by photometric and spectroscopic methods roughly agree with each other as shown in Figure 9 for $L_{1350}$ and $L_{5100}$, though there are not enough data points (N=2) to plot for $L_{3000}$. Likewise to $L_{5100}$, we calculated the level of {\ion{C}{4}} or {\ion{Mg}{2}} line contamination to the $L_{1350}$ and $L_{3000}$ from broad-band photometry. The {\ion{C}{4}} and {\ion{Mg}{2}} elevates the observed broad-band flux by up to 0.06 and 0.04\,dex. Thus, it is possible that the photometry embracing the broad UV emission is boosted by more than the typical measurement error in $L_{1350}$ and $L_{3000}$, which are 0.02 and 0.03\,dex, respectively. Between the spectroscopically and photometrically derived continuum luminosities, we will use in the following discussion the spectroscopically derived $L_{1350}$ and the photometrically derived $L_{3000}$ and $L_{5100}$. We do so because the line contaminations near 1350\,$\text{\AA}$ through the broad-band photometry can be significantly larger than the measurement uncertainty, while it is not so around the {\ion{Mg}{2}} and H$\alpha$ lines. The large error in $L_{5100}$ from $AKARI$ spectra is also another reason why we opt to use $L_{5100}$ from the broad-band SED fitting. The fitted properties from this section, and the $M_{\rm BH}$ to appear in section 4, are listed in Table 3. For objects with both NG and NP observations, we use the $z_{\mathrm{H}\alpha}$ and FWHM$_{\mathrm{H}\alpha}$ from the NG and the $L_{\mathrm{H}\alpha}$ from the NP when S/N$_{\mathrm{H}\alpha}$\,$>$\,3, while we list only the values from the NG otherwise.
\begin{figure*}
\centering
\includegraphics[scale=.63]{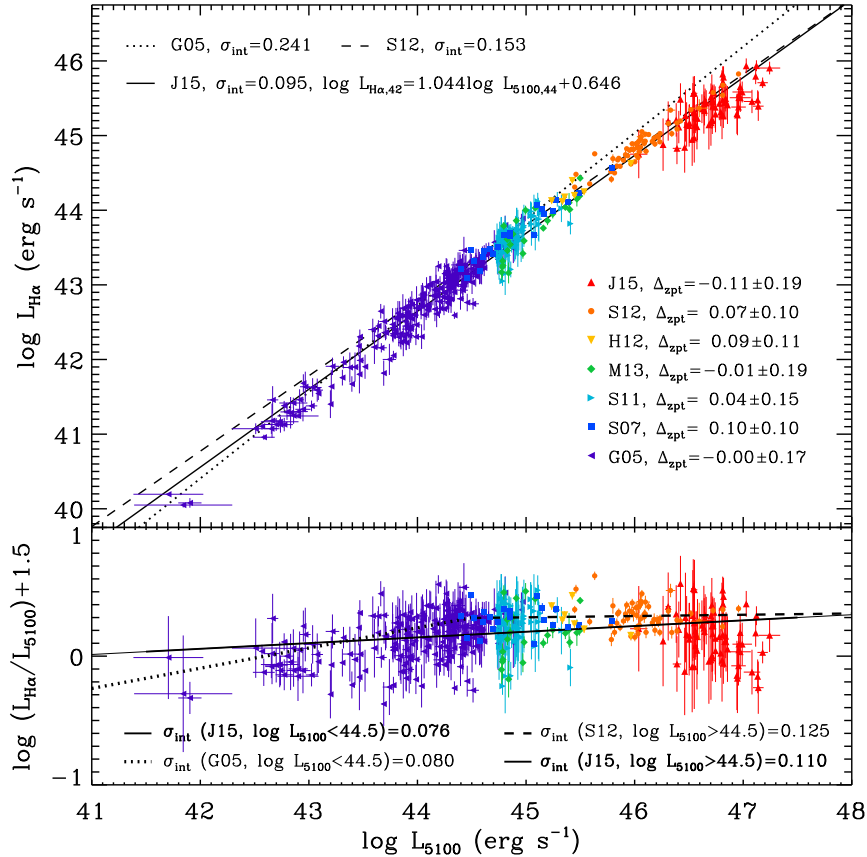}
\caption{The $L_{5100}$--$L_{\mathrm{H}\alpha}$ relation of AGNs (top), and its projection on the $L_{5100}$--$L_{\mathrm{H}\alpha}/L_{5100}$ (bottom), from combined references. The references abbreviated on the plot are summarized in Table 4, where all measurements from references are converted to our adopted cosmology. We limited the G05 data to $L_{5100}$\,$<$\,10$^{44.73}$\,ergs\,s$^{-1}$ to avoid overlap with the S11 data. The $L_{5100}$--$L_{\mathrm{H}\alpha}$ relation from G05, S12, and J15 are shown in dotted, dashed, and solid lines respectively. The zeropoint offset and rms scatter of each literature data with respect to our relation are denoted as $\Delta_{\text{zpt}}$. The intrinsic scatter ($\sigma_{\text{int}}$, dex) of the entire data with respect to the G05, S12, and J15 relations (top), and of the data divided by $L_{5100}$\,$=$\,10$^{44.5}$\,ergs\,s$^{-1}$ (bottom), are shown. The $L_{5100}$ in M13 and J15 are from photometric SED fitting (section 3.2), while the rest are spectroscopically derived from each reference. The $L_{\mathrm{H}\alpha}$ are from broad emission for all references but for the G05, S07, and S11 data, where the narrow component is included.} 
\end{figure*}

\section{Results}

\subsection{Composite spectra}
To investigate the overall rest-optical spectral properties of the sample, we construct the composite $AKARI$ spectra. Out of 154 objects observed with NG, 127 are used for the composite construction after removing 27 spectra due to a mild level of confusion from neighbor source spectra, negative continuum levels, or strong fluctuations near the H$\alpha$ due to warm pixels. The composite is constructed by normalizing the spectra at 5100\,$\text{\AA}$ and taking their error weighted mean to maximize the S/N. Each spectral flux and error were de-redshifted, and rebinned to a common wavelength scale of 18\,$\text{\AA}$ per bin which is equal to that of the $AKARI$ at rest-frame H$\alpha$. In Figures 10a--10c, we plot the composite spectrum, and zoomed-in fit to the H$\alpha$ and H$\beta$ regions, respectively. The H$\alpha$ emission is prominent in the composite spectrum, and the spiky feature at 8400\,$\text{\AA}$ is an artifact from a single spectrum with high S/N.

We determine the continuum slope from the 0.5--0.9\,$\mu$m region from Figure 10a, where the number of spectra used to construct the composite exceeds 60. The slope $\alpha$\,=\,$-$0.52\,$\pm$\,0.06 (where $f_{\nu} \propto \nu^{\alpha}$) is close to $\alpha$\,=\,$-$(0.37--0.48) of \citet{Gli06} determined through the composite of local luminous quasars at similar wavelengths, indicating a similarity in the rest-optical continuum shape of luminous type-1 quasars with respect to redshift. Interestingly, we detect a sign of the H$\beta$ emission from Figure 10c. The H$\beta$ region was fitted with the \citet{Bor92} {\ion{Fe}{2}} template, following the method of \citet{She08}. We find $L_{\mathrm{H}\alpha}$/$L_{\mathrm{H}\beta}$\,=\,4.5\,$\pm$\,1.6, which is roughly consistent with 3.6\,$\pm$\,1.4 from luminous $z$\,$\sim$\,2 quasars (S12) or the model broad Balmer line decrement of AGNs at T=10,000--12,000\,K, $L_{\mathrm{H}\alpha}$/$L_{\mathrm{H}\beta}$\,=\,3.6--8.8 \citep{Ost89}. Since the Balmer decrement value and the S/N$_{\mathrm{H}\alpha}$ of each $AKARI$ spectrum suggest that the strongest H$\beta$ in our individual spectrum would appear as S/N$_{\mathrm{H}\beta}$=1--2, we do not expect the H$\beta$ emission to be individually detected, consistent with the visual inspection in section 3.1. Apart from this, we do not find hint of other emission lines in the composite spectra.

\subsection{Luminosity and Line Width Scaling Relations}
\begin{deluxetable}{ccc}
\tablecolumns{6}
\tablecaption{Dynamic Range of References}
\tablewidth{0.45\textwidth}
\tablehead{
\colhead{Reference} & \colhead{$z$} & \colhead{log $L_{5100}$\,(ergs\,s$^{-1}$)}}
\startdata
G05  & $<$0.35     & 41.7--45.0\\
N07  & 2.3--3.5     & 45.2--46.3\\
S07   & 0.08--0.40 & 44.4--45.8\\
M08 & 0.35--0.37 & 43.6--44.4\\
D09  & 1.1--2.2     & 46.1--46.7\\
W09  & 0.00--0.16 & 42.0--45.9\\
A11  & 1.4--3.6     & 44.8--46.5\\
S11  & 0.08--0.80  & 43.8--46.4\\
H12 & 1.4--1.5      & 45.2--46.0\\
S12  & 1.5--2.2     & 45.4--47.0\\
B13 & 0.00--0.29 & 42.0--45.9\\
M13 & 0.7--1.7 & 44.8--45.5\\
P13  & 0.01--0.23 & 42.6--45.9\\
J15  & 3.3--6.2 & 46.0--47.2
\enddata
\tablecomments{The abbreviated references are \citealt{Gre05} (G05); \citealt{Net07} (N07); \citealt{Sha07} (S07); \citealt{McG08} (M08); \citealt{Die09} (D09); \citealt{Wan09} (W09); \citealt{Ass11} (A11); \citealt{She11} (S11); \citealt{Ho12} (H12); \citealt{She12} (S12); \citealt{Ben13} (B13); \citealt{Mat13} (M13); \citealt{Par13} (P13); and this work (J15).} 
\end{deluxetable} 
\begin{figure*}
\centering
\includegraphics[scale=.9]{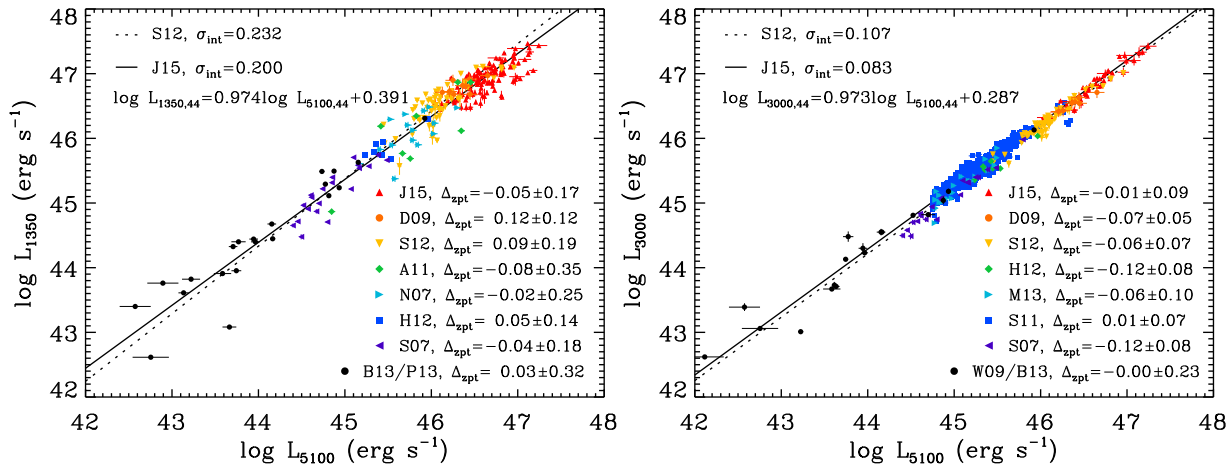}
\caption{The $L_{5100}$--$L_{1350}$ and $L_{5100}$--$L_{3000}$ relations of AGNs from combined references. The references abbreviated on the plot are summarized in Table 4, where all measurements from references are converted to our adopted cosmology. The $L_{1350}$ of the N07 sample are searched from S11. The $L_{5100}$, $L_{3000}$, and $L_{1350}$ are derived from spectra, except for the $L_{5100}$, $L_{3000}$ from J15 where they are from photometric SED fitting (section 3.2). We assign a modest 10\% error for $L_{5100}$ of the N07 data and $L_{1350}$, $L_{5100}$ of the A11 data, and 20\% error for the $L_{1350}$, $L_{3000}$, and $L_{5100}$ from D09 data, from visual inspection of their spectra. We removed the highly variable object 3C 390.3 from the B13 data, and additional two objects in S07 data that overlap with B13. The $L_{5100}$--$L_{1350}$, $L_{5100}$--$L_{3000}$ relations from S12 and J15 are shown in dotted and solid lines respectively. The zeropoint offset and rms scatter of each literature data with respect to our relations, are indicated.} 
\end{figure*}

The derivation of continuum and line luminosities for distant, luminous quasars allows us to examine the universality of the luminosity scaling relations over a wide range of redshifts and luminosities. Starting from the $L_{5100}$--$L_{\mathrm{H}\alpha}$ relation, we plot in Figure 11 our derived data points and the data taken from literatures (\citealt{Gre05}, hereafter G05; \citealt{Sha07}, hereafter S07; S11; \citealt{Ho12}; \citealt{Mat13}; S12) that cover a range of $L_{5100}$ and $z$, as summarized in Table 4. Our $AKARI$ data extends the relation at the high redshift ($z$\,$>$\,3.3) and high luminosity end ($L_{5100}$\,$>$\,10$^{46}$\,ergs\,s$^{-1}$). To minimize the host galaxy contribution to the AGN luminosities, we chose AGNs with host contamination $<$20\% in $L_{5100}$, $L_{\mathrm{H}\alpha}$ for some datasets (G05; S07), while we plotted only the $L_{5100}$\,$>$\,$10^{44.73}\,$ergs\,s$^{-1}$ data for the rest of references that meet $<$10\% in host contamination (S11). Meanwhile, the broad $L_{\mathrm{H}\alpha}$ could contain the narrow component for $AKARI$ data, while the broad and narrow line luminosities are combined for the G05, S07, and S11 data too. We find that the contribution from the narrow component to $L_{\mathrm{H}\alpha}$ estimated from section 3.1 and the references, is negligible (2\% and $<$\,10\%, respectively), allowing us to consider $L_{\mathrm{H}\alpha}$ to be approximately the line luminosity of the broad line component. 

\begin{figure*}
\centering
\includegraphics[scale=.67]{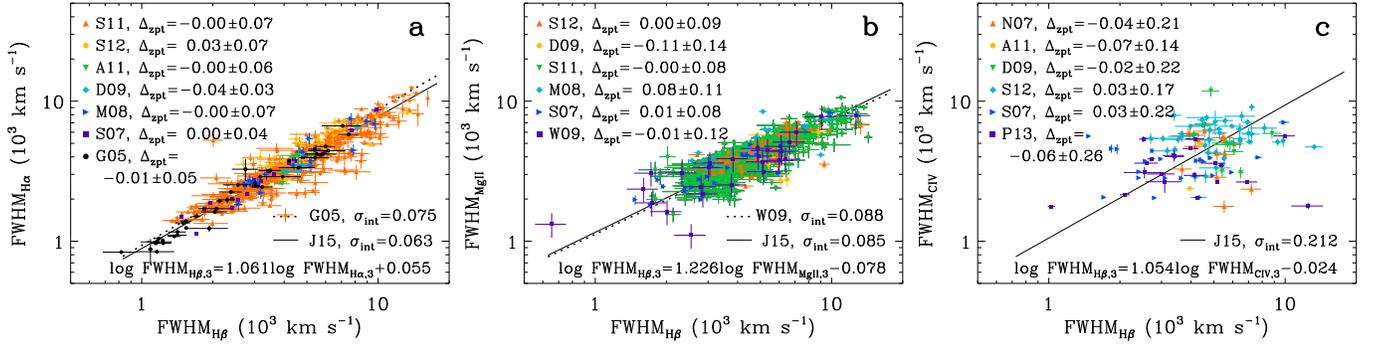}
\caption{The FWHM$_{\mathrm{H}\beta}$--FWHM$_{\mathrm{H}\alpha}$, FWHM$_{\mathrm{H}\beta}$--FWHM$_{\mathrm{Mg_{II}}}$, and FWHM$_{\mathrm{H}\beta}$--FWHM$_{\mathrm{C_{IV}}}$ relations of AGNs from combined references. The references abbreviated on the plot are summarized in Table 4. The $AKARI$ data are not present since the sample do not have simultaneous coverage of the FWHMs plotted. We limited the G05 data to $L_{5100}$\,$<$\,10$^{43.83}$\,ergs\,s$^{-1}$ to avoid overlap with the S11 data. We assign a 10\% error for FWHM$_{\mathrm{H}\beta}$, FWHM$_{\mathrm{C_{IV}}}$ of the N07 data, following their argument. We removed one object in S07 data that overlaps with W09, and two objects in S07 that overlap with P13. The FWHM$_{\mathrm{H}\beta}$--FWHM$_{\mathrm{H}\alpha}$ relation of G05, and FWHM$_{\mathrm{H}\beta}$--FWHM$_{\mathrm{Mg_{II}}}$ relation of W09 are shown in dotted lines, while the relations from this work are shown as solid lines. The zeropoint offset and rms scatter of each literature data with respect to our relations, are indicated. The FWHMs are of the broad emission line.} 
\end{figure*}

Figure 11 shows a remarkable correlation between $L_{5100}$ and $L_{\mathrm{H}\alpha}$ even when AGNs are drawn from various samples covering a wide range of redshifts and luminosities. This strongly suggests that the physics governing the correlation is the same for low and high luminosity AGNs, and there is no strong evolution in the relation from $z$\,=\,0 to $z$\,=\,6 over the range explored here. The $AKARI$ data points are mildly below the relation where the offset could indicate a growing population of weak emission line quasars at high redshift (\citealt{Fan99}; \citealt{Dia09}), but the overall deviations are within the scatter of the data. The possible downturn in the correlation at very high luminosity ($L_{5100}$\,$\sim$\,10$^{47}$\,ergs\,s$^{-1}$) produces a $\sim$\,0.2\,dex offset in $\log\,(L_{\mathrm{H}\alpha}$/$L_{5100})$, but this downturn affects the $M_{\rm BH}$ estimates only by $\sim$\,0.1\,dex. The possible downturn could be caused by cold accretion disks of slowly spinning, extremely high mass BHs (\citealt{Lao11}; \citealt{Wan14}), but we will leave the investigation of the possible downturn as a subject of a future work as its effect on $M_{\rm BH}$ estimates is small. We fitted the relation using the linear regression with bivariate correlated errors and intrinsic scatter (BCES, \citealt{Akr96})\footnote{Throughout this paper we use the BCES fit to derive the linear relations.}, to find the following result,\footnote{Throughout this paper we use subscript numbers to the luminosity to indicate its wavelength and unit, such as $L_{5100,44}\,$=$\,L_{5100\, \text{\AA}}/10^{44}$\,ergs\,s$^{-1}$.}
\begin{eqnarray}\begin{aligned}
\log L_{\mathrm{H}\alpha,42}=(1.044\pm0.008)\,\log L_{5100,44}\\+(0.646\pm0.011).
\end{aligned}\end{eqnarray}
The best-fit relation fits the entire data with $\sigma_{\text{int}}$\,=\,0.095\,dex. The flux--flux relation of 5100\,$\text{\AA}$ continuum and H$\alpha$ shows almost identical slope and the intrinsic scatter, suggesting that the tight correlation in Equation (2) is not due to a sample selection effect. The data points at three redshift intervals, namely  $0 < z < 0.8$  (G05, S07, and S11) for $41.7 < {\rm log}\, L_{\rm 5100} < 46.4$,   $0.7 < z < 2.2$ for $44.8 < {\rm log}\, L_{\rm 5100} < 47.0$ (H12, S12, and M13), and  $3.3 < z < 6.2$ for  $46.0 < {\rm log}\, L_{\rm 5100} < 47.2$ (J15), overlap with each other and show no evolution. This suggests that Equation (2) is universal, and not due to a distance effect like the Malmquist bias.

To examine the universality of the relation further, we discuss how the $L_{5100}$--$L_{\mathrm{H}\alpha}$ relation of G05 at $z$\,$\sim$\,0 of lower luminosity AGNs and S12 at $z$\,$\sim$\,2 of higher luminosity AGNs fare with each literature values. First, the G05 relation can describe the $L_{5100}$--$L_{\mathrm{H}\alpha}$ relation of S07, S11, and M13 AGNs over the overlapping luminosity interval ($L_{5100}$\,$\lesssim$\,10$^{45}$\,ergs\,s$^{-1}$). When extrapolated to higher luminosity, it starts to deviate from the data points regardless of redshift. Likewise, the S12 relation can describe the $L_{5100}$--$L_{\mathrm{H}\alpha}$ relation down to $L_{5100}$\,$\lesssim$\,10$^{44.5}$\,ergs\,s$^{-1}$ including the S07 data at $z$\,$<$\,0.4 and our AGNs at $z$\,$>$\,3.3. However, when applied to the entire datasets, the G05 and S12 relations show deviations from the data at high and low luminosity regions, respectively. Consequently, both relations produce $\sigma_{\text{int}}$\,=\,0.15--0.24\,dex against the data which is worse than 0.095\,dex of our $L_{5100}$--$L_{\mathrm{H}\alpha}$ relation. 

To check if the inconsistency in the G05 and S12 relations at the faint and luminous end arises from a possible break in the relation itself, we considered the case where the slope changes at $L_{5100}$\,$\sim$\,10$^{44.5}$\,ergs\,s$^{-1}$, where the G05 and S12 relations meet. For this, we computed the $\sigma_{\text{int}}$ of the $L_{5100}$\,$<$\,10$^{44.5}$\,ergs\,s$^{-1}$ and $L_{5100}$\,$>$\,10$^{44.5}$\,ergs\,s$^{-1}$ data against the G05, S12, and our relation in Equation (2), and examined if our simple relation is any worse than the combination of G05 and S12 relations with a break at $L_{5100}$\,=\,10$^{44.5}$\,ergs\,s$^{-1}$. As indicated in the lower panel of Figure 11, the $\sigma_{\text{int}}$ values against our relation is comparable to or slightly smaller than the $\sigma_{\text{int}}$ against the G05 or S12 relations at low and high luminosities respectively. This suggests that there is no strong need for a broken power-law form of the $L_{5100}$--$L_{\mathrm{H}\alpha}$ relation, and a simple relation of Equation (2) can be employed to describe the response of broad line region to the incident continuum emission over the covered redshift and luminosity ranges in Figure 11 and Table 4.

Likewise, we plot in Figure 12 the $L_{5100}$--$L_{1350}$ and $L_{5100}$--$L_{3000}$ relations from the luminosities derived in section 3 and taken from references. Apart from the literature data where the host contamination in $L_{5100}$ is estimated to be $<$20\% (S07), or minimized by Hubble Space Telescope (HST) observations, we limit the literature sample with $L_{5100}$\,$>$\,10$^{44.73}$\,ergs\,s$^{-1}$ to keep the host galaxy contamination below 10\% (S11). Meanwhile, we replaced the $L_{5100}$ of \citet{Wan09} and \citet{Par13} with the HST data from \citet{Ben13}, while only including the AGNs with $<$20\% host contamination in $L_{5100}$. Similarly to the $L_{5100}$--$L_{\mathrm{H}\alpha}$ relation, we do not find any evolution in the $L_{5100}$--$L_{1350}$ and $L_{5100}$--$L_{3000}$ relations for a particular set of data, and we find the best-fit correlation to be,

\begin{eqnarray}\begin{aligned}
\log L_{1350,44}=(0.974\pm0.023)\,\log L_{5100,44}\\+(0.391\pm0.053)\\
\log L_{3000,44}=(0.973\pm0.010)\,\log L_{5100,44}\\+(0.287\pm0.013). 
\end{aligned}\end{eqnarray}
The $\sigma_{\text{int}}$ values (dex) to these best-fit relation, as well as the relation with respect to S12 are presented in Figure 12.
Like for the $L_{\mathrm{H}\alpha}$--$L_{5100}$ relation, the flux--flux relations show virtually identical slopes and intrinsic scatter to the luminosity--luminosity relations, showing again that sample selection effect is not a main driver for the relations in Equation (3).

\begin{figure*}
\centering
\includegraphics[scale=.67]{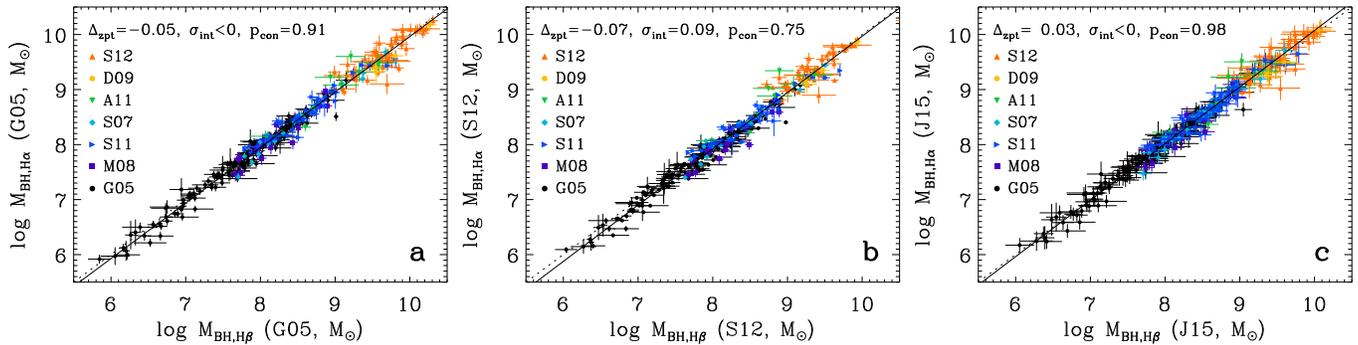}
\caption{The comparison of Balmer $M_{\rm BH}$ relations of AGNs, from combined references. The references abbreviated on the plot are summarized in Table 4. All measurements are converted to our adopted cosmology and $f$--factor. We limited the G05 data to $L_{5100}$\,$<$\,10$^{44.73}$\,ergs\,s$^{-1}$ and the S11 data to $L_{5100}$\,$>$\,10$^{44.73}$\,ergs\,s$^{-1}$, in order to avoid overlap. The linear fit to the $M_{\rm BH}$ relation and a 1--1 relation are represented by solid and dotted lines respectively, while the zeropoint offset between the masses ($\Delta_{\text{zpt}}$), intrinsic scatter with respect to a 1-1 relation ($\sigma_{\text{int}}$, dex), and the fraction where the masses overlap within error (p$_{\text{con}}$) are printed.} 
\end{figure*}

Finally, we compare the broad line FWHM of H$\beta$, H$\alpha$, {\ion{Mg}{2}}, and {\ion{C}{4}} in Figure 13 in order to calibrate the $M_{\rm BH}$ from multiple line based recipes and to check for any evolution in the FWHM relations. For good comparison of FWHMs, we restricted the mixed samples to have the fractional errors of FWHM less than 20\%, while additionally limiting the S/N and reduced chi-square of the SDSS spectra to be S/N$>$20 and $\chi^{2}_{\nu}$\,$<$\,2. In Figure 13a, we fit the FWHM$_{\mathrm{H}\beta}$--FWHM$_{\mathrm{H}\alpha}$ relation from the collected data, where we find the offset of each reference data to this relation to fall within each scatter. The $\sigma_{\text{int}}$ of all the data to our relation, 0.063\,dex, is smaller than when using the relation from G05.

Second, we derive the FWHM$_{\mathrm{H}\beta}$--FWHM$_{\mathrm{Mg_{II}}}$ relation. Since it is debatable whether to subtract the narrow component for the {\ion{Mg}{2}} line width measurement (e.g., S11), we followed \citet{Jun13} to average the FWHM$_{\mathrm{Mg_{II}}}$ derived with and without the subtraction of the narrow component. Also, the slope and constant of the relation are consistent within uncertainty with those derived with and without the subtraction of the narrow FWHM$_{\mathrm{Mg_{II}}}$ component. Therefore, we combined the literature data with respective treatment of the narrow {\ion{Mg}{2}} and the averaged FWHM$_{\mathrm{Mg_{II}}}$ in S11, to derive the FWHM$_{\mathrm{H}\beta}$--FWHM$_{\mathrm{Mg_{II}}}$ relation altogether. We note that the relative FWHM offset of the literature data to our relation shown in Figure 13b are within the scatter of data points, indicating that the details of fitting to exclude the narrow {\ion{Mg}{2}} component (S07; \citealt{McG08}; \citealt{Die09}) or to include but subtract it (\citealt{Wan09}; S12) do not affect the line widths significantly when compared overall. The rms of all the data to our relation, 0.085 dex, is similar to that from \citet{Wan09}, and small enough to regard the FWHM$_{\mathrm{Mg_{II}}}$ as a marginally good substitute of FWHM$_{\mathrm{H}\beta}$ as much as FWHM$_{\mathrm{H}\alpha}$. 

Third, we derive the FWHM$_{\mathrm{H}\beta}$--FWHM$_{\mathrm{C_{IV}}}$ relation in Figure 13c. The data can be fitted altogether with a log-linear relation, but the $\sigma_{\text{int}}$ of the data to the relation, 0.212 dex, is large and comparable to the systematic uncertainty of single-epoch $M_{\rm BH}$ estimators when scaled as the FWHM squared. We checked the effect of fitting methodology by deriving all the $L$--$L$, FWHM--FWHM relations with the FITEXY method \citep{Tre02}, to find that the slope of the FWHM$_{\mathrm{H}\beta}$--FWHM$_{\mathrm{C_{IV}}}$ relation, 1.798\,$\pm$\,0.026, shows a meaningfully large difference to the BCES results. Still, we keep the BCES slope since it brings the Balmer and {\ion{C}{4}} $M_{\rm BH}$ estimates more consistent (section 4.3), and we note that the large $\sigma_{\text{int}}$ between the FWHM$_{\mathrm{H}\beta}$--FWHM$_{\mathrm{C_{IV}}}$ relation represents a poor correlation at best. 

Overall, we find the relations of FWHM$_{\mathrm{H}\alpha}$, FWHM$_{\mathrm{Mg_{II}}}$ and FWHM$_{\mathrm{C_{IV}}}$ against the FWHM$_{\mathrm{H}\beta}$ without any noticeable evolution for the samples considered, covering a wide range of luminosity or redshift. Therefore, although the data for calibration is missing at $z$\,$\gtrsim$\,3, we use our FWHM relations to calibrate the $M_{\rm BH}$ estimators for general usage in terms of redshift. Our derived FWHM relations are as follows,\footnote{Throughout this paper we use subscript numbers to the line width to indicate its unit, such as FWHM$_{\mathrm{H}\beta,3}$ = FWHM$_{\mathrm{H}\beta}/10^{3}\,\mathrm{km\,s^{-1}}$.}
\begin{eqnarray}\begin{aligned}
&\log \mathrm{FWHM}_{\mathrm{H}\beta,3}\\&= (1.061\pm0.013)\,\log \mathrm{FWHM}_{\mathrm{H}\alpha,3}+(0.055\pm0.008)\\
&= (1.226\pm0.032)\,\log \mathrm{FWHM}_{\mathrm{Mg_{II}},3}-(0.078\pm0.021)\\
&=(1.054\pm0.057)\,\log \mathrm{FWHM}_{\mathrm{C_{IV}},3}-(0.024\pm0.045).
\end{aligned}\end{eqnarray}
We note the possibility that the FWHM relations established with spectra taken in different epochs could be affected by variability of the emission line shape, especially when the dynamic range of the probed FWHMs are narrower than that of the continuum luminosities. Still, the offset or scatter of the references with and without simultaneously acquired data in Figure 13 do not differ with each other. Also, we find the average and rms scatter in the ratio of the SDSS-III BOSS over SDSS-I/SDSS-II FWHM$_{\mathrm{C_{IV}}}$ from 12, S/N\,$>$\,15 objects matched with the $AKARI$ sample to be 0.99\,$\pm$\,0.08, which implies the effect of variability to the shift or broadening of the FWHM$_{\mathrm{H}\beta}$--FWHM$_{\mathrm{C_{IV}}}$ relation is negligible compared to the $\sigma_{\text{int}}$ of the relation by an order of magnitude. 

\begin{figure*}
\centering
\includegraphics[scale=.67]{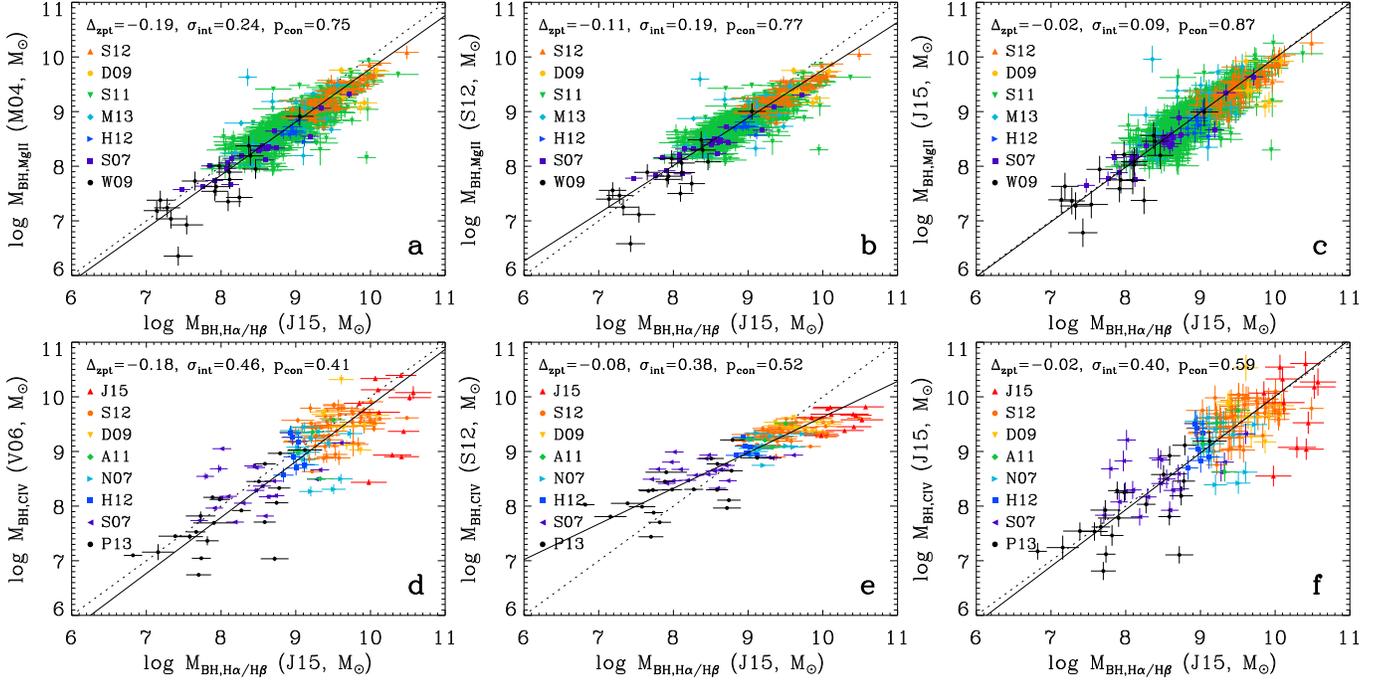}
\caption{The comparison of Balmer $M_{\rm BH}$ to the UV $M_{\rm BH}$ of AGNs, from combined references. The references abbreviated on the plot are summarized in Table 4, with M04, V06 estimators from \citet{Mcl04} and \citet{Ves06}. All measurements are converted to our adopted cosmology and $f$--factor. We use the $M_{\mathrm{BH}}(L_{5100}, \mathrm{FWHM}_{\mathrm{H}\alpha})$ for the Balmer masses, unless when the H$\alpha$ line was uncovered and the $M_{\mathrm{BH}}(L_{5100}, \mathrm{FWHM}_{\mathrm{H}\beta})$ was used. The rest of the figure format follows that of Figure 14. We removed the objects unused in Figures 13 and 14.} 
\end{figure*}

\subsection{Updated $M_{\rm BH}$ estimators}
Using the scaling relations obtained above, we now present an updated set of $M_{\rm BH}$ estimators based on various lines. Since the reverberation mapping of H$\beta$ and $L_{5100}$ forms the basis of mass estimation for AGNs, we start from the $M_{\rm BH}$ estimator that uses the 5100\,$\text{\AA}$ luminosity and H$\beta$ line width to take the following form and derive other estimators using scaling relations discussed above: 
\begin{equation} M_{\mathrm{BH}}=\frac{f}{G}R_{\mathrm{BLR}}(L_{5100})\Big(\frac{\mathrm{FWHM}_{\mathrm{H}\beta}}{2}\Big)^{2}. \end{equation}
Previous secondary calibrations to the $M_{\rm BH}$ were performed through replacing the ($L_{5100}$, FWHM$_{\mathrm{H}\beta}$) of the local reverberation mapped sample or SDSS AGNs, by ($L_{\mathrm{H}\alpha}$, FWHM$_{\mathrm{H}\alpha}$) (e.g., \citealt{Gre05}), ($L_{3000}$, FWHM$_{\mathrm{MgII}}$) (e.g., \citealt{Mcl02}), or ($L_{1350}$, FWHM$_{\mathrm{C_{IV}}}$) (e.g., \citealt{Ves06}). Using the latest $R$--$L$ relation from \citet{Ben13}\footnote{We shifted the relation to Hubble parameter of $H_{0}=\mathrm{70\,km\,s^{-1}\,Mpc^{-1}}$ which is our adopted value throughout this work.} and the constant for the mass equation ($f$--factor, $f=5.1\pm1.3$) from \citet{Woo13}, we derive first the H$\beta$ $M_{\rm BH}$ estimator as, 
\begin{eqnarray}\begin{aligned}
&R_{\mathrm{BLR}}=(34.7\pm2.5)\,L_{5100,44}\,^{(0.533\pm0.034)}\,\mathrm{lt-day}\\
&M_{\mathrm{BH}}(L_{5100}, \mathrm{FWHM}_{\mathrm{H}\beta})=(8.63\pm2.29)\times10^{6}\\
&\times L_{5100,44}\,^{(0.533\pm0.034)}\,\mathrm{FWHM}_{\mathrm{H}\beta,3}\,^{2}M_{\odot}.
\end{aligned}\end{eqnarray}
Replacing the H$\beta$ line width through the $\mathrm{FWHM}_{\mathrm{H}\beta}$--$\mathrm{FWHM}_{\mathrm{H}\alpha}$ 
relation from Equation (4) we get the $M_{\rm BH}$ from ($L_{5100}$, FWHM$_{\mathrm{H}\alpha}$),
\begin{eqnarray}\begin{aligned}
&M_{\mathrm{BH}}(L_{5100}, \mathrm{FWHM}_{\mathrm{H}\alpha})=(1.11\pm0.30)\times10^{7}\\
&\times L_{5100,44}\,^{(0.533\pm0.034)}\,\mathrm{FWHM}_{\mathrm{H}\alpha,3}\,^{(2.12\pm0.03)}M_{\odot}.
\end{aligned}\end{eqnarray}
Finally, the application of Equation (2) yields the $M_{\rm BH}$ from ($L_{\mathrm{H}\alpha}$, FWHM$_{\mathrm{H}\alpha}$),
\begin{eqnarray}\begin{aligned}
&M_{\mathrm{BH}}(L_{\mathrm{H}\alpha}, \mathrm{FWHM}_{\mathrm{H}\alpha})=(5.20\pm1.41)\times10^{6}\\
&\times L_{\mathrm{H}\alpha,42}\,^{(0.511\pm0.033)}\,\mathrm{FWHM}_{\mathrm{H}\alpha,3}\,^{(2.12\pm0.03)}M_{\odot}.
\end{aligned}\end{eqnarray}
Likewise, replacing the continuum luminosity and line width in Equation (6) to those of {\ion{Mg}{2}} 
and {\ion{C}{4}} using Equations (3) and (4) yields,
\begin{eqnarray}\begin{aligned}
&M_{\mathrm{BH}}(L_{3000}, \mathrm{FWHM}_{\mathrm{Mg_{II}}})=(4.19\pm1.19)\times10^{6}\\
&\times L_{3000,44}\,^{(0.548\pm0.035)}\,\mathrm{FWHM}_{\mathrm{Mg_{II}},3}\,^{(2.45\pm0.06)}M_{\odot}
\end{aligned}\end{eqnarray}
\begin{eqnarray}\begin{aligned}
&M_{\mathrm{BH}}(L_{1350}, \mathrm{FWHM}_{\mathrm{C_{IV}}})=(4.72\pm1.63)\times10^{6}\\
&\times L_{1350,44}\,^{(0.547\pm0.037)}\,\mathrm{FWHM}_{\mathrm{C_{IV}},3}\,^{(2.11\pm0.11)}M_{\odot}.
\end{aligned}\end{eqnarray}

\begin{deluxetable*}{ccccccccc}
\tablecaption{$L_{5100}$--$L_{{\rm H\alpha}}$ relation\label{tbl-control}}
\tablewidth{0.95\textwidth}
\tablehead{
\colhead{Reference} & \colhead{$\alpha$} & \colhead{$\beta$} & \colhead{$\sigma_{\rm int}$} & \colhead{$\sigma_{\rm int, all}$}  & \colhead{N} & \colhead{$z$}  & \colhead{log $L_{5100}$} & \colhead{Dataset} \\
\colhead{(1)} & \colhead{(2)} & \colhead{(3)} & \colhead{(4)} & \colhead{(5)} & \colhead{(6)} & \colhead{(7)} & \colhead{(8)} & \colhead{(9)} 
}
\startdata
\citet{Gre05} & 0.720\,$\pm$\,0.002 & 1.157\,$\pm$\,0.005 & 0.078 & 0.241 & 229 & $<$\,0.35 &  41.7--45.0    &  G05\\
\citet{She12} &  0.791\,$\pm$\,0.093 &  1.010\,$\pm$\,0.042 & 0.088 &  0.153 & 60   & 1.5--2.2  &  45.4--47.0 &  S12    \\
\multirow{2}{*}{This work}     &  \multirow{2}{*}{0.646\,$\pm$\,0.011}  &  \multirow{2}{*}{1.044\,$\pm$\,0.008} & \multirow{2}{*}{0.095} &  \multirow{2}{*}{0.095} & \multirow{2}{*}{464} & \multirow{2}{*}{0.0--6.2}     &  \multirow{2}{*}{41.7--47.2}  &  G05, S07, S11, H12, \\ 
     &  &  &  &  &  &  &  &  S12, M13, J15
\enddata
\tablecomments{The $L_{5100}$--$L_{{\rm H\alpha}}$ relations from previous studies and this work, where $\alpha$ and $\beta$ are defined as 
${\rm log}\,L_{{\rm H\alpha,42}}=\alpha+\beta\, {\rm log}\,L_{5100,44}$. Column 1: Reference; Column 2: $\alpha$ and its uncertainty (1$\sigma$); Column 3: $\beta$ and its uncertainty; Column 4: Intrinsic scatter (dex) of the relation over the $L_{5100}$ range covered; Column 5: Intrinsic scatter of the relation over the range $41.7 < {\rm log}\,L_{\rm 5100}< 47.2$; Column 6: Number of objects used; Column 7: Redshift range; Column 8: Range of $L_{\rm 5100}$ (ergs\,s$^{-1}$); Column 9: References of AGN dataset with abbreviations from Table 4.}
\end{deluxetable*}
\begin{deluxetable*}{cccccccc}
\tablecaption{Virial Mass Estimators}
\tablewidth{0.95\textwidth}
\tablehead{
\colhead{Reference} & \colhead{$\alpha$} & \colhead{$\beta$} & \colhead{$\gamma$} & \colhead{N} & \colhead{$z$} & \colhead{${\rm log}\,M_{\rm BH}$} & \colhead{Method}\\
\colhead{(1)} & \colhead{(2)} & \colhead{(3)} & \colhead{(4)} & \colhead{(5)} & \colhead{(6)} & \colhead{(7)} & \colhead{(8)}}
\startdata
\sidehead{$M_{\mathrm{BH}}(L_{5100,44}, \mathrm{FWHM}_{\mathrm{H}\beta,3})$}
\citet{Gre05} & 6.64\,$\pm$\,0.02 & 0.64\,$\pm$\,0.02 & 2 & 35 & $<$\,0.37 & 5.5--9.0 & RM\\
\citet{Ves06}  & 6.91\,$\pm$\,0.02 & 0.5 & 2 & 25 & 0.00--0.29 & 7.2--9.3 & RM\\
\citet{Ben13}, this work  & 6.94\,$\pm$\,0.12 & 0.533\,$\pm$\,0.034 & 2 & 41 & 0.00--0.29 & 6.0--10.7 & RM\\
\hline\sidehead{$M_{\mathrm{BH}}(L_{5100,44}, \mathrm{FWHM}_{\mathrm{H}\alpha,3})$}
\citet{Gre05} & 6.70\,$\pm$\,0.06 & 0.64\,$\pm$\,0.02 & 2.06\,$\pm$\,0.06 & 162 & $<$\,0.37 & 5.2--9.0 & SE\\
\citet{She12} & 7.01 & 0.555 & 1.87 & 60 & 1.5--2.2 & 8.8--10.3 & SE\\
This work  & 7.05\,$\pm$\,0.12 & 0.533\,$\pm$\,0.034 & 2.12\,$\pm$\,0.03 & 654 & 0.0--2.4 & 5.8--10.6 & SE\\
\hline\sidehead{$M_{\mathrm{BH}}(L_{\mathrm{H}\alpha,42}, \mathrm{FWHM}_{\mathrm{H}\alpha,3})$}
\citet{Gre05} & 6.30\,$\pm$\,0.08 & 0.55\,$\pm$\,0.02 & 2.06\,$\pm$\,0.06 & 243 & $<$\,0.37 & 5.1--9.0 & SE\\
\citet{She12} & 6.55 & 0.564 & 1.82 & 60 & 1.5--2.2 & 8.8--10.2 & SE\\
This work  & 6.72\,$\pm$\,0.12 & 0.511\,$\pm$\,0.033 & 2.12\,$\pm$\,0.03 & 969 & 0.0--6.2 & 5.6--10.5 & SE\\
\hline\sidehead{$M_{\mathrm{BH}}(L_{3000,44}, \mathrm{FWHM}_{\mathrm{Mg_{II}},3})$}
\citet{Mcl04} & 6.5 & 0.62 & 2 & 22 & 0.03--0.37 & 7.5--8.9 & RM\\
\citet{Ves09} & 6.86 & 0.5 & 2 & -- & -- & -- & SE\\
\citet{Wan09} & 7.15\,$\pm$\,0.27 & 0.46\,$\pm$\,0.08 & 1.48\,$\pm$\,0.49 & 29 & 0.00--0.16 & 6.3--9.0 & RM\\
\citet{She12} & 6.95 & 0.584 & 1.71 & 60 & 1.5--2.2 & 8.8--10.1 & SE\\
This work  & 6.62\,$\pm$\,0.12 & 0.548\,$\pm$\,0.035 & 2.45\,$\pm$\,0.06 & 1010 & 0.0--6.4 & 6.8--10.3 & SE\\
\hline\sidehead{$M_{\mathrm{BH}}(L_{1350,44}, \mathrm{FWHM}_{\mathrm{C_{IV}},3})$}
\citet{Ves06}  & 6.66\,$\pm$\,0.01 & 0.53 & 2 & 27 & 0.00--0.23 & 5.5--9.3 & RM\\
\citet{She12}  & 8.02 & 0.471 & 0.24 & 60 & 1.5--2.2 & 8.9--9.6 & SE\\
\citet{Par13} & 7.48\,$\pm$\,0.24 & 0.52\,$\pm$\,0.09 & 0.56\,$\pm$\,0.48 & 25 & 0.01--0.23 & 7.0--9.0 & RM\\
This work  & 6.67\,$\pm$\,0.15 & 0.547\,$\pm$\,0.037 & 2.11\,$\pm$\,0.11 & 258 & 0.0--5.4 & 6.8--10.7 & SE
\enddata
\tablecomments{The  $M_{\rm BH}$ relations from previous studies and this work, where $\alpha$, $\beta$, $\gamma$ are defined as 
${\rm log}\,M_{\rm BH}=\alpha+\beta\,{\rm log}\,L+\gamma\,{\rm log\,FWHM}$. Column 1: Reference; Column 2: $\alpha$ and its uncertainty (1$\sigma$); Column 3: $\beta$ and its uncertainty; Column 4: Column 3: $\gamma$ and its uncertainty; Column 5: Number of objects used for luminosity or line width calibration; Column 6: Redshift range of objects used for luminosity or line width calibration; Column 7: Range of $M_{\rm BH}$ from reference; Column 8: Method of calibration, where RM, SE denote calibrations of the luminosity and line width using the reverberation mapped and single epoch samples, respectively.} 
\end{deluxetable*}

The $M_{\rm BH}$'s based on H$\alpha$ and {\ion{C}{4}} from our estimators for the $AKARI$ sample are given in Table 3. It is worth noting that Equation (5) may not hold if the FWHM$_{\mathrm{H}\beta}$ is not exactly proportional to the velocity dispersion, $\sigma$ (\citealt{Pet04}; \citealt{Col06}), or if the $R$--$L$ relation breaks down at high luminosity where the relation has not been tested extensively with the reverberation technique ($L_{5100}>10^{46}\mathrm{ergs\,s^{-1}}$, \citealt{Ben13}). Nevertheless, considering the advantages of using the FWHM (to be robust under poor sensitivity, wings in the line profile, or deblending, see S12) for the single epoch mass estimation, and the current expectations in the high luminosity $R$--$L$ relation (see section 5.1), our set of calibrations have the merit where the rest-UV to optical $M_{\rm BH}$ estimations are mutually consistent through a wide range of redshifts, luminosities, and fitting methodologies. 

To check the consistency between our $M_{\rm BH}$ estimators, we first compare the Balmer $M_{\rm BH}$ using the estimator in this work and existing estimators in Figure 14. The $M_{\mathrm{BH, H\beta}}$ and $M_{\mathrm{BH, H\alpha}}$ are compared, where a $\Delta M_{\mathrm{BH}}<0.3$\,dex error cut is applied to the data out of various references. The $M_{\rm BH}$ values are derived from the G05, S11 estimators and from this work (Equations (6), (7)) in Figures 14a--14c respectively, where the zeropoint offset between the H$\alpha$ and H$\beta$ masses ($\Delta_{\text{zpt}}$), intrinsic scatter with respect to a 1-1 relation ($\sigma_{\text{int}}$), and the fraction of the data points where 1\,$\sigma$ errors of $M_{\rm BH}$ values from different estimators overlap with each other (p$_{\text{con}}$) are shown. We find that the masses from our estimators are closer to a 1--1 relation than of G05 or S11, throughout the range 10$^{6-10} M_{\odot}$. Also, fully considering the propagated errors in the mass equation, our estimators may overestimate the $M_{\rm BH}$ uncertainty when comparing the H$\beta$ and H$\alpha$ masses as reflected from the negative $\sigma_{\text{int}}$, which suggests that the actual difference in the $f$--factor or $R_{\text{BLR}}$ between the H$\beta$ and H$\alpha$ line emitting regions are likely to be smaller than their uncertainty. Therefore, we regard the H$\beta$ and H$\alpha$ $M_{\rm BH}$ from our estimators to be indistinguishable, which is supported by the high fraction (98\,\%) of H$\beta$/H$\alpha$ masses to be consistent within measurement error.

With the H$\beta$ and H$\alpha$ $M_{\rm BH}$ recipes checked to be mutually consistent, we further compare the Balmer to the UV-based $M_{\rm BH}$ with $\Delta M_{\mathrm{BH}}<0.3$\,dex, in Figure 15. We use $M_{\mathrm{BH, H\alpha}}$ as the Balmer mass due to the stronger H$\alpha$ emission than H$\beta$ and the availability of H$\alpha$ line in our $AKARI$ data, unless only the $M_{\mathrm{BH, H\beta}}$ can be estimated using the measurements in the literature. Compared to the conventional estimators calibrated at relatively low luminosity and redshift (\citealt{Mcl04}; \citealt{Ves06}) or relatively high luminosity and redshift (S12), our calibrations (Equations (6), (7), (9), (10)) bring the rest-UV ({\ion{Mg}{2}} in Figures 15a--15c, {\ion{C}{4}} in Figures 15d--15f) and Balmer $M_{\rm BH}$'s to be closer to a 1--1 relation in the range 10$^{7-10} M_{\odot}$ in terms of both zeropoint offset and intrinsic scatter. Also, we find that our $M_{\rm BH}$ estimators fully considering the propagated errors give more realistic value of the error bars compared to estimators that do not, and raise the p$_{\text{con}}$. Moreover, previous UV-based $M_{\rm BH}$'s are tilted to ours largely from the different slope of the FWHM term to that of our estimator (e.g., 0.24 in S12, 2.11 in this work for {\ion{C}{4}}, see also, \citealt{Par13}), and this produces a systematic offset in $M_{\rm BH}$ values. For example, we find that S12 estimators produce a systematic difference between {\ion{C}{4}} and Balmer $M_{\rm BH}$  in such a way that $M_{\mathrm{BH, C_{IV}}}$ are larger by 0.7 dex at $M_{\mathrm{BH, H\alpha/H\beta}}=10^{7} M_{\odot}$ and smaller by 0.4 dex at $M_{\mathrm{BH, H\alpha/H\beta}}=10^{10} M_{\odot}$ (Figure 15e). Previous UV-based $M_{\rm BH}$ calibrations from limited dynamic range yields $\sigma_{\text{int}}$ to the Balmer $M_{\rm BH}$ as small as our estimator, and may place the {\ion{C}{4}} and Balmer masses to be consistent within a narrow range of masses, but we caution on the usage of conventional estimators when extensively comparing the rest-optical and UV $M_{\rm BH}$.

Even when using our rest-UV mass equations with minimized zeropoint and systematic offsets from a large dynamic range, the intrinsic scatter is another issue. Although our $\sigma_{\text{int}}$ of the $M_{\mathrm{BH, Mg_{II}}}$ to the Balmer $M_{\rm BH}$ is relatively small (0.09\,dex), the $\sigma_{\text{int}}$ of the $M_{\mathrm{BH, C_{IV}}}$ to the Balmer $M_{\rm BH}$ (0.40\,dex) is comparable to the systematic uncertainty of the single epoch mass estimator itself (\citealt{Ben13}; \citealt{Woo13}). Moreover, since our estimators take into account the measurement error of $f$--factor, $R-L$ relation, scatter in the UV--optical luminosity and line width relations, the non-negligible fraction of rest-UV masses that deviate from the Balmer masses more than their overlapping measurement errors (13\% of the $M_{\mathrm{BH, Mg_{II}}}$ and 41\% of the $M_{\mathrm{BH, C_{IV}}}$) suggest that one needs to be cautious about the usage of the {\ion{C}{4}} line when deriving $M_{\rm BH}$ values.

\section{Discussion}
\subsection{Reliability of single-epoch AGN black hole masses}
Does the relations between continuum and line luminosities, and the FWHM relations that hold up to the highest luminosity AGNs, guarantee the accurate mass estimation of the most luminous AGNs? Unfortunately the answer is not clear yet, since the $R_{\mathrm{BLR,H\beta}}$--$L_{5100}$ relation is not observationally probed for $L_{5100}$\,$>$\,10$^{46}$\,ergs\,s$^{-1}$ \citep{Ben13}. A hint to estimate the high luminosity end behavior of the optical $R$--$L$ relation is to look at the $R_{\mathrm{BLR,C_{IV}}}$--$L_{1350}$ relation, where the $R_{\mathrm{BLR,C_{IV}}}$ traces the inner part to the H$\beta$ line region. Although the current number of {\ion{C}{4}} reverberation measurements is small (\citealt{Kas07}, \citealt{Slu11}, \citealt{Che12}), they cover up to luminosities of $L_{1350}=10^{47.0}$\,ergs\,s$^{-1}$ and $L_{5100}=10^{46.8}$\,ergs\,s$^{-1}$. The slope of the relation in \citet{Kas07}, (0.52--0.55)$\pm0.04$, is within the uncertainty to the slope of the $R_{\mathrm{BLR,H\beta}}$--$L_{5100}$ relation \citep{Ben13}, $0.533\pm0.034$. This suggests that the ratio of $R_{\mathrm{BLR,H\beta}}$-to-$R_{\mathrm{BLR,C_{IV}}}$ is nearly a constant at 2-4 over $L_{5100}$\,$<$\,10$^{46}$\,ergs\,s$^{-1}$, where the proportionality of $L_{5100}$ and $L_{1350}$ suggests that it is likely so at a higher luminosity range. Therefore, we do not expect a break in the luminous end optical $R$--$L$ relation unless the optical data will somehow fail to form a simple relation at high luminosity where the current UV relation holds. Future compilations of both the {\ion{C}{4}} and H$\beta$ broad line region sizes will help to better constrain the luminous end $R$--$L$ relation.

\begin{figure*}
\begin{center}$
\begin{array}{ccc}
\vspace*{-0.4cm} 
\includegraphics[scale=.659]{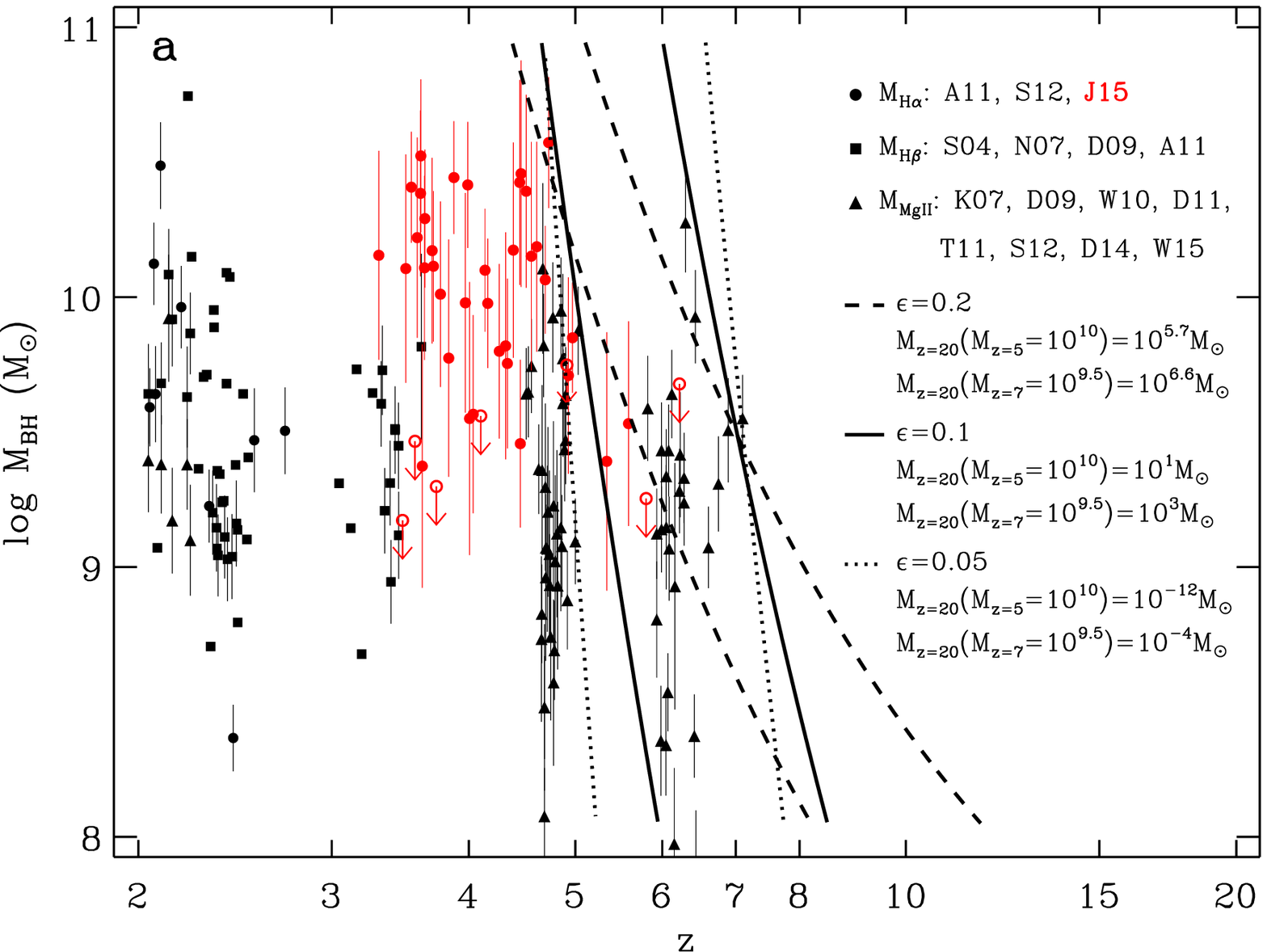} & \hspace*{-0.15cm}
\includegraphics[trim=0 5 0 0, scale=.667]{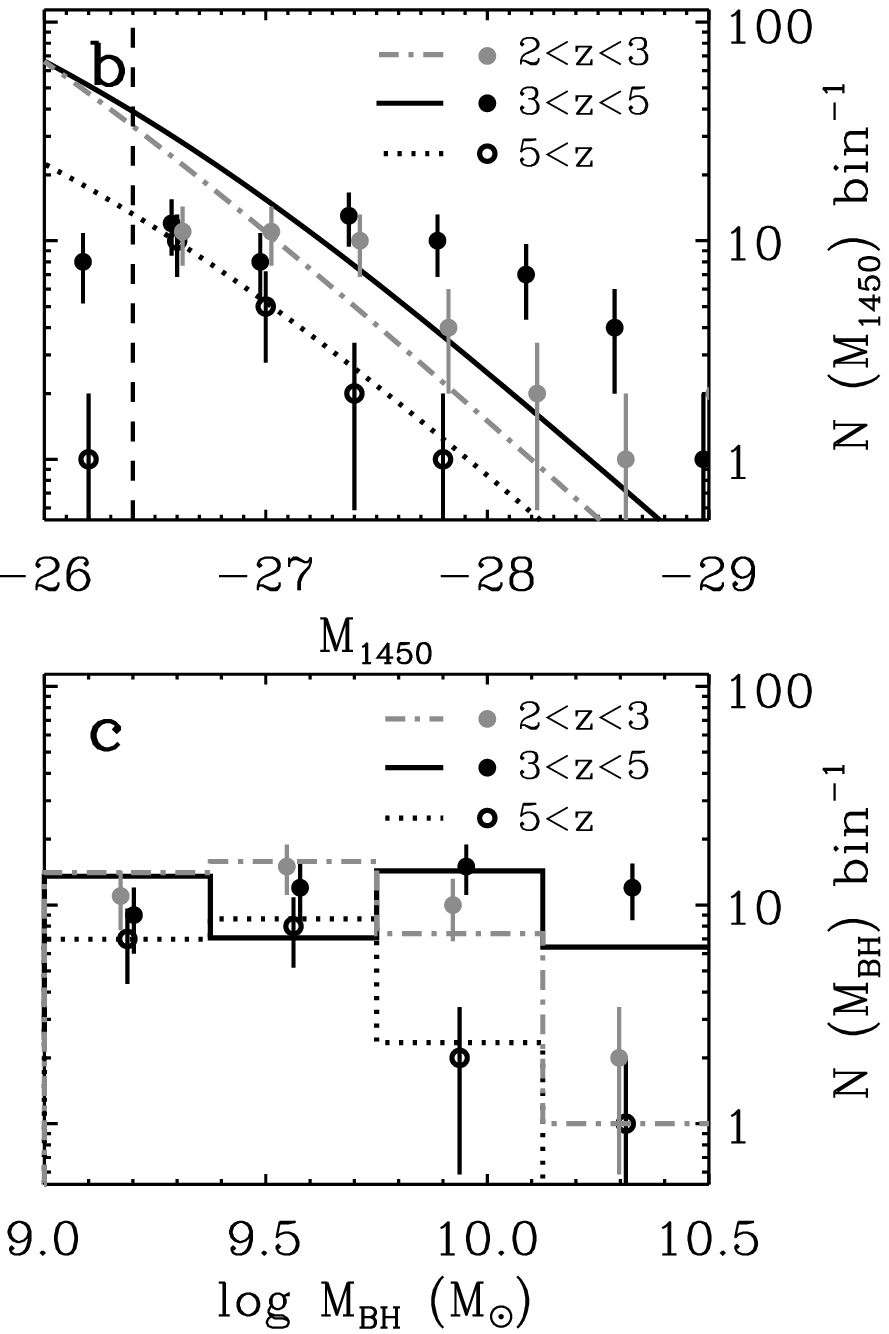}
\end{array}$
\end{center}
\caption{The $M_{\rm BH}$ of AGNs along redshift (left). The $M_{\rm BH}$ from H$\alpha$ are in circles, H$\beta$ in squares, and {\ion{Mg}{2}} in triangles, while our $AKARI$ data points are in red circles. For objects with multiple lines, we choose the H$\alpha$ over H$\beta$, and H$\beta$ over {\ion{Mg}{2}}, while for repeated measurements of an object we used the latest results. Most of the references abbreviated on the plot are summarized in Table 4, with S04, K07, W10, D11, T11, D14, W15 additionally from \citet{She04}, \citet{Kur07}, \citet{Wil10a}, \citet{DeR11}, \citet{Tra11}, \citet{DeR14}, and \citet{Wu15}. We do not plot the error for the S04 data as they are not available, and do not use a part of the T11 data with poor quality (flagged ``3''). The model tracks of exponential BH growth that matches the massive envelope of AGNs at $z\sim5$ and 7 are plotted, with each set of parameters describing the curve shown. On the right, we plot the histogram of $M_{1450}$ (top) and $M_{\rm BH}$ (bottom) at each denoted redshift bin in circles, with errors determined through Poisson statistics. The luminosity functions scaled to the observed number counts to yield the same number at $M_{1450}<-26.4$ at each redshift (top), and the resultant mass counts from the luminosity scaling and a FWHM\,$>$\,2500 km\,s$^{-1}$ cut (bottom), are overplotted in lines with styles marked next to the redshift.} 
\end{figure*}

Even if we find it plausible to assume that the $R_{\mathrm{BLR,H\beta}}$--$L_{5100}$ relation does not change its slope in the luminous end, the large scatter in between the Balmer and rest-UV $M_{\rm BH}$ imposes further limitations on the usage of UV $M_{\rm BH}$ estimators. As discussed in the introduction, it has been known that there is a large $\sigma_{\text{int}}$ in the {\ion{C}{4}} and Balmer $M_{\rm BH}$ relation in comparison to that of the {\ion{Mg}{2}} and Balmer. Our comparison of $M_{\mathrm{BH, H\alpha}}$ against $M_{\mathrm{BH, C_{IV}}}$ for high luminosity AGNs at $z>3.3$ further shows consistent results. Indeed, we find 4 out of 11 $M_{\mathrm{BH, C_{IV}}}$ of $AKARI$ quasars in Figure 15f to be scattered to the $M_{\mathrm{BH, H\alpha}}$ by an order of magnitude or larger, therefore the use of {\ion{C}{4}} masses can be questioned up to the highest redshifts even when considering the additional uncertainties in the $AKARI$ $M_{\mathrm{BH, H\alpha}}$ tested in Figure 5. Overall, our result supports that the $M_{\rm BH}$ estimate from {\ion{C}{4}} contains a rather large uncertainty at any redshift or luminosity.

The greatest uncertainties when calibrating the UV $M_{\rm BH}$ estimators in this work ($\sigma_{\text{int}}$\,$>$\,0.1 dex), come from the scatter in the $L_{5100}$--$L_{1350}$ and FWHM$_{\mathrm{H}\beta}$--FWHM$_{\mathrm{C_{IV}}}$ relations. Since a variety of obscuration in the rest-UV continuum of quasars may result in a scattered $L_{5100}$--$L_{1350}$ relation, we suggest to take into account the level of obscuration when establishing the $M_{\mathrm{BH, C_{IV}}}$ estimator, although careful treatment is required as it is controversial whether the color correction into the $M_{\rm BH}$ works to reduce the scatter between the {\ion{C}{4}} and Balmer masses (\citealt{Ass11}; S12). Meanwhile, the scatter in the FWHM relations which is the largest for the FWHM$_{\mathrm{H}\beta}$--FWHM$_{\mathrm{C_{IV}}}$ relation, could arise from broad absorption features or a non-reverberating component of the {\ion{C}{4}} broad emission \citep{Den12}, where future reverberation mapping will further enable us to test. Without detailed analysis on the origin of the scatter in between the Balmer and rest-UV $M_{\rm BH}$, we caution on the individual measurement of $M_{\rm BH}$ from UV, especially the one using  {\ion{C}{4}}.

\subsection{On the massive end black hole mass evolution}
Having cross-calibrated the $M_{\rm BH}$ estimators, we plot the Balmer and {\ion{Mg}{2}} $M_{\rm BH}$ measurements of quasars along redshift in Figure 16a, from our $AKARI$ observations and the literature indicated on the figure. First of all, we notice that the most massive envelope that stays at $\sim$\,10$^{10}M_{\odot}$ up to $z\sim5$, starts to disappear at higher redshift. To quantify how significant the $M_{\rm BH}$'s are evolving at the massive end for $z$\,$>$\,5 quasars, we performed the Kolmogorov--Smirnov (K--S) test to compare the mass distribution at above and below $z$\,=\,5, at $M_{\mathrm{BH}}>2\times10^{9}M_{\odot}$. This mass threshold for completeness is governed by the shallow $AKARI$ data, through the $L_{5100}\gtrsim10^{46.5}$\,ergs\,s$^{-1}$ limit at S/N$_{\mathrm{H}\alpha}$\,=\,3 and the FWHM limit of 2500\,km\,s$^{-1}$ from section 3.1. The K--S probability $p_{KS}=0.34\%$ turns out to be meaningfully small, supporting the massive end BHs at $z$\,$<$\,5 to be heavier than at $z$\,$>$\,5. We checked our results to be unchanged much by the inclusion of $AKARI$ data that could bias the masses by up to $\sim$\,0.1\,dex (section 3.1), by performing the same test while giving a $-$0.1\,dex correction to the $AKARI$ data, only to find the trend to be mildly weaker ($p_{KS}=0.64\%$).

We cross-check if this evolution in $M_{\rm BH}$ is also reflected through the $M_{\rm BH}$ histograms. In order to correct for the luminosity selection effect in number counting, we plot the 1450\,$\text{\AA}$ absolute magnitude ($M_{1450}$) histogram in Figure 16b. We overplot the quasar luminosity functions in redshift bins of 2\,$<$\,$z$\,$<$\,3, 3\,$<$\,$z$\,$<$\,5, $z$\,$>$\,5, scaled by a constant so that they match the $M_{1450}$ counts in total number at $M_{1450}<-26.4$. We use only the $M_{1450}<-26.4$ data for better completeness in scaling the luminosities. The break luminosity ($M_{1450}$), the faint end slope ($\alpha$), and the bright end slope ($\beta$) of the luminosity function are fixed to $M_{1450}$\,=\,$-26.39$ at $z$\,$>$\,3 and $M_{1450}$\,=\,$-25.5$ at $z$\,$<$\,3 while ($\alpha, \beta$)=($-1.80, -3.26$) at all redshifts (\citealt{McG13}; \citealt{Ros13}). The set of ($\alpha, \beta$) are fixed as the evolution in the $\alpha$ value does not affect the luminosity regime of interest much, and $\beta$ stays nearly constant with redshift. Overall, we find that the quasar samples at $z$\,$<$\,5 are in excess of luminous objects compared to the luminosity functions. 

Not only there is a luminosity selection effect in the quasars plotted in Figure 16a, there is a FWHM limit for each reference data which is the poorest for the $AKARI$ data. Taking into account of these effects, we plot the $M_{\rm BH}$ histograms in Figure 16c before (points) and after (histograms) applying the luminosity selection correction and a FWHM\,=\,2500\,km\,s$^{-1}$ cut. After correcting the luminosity selection and giving the FWHM cut the $M_{\rm BH}$ distribution is shifted to more 10$^{9}M_{\odot}$ BHs and less 10$^{10}M_{\odot}$ BHs at $z$\,$<$\,5. Still, the number of 10$^{10}M_{\odot}$ BHs start to drop at $z$\,$>$\,5, whereas the 10$^{9}M_{\odot}$ BHs do not exhibit an evolution up to Poisson error. 
Thus, we interpret the $M_{\rm BH}$ histogram that there is an evolution in the number of 10$^{10}M_{\odot}$ BHs to be increasing with time at $z$\,$>$\,5. The $z$\,$>$\,5, $\sim$\,10$^{9}M_{\odot}$ AGNs are likely to be in a rapidly growing state such that their masses can soon reach $\sim$\,10$^{10}M_{\odot}$, consistent with the higher Eddington ratio trend found in distant, luminous AGNs (\citealt{Wil10a}; \citealt{DeR11}). Note that the number density of quasars evolve strongly from $z$\,=\,2 to 6 (e.g., \citealt{Wil10a}). We caution that the mass and luminosity histograms in Figures 16b and 16c do not reflect this number density evolution, and that Figures 16b and 16c are used for comparing the shape of these functions. The fact that the numbers of $z$\,=\,2 through 6 quasars appear similar in these figures is due to the fact that the sampling of quasars is different at each redshift bin.

We further investigate the upper mass envelope of $z$\,$>$\,5 AGNs to check if their extremely high masses can be explained by the earliest BH growth from model seed masses. For this we follow \citet{Vol05} to assume a continuous, Eddington limited accretion of matter to a BH seed at $z$\,=\,20, where the BH will grow with time as
 Equation (1), with $t_{0}=t_{z=20}$ and $M_{0}=M_{z=20}$ for the age of the universe and the seed BH mass, respectively at $z$\,=\,20. 
  For each given $\epsilon=0.05, 0.1, 0.2$, we determined a pair of seed masses $M_{0}$ such that the growth curve reaches the observed massive limit at $z\sim5$ and $z\sim7$ respectively, shown in Figure 16a. 

When $\epsilon=0.05$, the seed masses to explain the most massive BHs at $z\sim5$ and 7 are $M_{0}\sim10^{-12}M_{\odot}$ and $10^{-4}M_{\odot}$ respectively, which in other words, gives plenty of time for seed BHs of any meaningful mass to grow up to the most massive quasars without a strictly continuous, Eddington limited accretion. One can expect to find fully grown massive BHs at $z\sim6$ detectable as high or low Eddington ratio AGNs, if such rapid growth is possible. Luminous AGNs at $z\sim 6$ have high Eddington ratios so far, but future studies may uncover fainter objects with weakly active BHs. If $\epsilon=0.1$ that is roughly consistent with the Soltan argument measurements (\citealt{Sol82}; \citealt{Yu02}), it predicts $M_{0}\sim10^{1-3}M_{\odot}$ which gives a reasonable estimate of the seed mass to be Population III stars. However, it may be difficult for stellar seed BHs to keep its Eddington limited accretion from $z$\,=\,20 to 7--5, which is 0.6--1.0\,Gyrs in duration and is longer than a typical quasar lifetime ($\lesssim$\,0.1\,Gyr, \citealt{Hop05}), perhaps requiring extended period of accretion or BH-BH mergers. Lastly, the $\epsilon=0.2$ model which corresponds to the case when BHs are rotating rapidly (e.g., \citealt{Ker63}; \citealt{Mar04}) only accepts very heavy seed BHs at $M_{0}\sim10^{5.7-6.6}M_{\odot}$, requiring supercritical accretion from lighter seed masses \citep{Vol05} or supports direct collapse of massive primordial gas \citep{Vol08}. With $\epsilon = 0.2$, we expect that the most massive $\sim$10$^{10}\,M_{\odot}$ BHs from $M_{0}\sim10^{6}M_{\odot}$ seeds can only start appearing at the redshifts between 5 and 6.

\section{Summary}
We measured the redshifted H$\alpha$ emission and rest-UV to rest-optical continuum properties of 155 luminous quasars at 
3\,$\lesssim$\,$z$\,$\lesssim$\,6 using the $AKARI$ spectra and other datasets, and estimated their $M_{\rm BH}$. We summarize our findings as the following.\\
1. The $L_{5100}$--$L_{\mathrm{H}\alpha}$ relation holds up to the most luminous quasars ($L_{5100}\sim10^{47}$\,ergs\,s$^{-1}$) 
with a single slope unchanging up to $z\sim6$, suggesting a consistent response of the broad line region to the incident 
continuum irrespective of AGN luminosity and redshift.\\
2. The relations between rest-optical and UV continuum and line luminosities, and the FWHM relations hold up to the highest 
luminosity AGNs. Together with predictions of an extended optical $R$--$L$ relation to the highest luminosities, 
it enables us to calibrate the rest-UV $M_{\rm BH}$ estimators to be consistent with the Balmer masses overall. However, 
some of the rest-UV and optical $M_{\rm BH}$ are scattered more than their uncertainties including the errors from the 
recipe, for only 13\% of {\ion{Mg}{2}} masses but for 41\% of the {\ion{C}{4}} estimation. The {\ion{C}{4}} masses have a 0.40\,dex intrinsic scatter to the Balmer masses, which places negative implications on its reliability.\\
3. The massive end envelope of $M_{\rm BH}$ steeply evolves at $z$\,$>$\,5, suggesting they are in a rapidly growing state 
from given seed masses. The most massive BHs at $z$\,=\,5--7 can be explained by the Eddington limited, continuous accretion onto $\sim10^{1-3}M_{\odot}$ seed masses at $z$\,=\,20 if $\epsilon=0.1$ and the formation redshift is $z$\,=\,20 for the seed BH, while there are a range of viable accretion rates and seed masses if $\epsilon$ is different.
 
We expect future observations to compile sensitive measurements of ($L_{5100}$, $L_{\mathrm{H}\alpha}$) to better identify the outliers in the $L_{5100}$--$L_{\mathrm{H}\alpha}$ relation, where an example would be a small population of weak emission line quasars. Also, future rest-optical reverberation mapping of high luminosity AGNs will verify if the prediction on the $R$--$L$ relation to extend with a single slope will hold. Most importantly, the discovery of the highest redshift quasars will further uncover the evolutionary tracks of the earliest BH growth, also improving the current understanding of $M_{\rm BH}$ growth at $z$\,$>$\,5 from small number statistics.

\acknowledgments
We thank Todd Boroson, Jenny Greene, Lisa Storrie-Lombardi, Celine P{\'e}roux, and Donald Schneider for kindly providing the iron template derived from I Zw 1, the rest-frame optical luminosities and line widths of local AGNs, and the optical spectra of APM-UKST quasars and Q0000--26. Also, we thank Eduardo Ba{\~n}ados, Hyunjin Shim, and Doosoo Yoon for useful communication. This research was supported by an appointment to the NASA Postdoctoral Program at the Jet Propulsion Laboratory, administered by Oak Ridge Associated Universities through a contract with NASA. This work was supported by the National Research Foundation of Korea (NRF) grant, No. 2008-0060544 (H.D.J. and M.I.), 2012R1A2A2A01006087 (J.H.W), 2012R1A4A1028713 (H.M.L. and M.G.L.), and NRF-2014-Fostering Core Leaders of Future Program, No. 2014-009728 (D.K.), funded by the Korea government (MSIP). This work was supported by grant MOST 100-2112-M-001-001-MY3 (Y.O.).

This research is based on observations with $AKARI$, a JAXA project with the participation of ESA. 
Funding for SDSS-III has been provided by the Alfred P. Sloan Foundation, the Participating Institutions, the National Science Foundation, and the U.S. Department of Energy Office of Science. The SDSS-III web site is http://www.sdss3.org/.
SDSS-III is managed by the Astrophysical Research Consortium for the Participating Institutions of the SDSS-III Collaboration including the University of Arizona, the Brazilian Participation Group, Brookhaven National Laboratory, Carnegie Mellon University, University of Florida, the French Participation Group, the German Participation Group, Harvard University, the Instituto de Astrofisica de Canarias, the Michigan State/Notre Dame/JINA Participation Group, Johns Hopkins University, Lawrence Berkeley National Laboratory, Max Planck Institute for Astrophysics, Max Planck Institute for Extraterrestrial Physics, New Mexico State University, New York University, Ohio State University, Pennsylvania State University, University of Portsmouth, Princeton University, the Spanish Participation Group, University of Tokyo, University of Utah, Vanderbilt University, University of Virginia, University of Washington, and Yale University. 
This publication makes use of data products from the Two Micron All Sky Survey, 
which is a joint project of the University of Massachusetts and the Infrared Processing 
and Analysis Center/California Institute of Technology, funded by the National Aeronautics and 
Space Administration and the National Science Foundation.
This publication makes use of data products from the United Kingdom Infrared Deep Sky Survey.
The UKIDSS project is defined in \citet{Law07}. UKIDSS uses the UKIRT Wide Field Camera (WFCAM; \citealt{Cas07}). The photometric system is described in \citet{Hew06}, and the calibration is described in \citet{Hod09}. The pipeline processing and science archive are described in Irwin et al (2009, in prep) and \citet{Ham08}.
This publication makes use of data products from the Wide-field Infrared Survey Explorer, 
which is a joint project of the University of California, Los Angeles, and the Jet Propulsion 
Laboratory/California Institute of Technology, funded by the National Aeronautics and Space Administration.

\clearpage
 \LongTables 
\renewcommand{\tabcolsep}{1.5pt}
\begin{deluxetable*}{ *{10}{c} }
\tablecolumns{10}
\tabletypesize{\scriptsize}
\tablecaption{Continum and Line Based Properties of the Sample}
\tablehead{
\colhead{Name} & \colhead{$z_{\mathrm{ref}}$} & \colhead{$z_{\mathrm{H}\alpha}$} & \colhead{log $L_{1350}$} & \colhead{log $L_{5100}$} & \colhead{log $L_{\mathrm{H}\alpha}$} &  \colhead{FWHM$_{3,\rm C_{IV}}$} & \colhead{FWHM$_{3,\rm H\alpha}$} & \colhead{log $M_{\mathrm{BH,C_{IV}}}$} & \colhead{log $M_{\mathrm{BH,H\alpha}}$}
\\ (1) & (2) & (3) & (4) & (5) & (6) & (7) & (8) & (9) & (10)}
\startdata
SDSS J000239.39+255034.8 & 5.80 & 5.79 & 99.00$\,\pm\,$99.00 & 46.56$\,\pm\,$0.02 & 45.14$\,\pm\,$0.11 & 99.00$\,\pm\,$99.00 & 2.50$\,\pm\,$-1.00 & 99.00$\,\pm\,$99.00 & 9.25$\,\pm\,$-1.00\\
Q 0000-26 & 4.10 & 4.11 & 47.40$\,\pm\,$0.01 & 47.13$\,\pm\,$0.01 & 45.48$\,\pm\,$0.13 & 4.33$\,\pm\,$0.28 & 2.50$\,\pm\,$-1.00 & 9.88$\,\pm\,$0.22 & 9.56$\,\pm\,$-1.00\\
SDSS J000552.34-000655.8 & 5.85 & 5.85 & 99.00$\,\pm\,$99.00 & 46.04$\,\pm\,$0.08 & 44.75$\,\pm\,$0.32 & 99.00$\,\pm\,$99.00 & 99.00$\,\pm\,$99.00 & 99.00$\,\pm\,$99.00 & 99.00$\,\pm\,$99.00\\
BR J0006-6208 & 4.45 & 4.49 & 46.90$\,\pm\,$0.02 & 46.71$\,\pm\,$0.00 & 45.46$\,\pm\,$0.06 & 11.33$\,\pm\,$1.33 & 2.85$\,\pm\,$0.85 & 10.48$\,\pm\,$0.24 & 9.46$\,\pm\,$0.31\\
SDSS J001115.23+144601.8 & 4.97 & 99.00 & 99.00$\,\pm\,$99.00 & 46.78$\,\pm\,$0.02 & 47.97$\,\pm\,$-1.00 & 99.00$\,\pm\,$99.00 & 99.00$\,\pm\,$99.00 & 99.00$\,\pm\,$99.00 & 99.00$\,\pm\,$99.00\\
BR J0018-3527 & 4.15 & 4.15 & 99.00$\,\pm\,$99.00 & 46.72$\,\pm\,$0.01 & 45.39$\,\pm\,$0.29 & 99.00$\,\pm\,$99.00 & 99.00$\,\pm\,$99.00 & 99.00$\,\pm\,$99.00 & 99.00$\,\pm\,$99.00\\
PMN J0022-0759 & 3.90 & 99.00 & 99.00$\,\pm\,$99.00 & 46.57$\,\pm\,$0.01 & 47.97$\,\pm\,$-1.00 & 99.00$\,\pm\,$99.00 & 99.00$\,\pm\,$99.00 & 99.00$\,\pm\,$99.00 & 99.00$\,\pm\,$99.00\\
BR 0019-1522 & 4.53 & 99.00 & 46.95$\,\pm\,$0.02 & 46.58$\,\pm\,$0.02 & 47.92$\,\pm\,$-1.00 & 3.48$\,\pm\,$0.33 & 99.00$\,\pm\,$99.00 & 9.43$\,\pm\,$0.21 & 99.00$\,\pm\,$99.00\\
BR J0030-5129 & 4.17 & 99.00 & 99.00$\,\pm\,$99.00 & 46.55$\,\pm\,$0.01 & 48.02$\,\pm\,$-1.00 & 99.00$\,\pm\,$99.00 & 99.00$\,\pm\,$99.00 & 99.00$\,\pm\,$99.00 & 99.00$\,\pm\,$99.00\\
BRI J0048-2442 & 4.15 & 99.00 & 99.00$\,\pm\,$99.00 & 99.00$\,\pm\,$99.00 & 48.08$\,\pm\,$-1.00 & 99.00$\,\pm\,$99.00 & 99.00$\,\pm\,$99.00 & 99.00$\,\pm\,$99.00 & 99.00$\,\pm\,$99.00\\
BRI J0048-244 & 4.15 & 99.00 & 99.00$\,\pm\,$99.00 & 46.07$\,\pm\,$0.03 & 47.90$\,\pm\,$-1.00 & 99.00$\,\pm\,$99.00 & 99.00$\,\pm\,$99.00 & 99.00$\,\pm\,$99.00 & 99.00$\,\pm\,$99.00\\
SDSS J010619.24+004823.3 & 4.45 & 4.44 & 46.99$\,\pm\,$0.01 & 46.80$\,\pm\,$0.02 & 45.69$\,\pm\,$0.19 & 3.00$\,\pm\,$0.20 & 7.75$\,\pm\,$2.92 & 9.32$\,\pm\,$0.11 & 10.43$\,\pm\,$0.38\\
BRI J0113-280 & 4.30 & 99.00 & 99.00$\,\pm\,$99.00 & 46.38$\,\pm\,$0.01 & 47.92$\,\pm\,$-1.00 & 99.00$\,\pm\,$99.00 & 99.00$\,\pm\,$99.00 & 99.00$\,\pm\,$99.00 & 99.00$\,\pm\,$99.00\\
SDSS J011351.96-093551.1 & 3.67 & 3.68 & 46.93$\,\pm\,$0.03 & 46.54$\,\pm\,$0.01 & 47.65$\,\pm\,$-1.00 & 8.64$\,\pm\,$0.36 & 99.00$\,\pm\,$99.00 & 10.25$\,\pm\,$0.22 & 99.00$\,\pm\,$99.00\\
SDSS J011453.81-103934.0 & 3.73 & 99.00 & 46.65$\,\pm\,$0.12 & 46.62$\,\pm\,$0.01 & 47.69$\,\pm\,$-1.00 & 5.11$\,\pm\,$0.35 & 99.00$\,\pm\,$99.00 & 9.62$\,\pm\,$0.22 & 99.00$\,\pm\,$99.00\\
SDSS J012211.11+150914.3 & 4.46 & 99.00 & 47.00$\,\pm\,$0.04 & 46.58$\,\pm\,$0.02 & 47.91$\,\pm\,$-1.00 & 7.59$\,\pm\,$0.82 & 99.00$\,\pm\,$99.00 & 10.17$\,\pm\,$0.23 & 99.00$\,\pm\,$99.00\\
SDSS J012403.77+004432.7 & 3.83 & 3.83 & 47.11$\,\pm\,$0.01 & 46.83$\,\pm\,$0.01 & 45.44$\,\pm\,$0.20 & 8.06$\,\pm\,$0.29 & 3.76$\,\pm\,$1.69 & 10.29$\,\pm\,$0.15 & 9.77$\,\pm\,$0.44\\
SDSS J012700.68-004559.1 & 4.08 & 4.07 & 47.01$\,\pm\,$0.01 & 46.68$\,\pm\,$0.01 & 45.14$\,\pm\,$0.33 & 6.96$\,\pm\,$0.27 & 99.00$\,\pm\,$99.00 & 10.10$\,\pm\,$0.13 & 99.00$\,\pm\,$99.00\\
BRI J0137-422 & 3.97 & 3.98 & 99.00$\,\pm\,$99.00 & 46.42$\,\pm\,$0.01 & 45.49$\,\pm\,$0.22 & 99.00$\,\pm\,$99.00 & 5.95$\,\pm\,$2.46 & 99.00$\,\pm\,$99.00 & 9.98$\,\pm\,$0.41\\
SDSS J014049.18-083942.5 & 3.71 & 3.69 & 47.24$\,\pm\,$0.01 & 46.96$\,\pm\,$0.01 & 45.51$\,\pm\,$0.12 & 4.84$\,\pm\,$0.07 & 5.05$\,\pm\,$1.27 & 9.89$\,\pm\,$0.21 & 10.11$\,\pm\,$0.28\\
SDSS J015048.83+004126.2 & 3.70 & 3.70 & 46.95$\,\pm\,$0.01 & 46.64$\,\pm\,$0.01 & 45.42$\,\pm\,$0.17 & 4.14$\,\pm\,$0.34 & 6.47$\,\pm\,$2.18 & 9.58$\,\pm\,$0.11 & 10.17$\,\pm\,$0.35\\
SDSS J021646.94-092107.3 & 3.72 & 99.00 & 47.01$\,\pm\,$0.01 & 46.74$\,\pm\,$0.01 & 45.47$\,\pm\,$0.11 & 9.87$\,\pm\,$0.96 & 99.00$\,\pm\,$99.00 & 10.42$\,\pm\,$0.18 & 99.00$\,\pm\,$99.00\\
SDSS J023137.65-072854.4 & 5.41 & 99.00 & 46.66$\,\pm\,$0.07 & 46.59$\,\pm\,$0.02 & 48.20$\,\pm\,$-1.00 & 3.03$\,\pm\,$0.33 & 99.00$\,\pm\,$99.00 & 9.15$\,\pm\,$0.22 & 99.00$\,\pm\,$99.00\\
BR J0234-1806 & 4.31 & 99.00 & 99.00$\,\pm\,$99.00 & 46.54$\,\pm\,$0.01 & 47.99$\,\pm\,$-1.00 & 99.00$\,\pm\,$99.00 & 99.00$\,\pm\,$99.00 & 99.00$\,\pm\,$99.00 & 99.00$\,\pm\,$99.00\\
SDSS J024447.79-081606.0 & 4.07 & 99.00 & 47.19$\,\pm\,$0.02 & 46.94$\,\pm\,$0.01 & 47.96$\,\pm\,$-1.00 & 9.76$\,\pm\,$0.54 & 99.00$\,\pm\,$99.00 & 10.51$\,\pm\,$0.23 & 99.00$\,\pm\,$99.00\\
BR J0301-5537 & 4.13 & 99.00 & 99.00$\,\pm\,$99.00 & 46.41$\,\pm\,$0.01 & 47.86$\,\pm\,$-1.00 & 99.00$\,\pm\,$99.00 & 99.00$\,\pm\,$99.00 & 99.00$\,\pm\,$99.00 & 99.00$\,\pm\,$99.00\\
BR J0307-4945 & 4.73 & 4.78 & 47.47$\,\pm\,$0.01 & 47.12$\,\pm\,$0.01 & 45.79$\,\pm\,$0.11 & 6.40$\,\pm\,$0.86 & 7.59$\,\pm\,$1.51 & 10.27$\,\pm\,$0.25 & 10.57$\,\pm\,$0.24\\
BR J0324-2918 & 4.62 & 4.60 & 99.00$\,\pm\,$99.00 & 46.94$\,\pm\,$0.01 & 45.29$\,\pm\,$0.27 & 99.00$\,\pm\,$99.00 & 99.00$\,\pm\,$99.00 & 99.00$\,\pm\,$99.00 & 99.00$\,\pm\,$99.00\\
SDSS J034402.85-065300.6 & 3.96 & 99.00 & 46.98$\,\pm\,$0.02 & 46.44$\,\pm\,$0.01 & 47.83$\,\pm\,$-1.00 & 7.23$\,\pm\,$0.24 & 99.00$\,\pm\,$99.00 & 10.11$\,\pm\,$0.21 & 99.00$\,\pm\,$99.00\\
BR J0426-2202 & 4.32 & 4.33 & 47.13$\,\pm\,$0.01 & 46.72$\,\pm\,$0.01 & 45.46$\,\pm\,$0.20 & 5.55$\,\pm\,$1.16 & 4.21$\,\pm\,$1.80 & 9.96$\,\pm\,$0.28 & 9.82$\,\pm\,$0.42\\
BR J0525-3343 & 4.38 & 99.00 & 47.01$\,\pm\,$0.02 & 46.84$\,\pm\,$0.01 & 47.89$\,\pm\,$-1.00 & 11.23$\,\pm\,$2.14 & 99.00$\,\pm\,$99.00 & 10.54$\,\pm\,$0.28 & 99.00$\,\pm\,$99.00\\
BR J0529-3526 & 4.41 & 4.41 & 99.00$\,\pm\,$99.00 & 46.64$\,\pm\,$0.01 & 45.35$\,\pm\,$0.33 & 99.00$\,\pm\,$99.00 & 99.00$\,\pm\,$99.00 & 99.00$\,\pm\,$99.00 & 99.00$\,\pm\,$99.00\\
BR J0714-6455 & 4.46 & 4.44 & 47.32$\,\pm\,$0.01 & 46.84$\,\pm\,$0.01 & 45.45$\,\pm\,$0.23 & 11.42$\,\pm\,$0.33 & 7.85$\,\pm\,$3.33 & 10.72$\,\pm\,$0.23 & 10.46$\,\pm\,$0.42\\
SDSS J073149.51+285448.7 & 3.68 & 3.69 & 46.96$\,\pm\,$0.02 & 46.50$\,\pm\,$0.01 & 45.13$\,\pm\,$0.28 & 7.08$\,\pm\,$0.63 & 99.00$\,\pm\,$99.00 & 10.09$\,\pm\,$0.22 & 99.00$\,\pm\,$99.00\\
SDSS J075303.33+423130.8 & 3.59 & 3.56 & 47.23$\,\pm\,$0.01 & 46.79$\,\pm\,$0.01 & 45.32$\,\pm\,$0.19 & 2.93$\,\pm\,$0.08 & 6.24$\,\pm\,$2.29 & 9.42$\,\pm\,$0.10 & 10.22$\,\pm\,$0.37\\
SDSS J080430.57+542041.1 & 3.76 & 3.75 & 46.99$\,\pm\,$0.01 & 46.74$\,\pm\,$0.01 & 45.29$\,\pm\,$0.24 & 7.13$\,\pm\,$0.32 & 99.00$\,\pm\,$99.00 & 10.11$\,\pm\,$0.15 & 99.00$\,\pm\,$99.00\\
SDSS J080849.43+521515.3 & 4.46 & 99.00 & 47.16$\,\pm\,$0.01 & 46.81$\,\pm\,$0.01 & 47.91$\,\pm\,$-1.00 & 8.41$\,\pm\,$0.31 & 99.00$\,\pm\,$99.00 & 10.35$\,\pm\,$0.13 & 99.00$\,\pm\,$99.00\\
SDSS J081754.52+413225.3 & 3.54 & 99.00 & 46.59$\,\pm\,$0.02 & 46.40$\,\pm\,$0.01 & 47.67$\,\pm\,$-1.00 & 4.94$\,\pm\,$0.12 & 99.00$\,\pm\,$99.00 & 9.55$\,\pm\,$0.11 & 99.00$\,\pm\,$99.00\\
SDSS J081827.40+172251.8 & 6.00 & 99.00 & 99.00$\,\pm\,$99.00 & 46.74$\,\pm\,$0.12 & 48.40$\,\pm\,$-1.00 & 99.00$\,\pm\,$99.00 & 99.00$\,\pm\,$99.00 & 99.00$\,\pm\,$99.00 & 99.00$\,\pm\,$99.00\\
SDSS J083118.52+424728.8 & 3.32 & 99.00 & 46.58$\,\pm\,$0.04 & 46.27$\,\pm\,$0.01 & 47.58$\,\pm\,$-1.00 & 7.21$\,\pm\,$0.48 & 99.00$\,\pm\,$99.00 & 9.90$\,\pm\,$0.16 & 99.00$\,\pm\,$99.00\\
SDSS J083700.82+350550.2 & 3.31 & 3.30 & 46.87$\,\pm\,$0.02 & 46.62$\,\pm\,$0.01 & 45.36$\,\pm\,$0.19 & 4.03$\,\pm\,$0.17 & 6.42$\,\pm\,$2.50 & 9.52$\,\pm\,$0.12 & 10.16$\,\pm\,$0.39\\
SDSS J083946.22+511202.8 & 4.39 & 4.39 & 46.84$\,\pm\,$0.03 & 46.71$\,\pm\,$0.01 & 45.61$\,\pm\,$0.21 & 5.78$\,\pm\,$0.21 & 6.22$\,\pm\,$2.50 & 9.83$\,\pm\,$0.15 & 10.17$\,\pm\,$0.40\\
SDSS J084119.52+290504.4 & 5.96 & 99.00 & 99.00$\,\pm\,$99.00 & 46.18$\,\pm\,$0.07 & 48.35$\,\pm\,$-1.00 & 99.00$\,\pm\,$99.00 & 99.00$\,\pm\,$99.00 & 99.00$\,\pm\,$99.00 & 99.00$\,\pm\,$99.00\\
SDSS J085837.95+052141.7 & 3.53 & 99.00 & 46.84$\,\pm\,$0.03 & 46.53$\,\pm\,$0.01 & 47.62$\,\pm\,$-1.00 & 6.82$\,\pm\,$0.68 & 99.00$\,\pm\,$99.00 & 9.98$\,\pm\,$0.22 & 99.00$\,\pm\,$99.00\\
87GB 090153.2+694215 & 5.47 & 99.00 & 99.00$\,\pm\,$99.00 & 46.26$\,\pm\,$0.03 & 48.11$\,\pm\,$-1.00 & 99.00$\,\pm\,$99.00 & 99.00$\,\pm\,$99.00 & 99.00$\,\pm\,$99.00 & 99.00$\,\pm\,$99.00\\
SDSS J090634.84+023433.8 & 4.51 & 99.00 & 47.03$\,\pm\,$0.02 & 46.73$\,\pm\,$0.02 & 47.99$\,\pm\,$-1.00 & 8.95$\,\pm\,$1.18 & 99.00$\,\pm\,$99.00 & 10.34$\,\pm\,$0.19 & 99.00$\,\pm\,$99.00\\
SDSS J092721.82+200123.7 & 5.77 & 99.00 & 99.00$\,\pm\,$99.00 & 46.20$\,\pm\,$0.06 & 48.32$\,\pm\,$-1.00 & 99.00$\,\pm\,$99.00 & 99.00$\,\pm\,$99.00 & 99.00$\,\pm\,$99.00 & 99.00$\,\pm\,$99.00\\
SDSS J093554.45+525616.5 & 4.01 & 3.99 & 46.90$\,\pm\,$0.02 & 46.78$\,\pm\,$0.01 & 45.43$\,\pm\,$0.22 & 5.74$\,\pm\,$0.37 & 3.04$\,\pm\,$1.59 & 9.86$\,\pm\,$0.21 & 9.55$\,\pm\,$0.51\\
BR 0951-0450 & 4.37 & 99.00 & 46.71$\,\pm\,$0.03 & 46.46$\,\pm\,$0.02 & 48.03$\,\pm\,$-1.00 & 5.64$\,\pm\,$0.60 & 99.00$\,\pm\,$99.00 & 9.74$\,\pm\,$0.22 & 99.00$\,\pm\,$99.00\\
SDSS J095511.32+594030.7 & 4.34 & 4.30 & 47.03$\,\pm\,$0.02 & 46.81$\,\pm\,$0.01 & 45.72$\,\pm\,$0.14 & 4.61$\,\pm\,$0.26 & 3.71$\,\pm\,$1.11 & 9.73$\,\pm\,$0.15 & 9.75$\,\pm\,$0.32\\
SDSS J095744.46+330820.7 & 4.23 & 99.00 & 47.17$\,\pm\,$0.02 & 46.67$\,\pm\,$0.01 & 48.05$\,\pm\,$-1.00 & 9.19$\,\pm\,$0.67 & 99.00$\,\pm\,$99.00 & 10.44$\,\pm\,$0.23 & 99.00$\,\pm\,$99.00\\
QUEST J101046.3-013104.1 & 5.09 & 99.00 & 99.00$\,\pm\,$99.00 & 47.70$\,\pm\,$0.01 & 48.02$\,\pm\,$-1.00 & 99.00$\,\pm\,$99.00 & 99.00$\,\pm\,$99.00 & 99.00$\,\pm\,$99.00 & 99.00$\,\pm\,$99.00\\
SDSS J101336.37+561536.4 & 3.63 & 3.64 & 46.95$\,\pm\,$0.03 & 46.99$\,\pm\,$0.09 & 45.58$\,\pm\,$0.14 & 6.90$\,\pm\,$0.91 & 6.65$\,\pm\,$1.88 & 10.06$\,\pm\,$0.24 & 10.38$\,\pm\,$0.31\\
SDSS J102622.89+471907.0 & 4.94 & 99.00 & 99.00$\,\pm\,$99.00 & 46.64$\,\pm\,$0.01 & 47.97$\,\pm\,$-1.00 & 99.00$\,\pm\,$99.00 & 99.00$\,\pm\,$99.00 & 99.00$\,\pm\,$99.00 & 99.00$\,\pm\,$99.00\\
SDSS J103221.11+092749.0 & 4.00 & 4.02 & 47.02$\,\pm\,$0.01 & 46.55$\,\pm\,$0.02 & 45.31$\,\pm\,$0.33 & 7.58$\,\pm\,$0.43 & 99.00$\,\pm\,$99.00 & 10.18$\,\pm\,$0.13 & 99.00$\,\pm\,$99.00\\
SDSS J103242.71+433605.4 & 3.46 & 99.00 & 46.50$\,\pm\,$0.07 & 46.40$\,\pm\,$0.01 & 47.62$\,\pm\,$-1.00 & 5.97$\,\pm\,$0.19 & 99.00$\,\pm\,$99.00 & 9.68$\,\pm\,$0.15 & 99.00$\,\pm\,$99.00\\
SDSS J104351.20+650647.6 & 4.47 & 99.00 & 46.85$\,\pm\,$0.05 & 46.46$\,\pm\,$0.01 & 47.91$\,\pm\,$-1.00 & 4.07$\,\pm\,$1.14 & 99.00$\,\pm\,$99.00 & 9.52$\,\pm\,$0.32 & 99.00$\,\pm\,$99.00\\
SDSS J104433.04-012502.2 & 5.80 & 99.00 & 99.00$\,\pm\,$99.00 & 46.51$\,\pm\,$0.03 & 48.35$\,\pm\,$-1.00 & 99.00$\,\pm\,$99.00 & 99.00$\,\pm\,$99.00 & 99.00$\,\pm\,$99.00 & 99.00$\,\pm\,$99.00\\
SDSS J104437.06+650645.0 & 3.65 & 99.00 & 46.88$\,\pm\,$0.03 & 46.60$\,\pm\,$0.01 & 47.81$\,\pm\,$-1.00 & 4.25$\,\pm\,$0.23 & 99.00$\,\pm\,$99.00 & 9.57$\,\pm\,$0.20 & 99.00$\,\pm\,$99.00\\
SDSS J104845.05+463718.3 & 6.19 & 99.00 & 99.00$\,\pm\,$99.00 & 46.55$\,\pm\,$0.03 & 48.50$\,\pm\,$-1.00 & 99.00$\,\pm\,$99.00 & 99.00$\,\pm\,$99.00 & 99.00$\,\pm\,$99.00 & 99.00$\,\pm\,$99.00\\
SDSS J105123.03+354534.3 & 4.91 & 4.91 & 46.86$\,\pm\,$0.05 & 46.70$\,\pm\,$0.01 & 45.52$\,\pm\,$0.18 & 3.35$\,\pm\,$0.17 & 3.94$\,\pm\,$-1.00 & 9.35$\,\pm\,$0.20 & 9.75$\,\pm\,$-1.00\\
SDSS J105756.28+455553.0 & 4.14 & 4.13 & 47.43$\,\pm\,$0.02 & 47.24$\,\pm\,$0.10 & 45.90$\,\pm\,$0.09 & 6.93$\,\pm\,$0.25 & 4.22$\,\pm\,$0.70 & 10.33$\,\pm\,$0.22 & 10.10$\,\pm\,$0.23\\
SDSS J110657.83+081643.3 & 4.27 & 99.00 & 46.79$\,\pm\,$0.02 & 46.45$\,\pm\,$0.03 & 47.95$\,\pm\,$-1.00 & 4.85$\,\pm\,$0.35 & 99.00$\,\pm\,$99.00 & 9.65$\,\pm\,$0.15 & 99.00$\,\pm\,$99.00\\
SDSS J111348.13+540940.0 & 3.78 & 3.77 & 99.00$\,\pm\,$99.00 & 46.57$\,\pm\,$0.01 & 45.12$\,\pm\,$0.30 & 99.00$\,\pm\,$99.00 & 99.00$\,\pm\,$99.00 & 99.00$\,\pm\,$99.00 & 99.00$\,\pm\,$99.00\\
BRI 1114-0822 & 4.49 & 99.00 & 46.58$\,\pm\,$0.03 & 46.19$\,\pm\,$0.02 & 48.07$\,\pm\,$-1.00 & 6.06$\,\pm\,$0.12 & 99.00$\,\pm\,$99.00 & 9.74$\,\pm\,$0.20 & 99.00$\,\pm\,$99.00\\
SDSS J113002.35+115438.3 & 3.40 & 99.00 & 46.91$\,\pm\,$0.02 & 46.49$\,\pm\,$0.02 & 47.64$\,\pm\,$-1.00 & 6.25$\,\pm\,$0.08 & 99.00$\,\pm\,$99.00 & 9.94$\,\pm\,$0.14 & 99.00$\,\pm\,$99.00\\
SDSS J113142.31+384854.6 & 4.13 & 99.00 & 99.00$\,\pm\,$99.00 & 46.74$\,\pm\,$0.01 & 47.94$\,\pm\,$-1.00 & 99.00$\,\pm\,$99.00 & 99.00$\,\pm\,$99.00 & 99.00$\,\pm\,$99.00 & 99.00$\,\pm\,$99.00\\
SDSS J113246.50+120901.6 & 5.17 & 99.00 & 99.00$\,\pm\,$99.00 & 46.52$\,\pm\,$0.03 & 48.10$\,\pm\,$-1.00 & 99.00$\,\pm\,$99.00 & 99.00$\,\pm\,$99.00 & 99.00$\,\pm\,$99.00 & 99.00$\,\pm\,$99.00\\
SDSS J113307.63+522835.5 & 3.74 & 3.74 & 46.71$\,\pm\,$0.04 & 46.64$\,\pm\,$0.01 & 45.33$\,\pm\,$0.20 & 7.14$\,\pm\,$0.37 & 2.50$\,\pm\,$-1.00 & 9.95$\,\pm\,$0.21 & 9.30$\,\pm\,$-1.00\\
SDSS J113418.96+574204.6 & 3.52 & 99.00 & 46.98$\,\pm\,$0.03 & 46.61$\,\pm\,$0.01 & 47.67$\,\pm\,$-1.00 & 7.90$\,\pm\,$0.84 & 99.00$\,\pm\,$99.00 & 10.20$\,\pm\,$0.23 & 99.00$\,\pm\,$99.00\\
SDSS J114514.18+394715.9 & 4.06 & 4.03 & 47.02$\,\pm\,$0.02 & 46.81$\,\pm\,$0.01 & 45.27$\,\pm\,$0.32 & 7.51$\,\pm\,$0.34 & 99.00$\,\pm\,$99.00 & 10.17$\,\pm\,$0.15 & 99.00$\,\pm\,$99.00\\
SDSS J114657.79+403708.6 & 5.01 & 99.00 & 46.54$\,\pm\,$0.08 & 46.57$\,\pm\,$0.01 & 48.07$\,\pm\,$-1.00 & 4.43$\,\pm\,$0.71 & 99.00$\,\pm\,$99.00 & 9.43$\,\pm\,$0.25 & 99.00$\,\pm\,$99.00\\
SDSS J114816.64+525150.2 & 6.42 & 99.00 & 99.00$\,\pm\,$99.00 & 46.70$\,\pm\,$0.02 & 48.19$\,\pm\,$-1.00 & 99.00$\,\pm\,$99.00 & 99.00$\,\pm\,$99.00 & 99.00$\,\pm\,$99.00 & 99.00$\,\pm\,$99.00\\
SDSS J115757.96+485655.7 & 4.25 & 99.00 & 46.72$\,\pm\,$0.05 & 46.47$\,\pm\,$0.01 & 47.93$\,\pm\,$-1.00 & 4.72$\,\pm\,$0.30 & 99.00$\,\pm\,$99.00 & 9.58$\,\pm\,$0.21 & 99.00$\,\pm\,$99.00\\
SDSS J115935.63+042420.0 & 3.45 & 99.00 & 46.53$\,\pm\,$0.04 & 46.12$\,\pm\,$0.03 & 47.62$\,\pm\,$-1.00 & 4.10$\,\pm\,$0.36 & 99.00$\,\pm\,$99.00 & 9.35$\,\pm\,$0.16 & 99.00$\,\pm\,$99.00\\
SDSS J120110.31+211758.5 & 4.58 & 99.00 & 47.00$\,\pm\,$0.04 & 46.63$\,\pm\,$0.01 & 48.03$\,\pm\,$-1.00 & 12.09$\,\pm\,$0.48 & 99.00$\,\pm\,$99.00 & 10.60$\,\pm\,$0.22 & 99.00$\,\pm\,$99.00\\
SDSS J120207.78+323538.8 & 5.29 & 5.27 & 47.05$\,\pm\,$0.03 & 46.54$\,\pm\,$0.02 & 45.35$\,\pm\,$0.35 & 10.06$\,\pm\,$1.83 & 99.00$\,\pm\,$99.00 & 10.46$\,\pm\,$0.28 & 99.00$\,\pm\,$99.00\\
SDSS J120441.71-002149.5 & 5.03 & 99.00 & 46.48$\,\pm\,$0.07 & 46.45$\,\pm\,$0.03 & 47.99$\,\pm\,$-1.00 & 5.71$\,\pm\,$1.75 & 99.00$\,\pm\,$99.00 & 9.63$\,\pm\,$0.34 & 99.00$\,\pm\,$99.00\\
SDSS J120447.15+330938.7 & 3.62 & 3.64 & 46.93$\,\pm\,$0.15 & 46.97$\,\pm\,$0.01 & 45.65$\,\pm\,$0.13 & 8.08$\,\pm\,$0.28 & 7.82$\,\pm\,$2.01 & 10.18$\,\pm\,$0.17 & 10.52$\,\pm\,$0.28\\
BR 1202-0725 & 4.69 & 4.69 & 47.32$\,\pm\,$0.02 & 46.96$\,\pm\,$0.01 & 45.56$\,\pm\,$0.07 & 9.44$\,\pm\,$0.45 & 4.78$\,\pm\,$0.65 & 10.55$\,\pm\,$0.23 & 10.06$\,\pm\,$0.20\\
SDSS J120934.54+553745.4 & 3.57 & 3.58 & 47.13$\,\pm\,$0.02 & 46.96$\,\pm\,$0.07 & 45.38$\,\pm\,$0.14 & 3.77$\,\pm\,$0.99 & 2.50$\,\pm\,$-1.00 & 9.60$\,\pm\,$0.31 & 9.47$\,\pm\,$-1.00\\
SDSS J122738.30+572748.9 & 3.99 & 99.00 & 99.00$\,\pm\,$99.00 & 46.79$\,\pm\,$0.01 & 47.93$\,\pm\,$-1.00 & 99.00$\,\pm\,$99.00 & 99.00$\,\pm\,$99.00 & 99.00$\,\pm\,$99.00 & 99.00$\,\pm\,$99.00\\
SDSS J123239.29+525250.9 & 4.29 & 4.27 & 47.08$\,\pm\,$0.02 & 46.74$\,\pm\,$0.01 & 45.38$\,\pm\,$0.28 & 7.71$\,\pm\,$0.19 & 99.00$\,\pm\,$99.00 & 10.23$\,\pm\,$0.21 & 99.00$\,\pm\,$99.00\\
SDSS J124306.55+530522.0 & 3.57 & 3.55 & 46.97$\,\pm\,$0.03 & 46.53$\,\pm\,$0.01 & 44.95$\,\pm\,$0.35 & 6.65$\,\pm\,$1.02 & 99.00$\,\pm\,$99.00 & 10.04$\,\pm\,$0.25 & 99.00$\,\pm\,$99.00\\
SDSS J125051.93+313021.9 & 6.13 & 99.00 & 99.00$\,\pm\,$99.00 & 46.46$\,\pm\,$0.01 & 48.45$\,\pm\,$-1.00 & 99.00$\,\pm\,$99.00 & 99.00$\,\pm\,$99.00 & 99.00$\,\pm\,$99.00 & 99.00$\,\pm\,$99.00\\
SDSS J125051.93+313021.9 & 6.13 & 6.09 & 99.00$\,\pm\,$99.00 & 46.46$\,\pm\,$0.01 & 44.84$\,\pm\,$0.34 & 99.00$\,\pm\,$99.00 & 99.00$\,\pm\,$99.00 & 99.00$\,\pm\,$99.00 & 99.00$\,\pm\,$99.00\\
SDSS J130002.16+011823.0 & 4.61 & 99.00 & 46.88$\,\pm\,$0.04 & 46.67$\,\pm\,$0.02 & 48.02$\,\pm\,$-1.00 & 5.44$\,\pm\,$0.87 & 99.00$\,\pm\,$99.00 & 9.80$\,\pm\,$0.21 & 99.00$\,\pm\,$99.00\\
SDSS J130348.94+002010.4 & 3.65 & 3.62 & 46.74$\,\pm\,$0.01 & 46.73$\,\pm\,$0.01 & 45.57$\,\pm\,$0.11 & 2.67$\,\pm\,$0.12 & 6.98$\,\pm\,$1.56 & 9.54$\,\pm\,$0.08 & 10.29$\,\pm\,$0.26\\
SDSS J130608.26+035626.3 & 5.99 & 99.00 & 99.00$\,\pm\,$99.00 & 46.36$\,\pm\,$0.11 & 48.47$\,\pm\,$-1.00 & 99.00$\,\pm\,$99.00 & 99.00$\,\pm\,$99.00 & 99.00$\,\pm\,$99.00 & 99.00$\,\pm\,$99.00\\
SDSS J131914.20+520200.0 & 3.90 & 3.93 & 47.22$\,\pm\,$0.02 & 46.85$\,\pm\,$0.01 & 45.41$\,\pm\,$0.25 & 4.65$\,\pm\,$0.20 & 99.00$\,\pm\,$99.00 & 9.84$\,\pm\,$0.21 & 99.00$\,\pm\,$99.00\\
SDSS J132420.83+422554.6 & 4.04 & 4.01 & 46.85$\,\pm\,$0.04 & 46.65$\,\pm\,$0.01 & 45.47$\,\pm\,$0.16 & 6.76$\,\pm\,$1.00 & 3.33$\,\pm\,$1.22 & 9.99$\,\pm\,$0.25 & 9.57$\,\pm\,$0.37\\
SDSS J132423.26+623342.1 & 3.63 & 99.00 & 46.48$\,\pm\,$0.06 & 46.35$\,\pm\,$0.01 & 47.68$\,\pm\,$-1.00 & 5.69$\,\pm\,$0.13 & 99.00$\,\pm\,$99.00 & 9.63$\,\pm\,$0.20 & 99.00$\,\pm\,$99.00\\
SDSS J132603.00+295758.1 & 3.77 & 99.00 & 46.93$\,\pm\,$0.03 & 46.23$\,\pm\,$0.02 & 47.72$\,\pm\,$-1.00 & 7.30$\,\pm\,$0.69 & 99.00$\,\pm\,$99.00 & 10.10$\,\pm\,$0.23 & 99.00$\,\pm\,$99.00\\
SDSS J133223.26+503431.3 & 3.81 & 99.00 & 47.01$\,\pm\,$0.02 & 46.56$\,\pm\,$0.01 & 47.72$\,\pm\,$-1.00 & 6.89$\,\pm\,$1.06 & 99.00$\,\pm\,$99.00 & 10.09$\,\pm\,$0.25 & 99.00$\,\pm\,$99.00\\
SDSS J133412.56+122020.7 & 5.13 & 99.00 & 46.68$\,\pm\,$0.06 & 46.36$\,\pm\,$0.03 & 48.16$\,\pm\,$-1.00 & 4.38$\,\pm\,$0.97 & 99.00$\,\pm\,$99.00 & 9.49$\,\pm\,$0.28 & 99.00$\,\pm\,$99.00\\
SDSS J133448.70+521317.9 & 3.61 & 3.60 & 46.71$\,\pm\,$0.05 & 46.51$\,\pm\,$0.01 & 45.17$\,\pm\,$0.25 & 3.68$\,\pm\,$0.25 & 99.00$\,\pm\,$99.00 & 9.35$\,\pm\,$0.20 & 99.00$\,\pm\,$99.00\\
SDSS J133529.45+410125.9 & 4.26 & 4.29 & 47.21$\,\pm\,$0.01 & 46.82$\,\pm\,$0.01 & 45.78$\,\pm\,$0.15 & 6.45$\,\pm\,$1.72 & 3.89$\,\pm\,$1.21 & 10.14$\,\pm\,$0.29 & 9.80$\,\pm\,$0.32\\
BRI 1335-0417 & 4.40 & 99.00 & 46.81$\,\pm\,$0.02 & 46.44$\,\pm\,$0.02 & 48.08$\,\pm\,$-1.00 & 10.33$\,\pm\,$0.29 & 99.00$\,\pm\,$99.00 & 10.35$\,\pm\,$0.22 & 99.00$\,\pm\,$99.00\\
SDSS J134015.03+392630.7 & 5.03 & 5.05 & 46.68$\,\pm\,$0.06 & 46.48$\,\pm\,$0.02 & 45.15$\,\pm\,$0.30 & 8.94$\,\pm\,$0.53 & 99.00$\,\pm\,$99.00 & 10.15$\,\pm\,$0.22 & 99.00$\,\pm\,$99.00\\
SDSS J134040.24+281328.1 & 5.34 & 5.38 & 99.00$\,\pm\,$99.00 & 46.54$\,\pm\,$0.02 & 45.56$\,\pm\,$0.23 & 99.00$\,\pm\,$99.00 & 2.94$\,\pm\,$1.45 & 99.00$\,\pm\,$99.00 & 9.39$\,\pm\,$0.48\\
SDSS J134743.29+495621.3 & 4.51 & 4.53 & 47.39$\,\pm\,$0.02 & 46.97$\,\pm\,$0.10 & 45.51$\,\pm\,$0.19 & 7.13$\,\pm\,$1.63 & 6.78$\,\pm\,$2.34 & 10.33$\,\pm\,$0.30 & 10.39$\,\pm\,$0.36\\
BRI 1346-0322 & 3.99 & 3.99 & 46.64$\,\pm\,$0.03 & 46.82$\,\pm\,$0.01 & 45.26$\,\pm\,$0.16 & 4.77$\,\pm\,$0.08 & 99.00$\,\pm\,$99.00 & 9.55$\,\pm\,$0.19 & 99.00$\,\pm\,$99.00\\
SDSS J140132.76+411150.4 & 4.02 & 99.00 & 46.48$\,\pm\,$0.04 & 46.51$\,\pm\,$0.01 & 47.88$\,\pm\,$-1.00 & 5.49$\,\pm\,$1.62 & 99.00$\,\pm\,$99.00 & 9.59$\,\pm\,$0.30 & 99.00$\,\pm\,$99.00\\
SDSS J140146.53+024434.7 & 4.44 & 99.00 & 47.01$\,\pm\,$0.02 & 46.65$\,\pm\,$0.02 & 48.00$\,\pm\,$-1.00 & 6.37$\,\pm\,$2.69 & 99.00$\,\pm\,$99.00 & 10.00$\,\pm\,$0.33 & 99.00$\,\pm\,$99.00\\
SDSS J140248.08+014634.0 & 4.16 & 99.00 & 47.09$\,\pm\,$0.03 & 46.62$\,\pm\,$0.01 & 48.07$\,\pm\,$-1.00 & 8.11$\,\pm\,$1.47 & 99.00$\,\pm\,$99.00 & 10.28$\,\pm\,$0.27 & 99.00$\,\pm\,$99.00\\
SDSS J141831.70+444937.5 & 4.31 & 4.33 & 47.04$\,\pm\,$0.02 & 46.54$\,\pm\,$0.01 & 45.32$\,\pm\,$0.29 & 6.10$\,\pm\,$0.39 & 99.00$\,\pm\,$99.00 & 10.00$\,\pm\,$0.16 & 99.00$\,\pm\,$99.00\\
SDSS J142144.98+351315.4 & 4.56 & 99.00 & 46.87$\,\pm\,$0.05 & 46.54$\,\pm\,$0.01 & 47.88$\,\pm\,$-1.00 & 6.98$\,\pm\,$1.33 & 99.00$\,\pm\,$99.00 & 10.03$\,\pm\,$0.27 & 99.00$\,\pm\,$99.00\\
SDSS J142242.13+461310.2 & 3.73 & 99.00 & 46.98$\,\pm\,$0.03 & 46.50$\,\pm\,$0.00 & 47.79$\,\pm\,$-1.00 & 7.76$\,\pm\,$0.57 & 99.00$\,\pm\,$99.00 & 10.18$\,\pm\,$0.22 & 99.00$\,\pm\,$99.00\\
SDSS J142243.02+441721.4 & 3.55 & 3.61 & 47.05$\,\pm\,$0.01 & 47.18$\,\pm\,$0.05 & 45.71$\,\pm\,$0.07 & 11.87$\,\pm\,$0.53 & 6.11$\,\pm\,$0.84 & 10.61$\,\pm\,$0.16 & 10.41$\,\pm\,$0.21\\
SDSS J143835.95+431459.2 & 4.61 & 4.67 & 47.35$\,\pm\,$0.01 & 47.14$\,\pm\,$0.05 & 45.39$\,\pm\,$0.20 & 9.84$\,\pm\,$0.42 & 4.92$\,\pm\,$1.89 & 10.60$\,\pm\,$0.14 & 10.19$\,\pm\,$0.39\\
SDSS J144144.76+472003.2 & 3.63 & 3.62 & 46.71$\,\pm\,$0.06 & 46.56$\,\pm\,$0.01 & 45.24$\,\pm\,$0.18 & 3.31$\,\pm\,$0.23 & 2.84$\,\pm\,$1.32 & 9.26$\,\pm\,$0.20 & 9.37$\,\pm\,$0.45\\
SDSS J144213.09+391856.0 & 3.63 & 99.00 & 47.01$\,\pm\,$0.02 & 46.43$\,\pm\,$0.01 & 47.58$\,\pm\,$-1.00 & 7.53$\,\pm\,$0.76 & 99.00$\,\pm\,$99.00 & 10.17$\,\pm\,$0.17 & 99.00$\,\pm\,$99.00\\
SDSS J144340.70+585653.2 & 4.28 & 99.00 & 47.24$\,\pm\,$0.02 & 46.75$\,\pm\,$0.10 & 47.96$\,\pm\,$-1.00 & 9.19$\,\pm\,$0.23 & 99.00$\,\pm\,$99.00 & 10.48$\,\pm\,$0.22 & 99.00$\,\pm\,$99.00\\
SDSS J144350.66+362315.1 & 5.27 & 99.00 & 46.78$\,\pm\,$0.03 & 46.58$\,\pm\,$0.01 & 48.08$\,\pm\,$-1.00 & 9.46$\,\pm\,$1.49 & 99.00$\,\pm\,$99.00 & 10.25$\,\pm\,$0.21 & 99.00$\,\pm\,$99.00\\
SDSS J144542.75+490248.9 & 3.88 & 3.87 & 47.30$\,\pm\,$0.02 & 47.12$\,\pm\,$0.01 & 45.94$\,\pm\,$0.07 & 3.14$\,\pm\,$0.12 & 6.57$\,\pm\,$0.97 & 9.52$\,\pm\,$0.20 & 10.44$\,\pm\,$0.21\\
SDSS J144733.46+465723.6 & 3.59 & 99.00 & 46.59$\,\pm\,$0.06 & 46.29$\,\pm\,$0.01 & 47.58$\,\pm\,$-1.00 & 4.79$\,\pm\,$0.90 & 99.00$\,\pm\,$99.00 & 9.53$\,\pm\,$0.26 & 99.00$\,\pm\,$99.00\\
SDSS J145408.95+511443.7 & 3.64 & 3.61 & 47.22$\,\pm\,$0.02 & 47.08$\,\pm\,$0.08 & 45.46$\,\pm\,$0.16 & 4.64$\,\pm\,$0.27 & 4.68$\,\pm\,$1.52 & 9.84$\,\pm\,$0.21 & 10.11$\,\pm\,$0.34\\
SDSS J150620.48+460642.4 & 3.50 & 3.52 & 46.84$\,\pm\,$0.02 & 46.38$\,\pm\,$0.01 & 45.21$\,\pm\,$0.22 & 7.21$\,\pm\,$0.21 & 6.98$\,\pm\,$3.02 & 10.03$\,\pm\,$0.15 & 10.11$\,\pm\,$0.42\\
SDSS J150654.54+522004.6 & 4.07 & 99.00 & 47.11$\,\pm\,$0.02 & 46.92$\,\pm\,$0.01 & 47.16$\,\pm\,$-1.00 & 9.24$\,\pm\,$1.43 & 99.00$\,\pm\,$99.00 & 10.41$\,\pm\,$0.21 & 99.00$\,\pm\,$99.00\\
SDSS J151035.29+514841.0 & 5.03 & 5.02 & 99.00$\,\pm\,$99.00 & 46.47$\,\pm\,$0.01 & 45.27$\,\pm\,$0.35 & 99.00$\,\pm\,$99.00 & 99.00$\,\pm\,$99.00 & 99.00$\,\pm\,$99.00 & 99.00$\,\pm\,$99.00\\
SDSS J151442.74+532412.0 & 3.57 & 99.00 & 46.65$\,\pm\,$0.06 & 46.42$\,\pm\,$0.01 & 47.58$\,\pm\,$-1.00 & 5.77$\,\pm\,$0.11 & 99.00$\,\pm\,$99.00 & 9.73$\,\pm\,$0.20 & 99.00$\,\pm\,$99.00\\
SDSS J152034.53+383906.5 & 3.40 & 99.00 & 46.47$\,\pm\,$0.05 & 46.11$\,\pm\,$0.01 & 47.46$\,\pm\,$-1.00 & 6.92$\,\pm\,$0.46 & 99.00$\,\pm\,$99.00 & 9.79$\,\pm\,$0.15 & 99.00$\,\pm\,$99.00\\
SDSS J152413.35+430537.4 & 3.93 & 3.95 & 46.83$\,\pm\,$0.02 & 46.58$\,\pm\,$0.01 & 45.23$\,\pm\,$0.29 & 5.07$\,\pm\,$0.89 & 99.00$\,\pm\,$99.00 & 9.71$\,\pm\,$0.21 & 99.00$\,\pm\,$99.00\\
SDSS J153650.25+500810.3 & 4.93 & 4.92 & 47.01$\,\pm\,$0.07 & 46.84$\,\pm\,$0.01 & 45.49$\,\pm\,$0.17 & 2.65$\,\pm\,$0.65 & 3.49$\,\pm\,$1.25 & 9.21$\,\pm\,$0.30 & 9.71$\,\pm\,$0.37\\
SDSS J153725.35-014650.3 & 3.45 & 99.00 & 46.55$\,\pm\,$0.06 & 46.04$\,\pm\,$0.04 & 47.56$\,\pm\,$-1.00 & 5.62$\,\pm\,$0.15 & 99.00$\,\pm\,$99.00 & 9.65$\,\pm\,$0.20 & 99.00$\,\pm\,$99.00\\
SDSS J153825.74+420933.3 & 3.83 & 99.00 & 46.82$\,\pm\,$0.02 & 46.52$\,\pm\,$0.01 & 47.79$\,\pm\,$-1.00 & 7.78$\,\pm\,$0.60 & 99.00$\,\pm\,$99.00 & 10.10$\,\pm\,$0.16 & 99.00$\,\pm\,$99.00\\
SDSS J154340.38+341744.4 & 4.41 & 99.00 & 47.15$\,\pm\,$0.02 & 47.01$\,\pm\,$0.01 & 48.02$\,\pm\,$-1.00 & 5.16$\,\pm\,$0.15 & 99.00$\,\pm\,$99.00 & 9.90$\,\pm\,$0.15 & 99.00$\,\pm\,$99.00\\
SDSS J154914.45+364822.0 & 3.58 & 3.60 & 46.59$\,\pm\,$0.03 & 46.26$\,\pm\,$0.01 & 44.88$\,\pm\,$0.35 & 6.96$\,\pm\,$0.51 & 99.00$\,\pm\,$99.00 & 9.87$\,\pm\,$0.16 & 99.00$\,\pm\,$99.00\\
SDSS J160254.18+422822.9 & 6.07 & 99.00 & 99.00$\,\pm\,$99.00 & 46.65$\,\pm\,$0.01 & 48.47$\,\pm\,$-1.00 & 99.00$\,\pm\,$99.00 & 99.00$\,\pm\,$99.00 & 99.00$\,\pm\,$99.00 & 99.00$\,\pm\,$99.00\\
SDSS J161140.13+273029.6 & 3.33 & 99.00 & 46.70$\,\pm\,$0.04 & 46.30$\,\pm\,$0.02 & 47.58$\,\pm\,$-1.00 & 4.52$\,\pm\,$0.38 & 99.00$\,\pm\,$99.00 & 9.53$\,\pm\,$0.16 & 99.00$\,\pm\,$99.00\\
SDSS J161425.13+464028.9 & 5.31 & 99.00 & 99.00$\,\pm\,$99.00 & 46.59$\,\pm\,$0.01 & 48.07$\,\pm\,$-1.00 & 99.00$\,\pm\,$99.00 & 99.00$\,\pm\,$99.00 & 99.00$\,\pm\,$99.00 & 99.00$\,\pm\,$99.00\\
SDSS J162100.70+515544.8 & 5.59 & 5.59 & 99.00$\,\pm\,$99.00 & 46.72$\,\pm\,$0.01 & 45.20$\,\pm\,$0.18 & 99.00$\,\pm\,$99.00 & 3.09$\,\pm\,$1.17 & 99.00$\,\pm\,$99.00 & 9.53$\,\pm\,$0.38\\
SDSS J162331.81+311200.5 & 6.22 & 6.23 & 99.00$\,\pm\,$99.00 & 46.39$\,\pm\,$0.04 & 44.83$\,\pm\,$0.21 & 99.00$\,\pm\,$99.00 & 4.37$\,\pm\,$-1.00 & 99.00$\,\pm\,$99.00 & 9.68$\,\pm\,$-1.00\\
SDSS J162520.31+225832.9 & 3.77 & 3.78 & 47.11$\,\pm\,$0.01 & 46.66$\,\pm\,$0.06 & 45.44$\,\pm\,$0.17 & 5.02$\,\pm\,$0.18 & 5.37$\,\pm\,$1.80 & 9.85$\,\pm\,$0.15 & 10.01$\,\pm\,$0.34\\
SDSS J162623.38+484136.4 & 4.89 & 99.00 & 47.09$\,\pm\,$0.05 & 46.69$\,\pm\,$0.01 & 47.92$\,\pm\,$-1.00 & 2.70$\,\pm\,$0.20 & 99.00$\,\pm\,$99.00 & 9.27$\,\pm\,$0.21 & 99.00$\,\pm\,$99.00\\
SDSS J162943.43+391211.4 & 3.91 & 99.00 & 46.71$\,\pm\,$0.03 & 46.32$\,\pm\,$0.01 & 47.93$\,\pm\,$-1.00 & 7.19$\,\pm\,$0.44 & 99.00$\,\pm\,$99.00 & 9.97$\,\pm\,$0.16 & 99.00$\,\pm\,$99.00\\
SDSS J163636.92+315717.0 & 4.56 & 4.57 & 46.95$\,\pm\,$0.03 & 46.55$\,\pm\,$0.03 & 45.57$\,\pm\,$0.23 & 4.90$\,\pm\,$0.85 & 6.66$\,\pm\,$2.87 & 9.74$\,\pm\,$0.21 & 10.15$\,\pm\,$0.42\\
SDSS J163847.42+232716.4 & 3.82 & 3.84 & 46.88$\,\pm\,$0.06 & 46.95$\,\pm\,$0.01 & 45.28$\,\pm\,$0.27 & 5.09$\,\pm\,$0.45 & 99.00$\,\pm\,$99.00 & 9.74$\,\pm\,$0.17 & 99.00$\,\pm\,$99.00\\
SDSS J163950.52+434003.7 & 3.99 & 99.00 & 47.28$\,\pm\,$0.01 & 46.95$\,\pm\,$0.01 & 47.87$\,\pm\,$-1.00 & 11.53$\,\pm\,$0.74 & 99.00$\,\pm\,$99.00 & 10.71$\,\pm\,$0.17 & 99.00$\,\pm\,$99.00\\
SDSS J164248.71+240303.3 & 3.48 & 3.48 & 46.96$\,\pm\,$0.02 & 46.41$\,\pm\,$0.01 & 45.11$\,\pm\,$0.22 & 6.53$\,\pm\,$0.45 & 2.50$\,\pm\,$-1.00 & 10.01$\,\pm\,$0.16 & 9.17$\,\pm\,$-1.00\\
SDSS J165354.61+405402.1 & 4.98 & 4.97 & 47.05$\,\pm\,$0.01 & 46.71$\,\pm\,$0.01 & 45.45$\,\pm\,$0.08 & 6.60$\,\pm\,$0.34 & 4.36$\,\pm\,$0.66 & 10.07$\,\pm\,$0.13 & 9.85$\,\pm\,$0.20\\
SDSS J165436.85+222733.7 & 4.70 & 4.68 & 47.09$\,\pm\,$0.02 & 46.81$\,\pm\,$0.01 & 45.48$\,\pm\,$0.24 & 6.51$\,\pm\,$0.61 & 99.00$\,\pm\,$99.00 & 10.08$\,\pm\,$0.17 & 99.00$\,\pm\,$99.00\\
SDSS J165902.11+270935.1 & 5.31 & 99.00 & 99.00$\,\pm\,$99.00 & 46.58$\,\pm\,$0.02 & 48.04$\,\pm\,$-1.00 & 99.00$\,\pm\,$99.00 & 99.00$\,\pm\,$99.00 & 99.00$\,\pm\,$99.00 & 99.00$\,\pm\,$99.00\\
SDSS J172100.76+601721.0 & 5.80 & 99.00 & 99.00$\,\pm\,$99.00 & 46.75$\,\pm\,$0.01 & 47.81$\,\pm\,$-1.00 & 99.00$\,\pm\,$99.00 & 99.00$\,\pm\,$99.00 & 99.00$\,\pm\,$99.00 & 99.00$\,\pm\,$99.00\\
SDSS J173744.87+582829.5 & 4.92 & 99.00 & 99.00$\,\pm\,$99.00 & 46.42$\,\pm\,$0.01 & 48.07$\,\pm\,$-1.00 & 99.00$\,\pm\,$99.00 & 99.00$\,\pm\,$99.00 & 99.00$\,\pm\,$99.00 & 99.00$\,\pm\,$99.00\\
PSS J1745+6848 & 4.13 & 99.00 & 99.00$\,\pm\,$99.00 & 46.31$\,\pm\,$0.01 & 47.99$\,\pm\,$-1.00 & 99.00$\,\pm\,$99.00 & 99.00$\,\pm\,$99.00 & 99.00$\,\pm\,$99.00 & 99.00$\,\pm\,$99.00\\
RX J1759.4+6638 & 4.32 & 99.00 & 99.00$\,\pm\,$99.00 & 46.22$\,\pm\,$0.01 & 47.96$\,\pm\,$-1.00 & 99.00$\,\pm\,$99.00 & 99.00$\,\pm\,$99.00 & 99.00$\,\pm\,$99.00 & 99.00$\,\pm\,$99.00\\
PSS J1802+5616 & 4.16 & 99.00 & 99.00$\,\pm\,$99.00 & 46.10$\,\pm\,$0.01 & 48.17$\,\pm\,$-1.00 & 99.00$\,\pm\,$99.00 & 99.00$\,\pm\,$99.00 & 99.00$\,\pm\,$99.00 & 99.00$\,\pm\,$99.00\\
BR 2212-1626 & 3.99 & 3.98 & 46.92$\,\pm\,$0.03 & 47.03$\,\pm\,$0.01 & 45.93$\,\pm\,$0.10 & 2.31$\,\pm\,$0.10 & 6.73$\,\pm\,$1.26 & 9.04$\,\pm\,$0.19 & 10.42$\,\pm\,$0.23\\
BRI 2235-0301 & 4.25 & 99.00 & 47.02$\,\pm\,$0.02 & 47.12$\,\pm\,$0.01 & 47.92$\,\pm\,$-1.00 & 10.29$\,\pm\,$0.39 & 99.00$\,\pm\,$99.00 & 10.46$\,\pm\,$0.22 & 99.00$\,\pm\,$99.00\\
SDSS J223841.81-000105.2 & 3.49 & 3.49 & 46.86$\,\pm\,$0.02 & 46.55$\,\pm\,$0.02 & 44.97$\,\pm\,$0.33 & 6.10$\,\pm\,$0.36 & 99.00$\,\pm\,$99.00 & 9.90$\,\pm\,$0.15 & 99.00$\,\pm\,$99.00\\
BR 2248-1242 & 4.16 & 4.14 & 46.66$\,\pm\,$0.07 & 46.81$\,\pm\,$0.01 & 45.82$\,\pm\,$0.10 & 1.58$\,\pm\,$0.07 & 4.74$\,\pm\,$0.96 & 8.55$\,\pm\,$0.19 & 9.98$\,\pm\,$0.24\\
BR J2328-4513 & 4.36 & 99.00 & 99.00$\,\pm\,$99.00 & 46.41$\,\pm\,$0.01 & 47.92$\,\pm\,$-1.00 & 99.00$\,\pm\,$99.00 & 99.00$\,\pm\,$99.00 & 99.00$\,\pm\,$99.00 & 99.00$\,\pm\,$99.00\\
SDSS J233330.17+152538.7 & 3.68 & 3.68 & 46.61$\,\pm\,$0.06 & 46.31$\,\pm\,$0.02 & 45.19$\,\pm\,$0.26 & 6.07$\,\pm\,$0.88 & 99.00$\,\pm\,$99.00 & 9.75$\,\pm\,$0.24 & 99.00$\,\pm\,$99.00\\
BR J2349-3712 & 4.21 & 99.00 & 46.86$\,\pm\,$0.02 & 46.53$\,\pm\,$0.01 & 47.95$\,\pm\,$-1.00 & 5.73$\,\pm\,$0.37 & 99.00$\,\pm\,$99.00 & 9.84$\,\pm\,$0.21 & 99.00$\,\pm\,$99.00
\enddata
\tablecomments{Catalog of the properties derived for the $AKARI$ quasars, sorted by right ascension. 
Column 1: Target name; Column 2: Redshift from references; Column 3: Redshift measured from H$\alpha$; Column 4: 1350\,$\text{\AA}$ luminosity and its uncertainty; Column 5: 5100\,$\text{\AA}$ luminosity and its uncertainty; Column 6: H$\alpha$ luminosity and its uncertainty; Column 7: FWHM of the {\ion{C}{4}} line and its uncertainty; Column 8: FWHM of the H$\alpha$ line and its uncertainty; Column 9: $M_{\rm BH}$ from the {\ion{C}{4}} line and its uncertainty; Column 10: $M_{\rm BH}$ from the H$\alpha$ line and its uncertainty.
The units for $L$, FWHM, and $M_{\rm BH}$ are ergs\,s$^{-1}$, 1000\,km\,s$^{-1}$, and $M_{\odot}$. Columns 9 and 10 are from Equations (10) and (7), respectively. Empty parameters are entered as 99 and upper limits are given with errors of -1.}
\end{deluxetable*}
         \clearpage
\end{document}